\documentclass[a4paper,11pt]{article}
\pdfoutput=1 % if your are submitting a pdflatex (i.e. if you have
\usepackage{jheppub} % for details on the use of the package, please
\usepackage[T1]{fontenc} % if needed

\graphicspath{{figs/}}
\usepackage{epsf,verbatim}
\usepackage{hyperref}
\usepackage{comment}
\usepackage{color}
\usepackage{slashed}
\usepackage{subfigure}
\usepackage{comment}
\usepackage{mdwlist, paralist}
\usepackage{rotating}
\usepackage{bm}
\usepackage{multirow}
\usepackage{cleveref}
\usepackage{enumitem}
\usepackage[section]{placeins}
\usepackage{fontawesome}
\usepackage[normalem]{ulem}

\newcommand{\bear}{\begin{array}}
\newcommand{\ear}{\end{array}}
\newcommand{\beq}{\begin{equation}}
\newcommand{\eeq}{\end{equation}}
\newcommand{\beqa}{\begin{eqnarray}}
\newcommand{\eeqa}{\end{eqnarray}}

\def\OMIT#1{{}}
\newcommand{\lsim}{\mathrel{\rlap{\lower4pt\hbox{\hskip1pt$\sim$}}
    \raise1pt\hbox{$<$}}}         %less than or approx. symbol
\newcommand{\gsim}{\mathrel{\rlap{\lower4pt\hbox{\hskip1pt$\sim$}}
    \raise1pt\hbox{$>$}}}         %greater than or approx. symbol
\newcommand{\bl}{\left}
\newcommand{\br}{\right}

\newcommand{\eg}{\textit{e.g.}\ }
\newcommand{\ie}{\textit{i.e.}\ }

\newcommand{\dd}{\mathrm{d}}

\newcommand{\mO}{\mathcal{O}}
\newcommand{\mH}{\mathcal{H}}

\newcommand{\tot}{\mathrm{tot}}

\newcommand{\mat}{\mathrm{m}}
\newcommand{\dm}{\mathrm{dm}}
\newcommand{\cdm}{\mathrm{cdm}}

\newcommand{\wdm}{\mathrm{wdm}}

\newcommand{\minn}{\mathrm{min}}
\newcommand{\maxx}{\mathrm{max}}

\newcommand{\LC}{\Lambda\mathrm{CDM}}

\newcommand{\si}{\sigma}
\newcommand{\sT}{\sigma_\text{T}}

\newcommand{\sn}{\sigma_n}
\newcommand{\sBn}{\sigma^{\chi B}_n}
\newcommand{\sbar}{\bar{\sigma}}
\newcommand{\sbarB}{\bar{\sigma}^{\chi B}}
\newcommand{\sbarBn}{\bar{\sigma}^{\chi B}_n}
\newcommand{\mx}{m_\chi}
\newcommand{\fx}{f_\chi}
\newcommand{\mB}{m_B}
\newcommand{\Rx}{R_\chi}
\newcommand{\Rxp}{R_\chi^\prime}
\newcommand{\VRMS}{V_\mathrm{RMS}}

\newcommand{\keV}{\mathrm{keV}}

\newcommand{\MeV}{\mathrm{MeV}}
\newcommand{\Mev}{\mathrm{MeV}}
\newcommand{\GeV}{\mathrm{GeV}}

\newcommand{\Mpc}{\mathrm{Mpc}}

\newcommand{\Sec}[1]{Sec.~\ref{#1}}

\newcommand{\App}[1]{App.~\ref{#1}}
\newcommand{\Tab}[1]{Table~\ref{#1}}
\newcommand{\Fig}[1]{Fig.~\ref{#1}}
\newcommand{\Figs}[2]{Figs.~\ref{#1} and \ref{#2}}
\newcommand{\Eq}[1]{Eq.~(\ref{#1})}
\newcommand{\Eqs}[2]{Eqs.~(\ref{#1}) and (\ref{#2})}

\newcommand{\githubmaster}{\href{https://github.com/ManuelBuenAbad/class\_dmb}{\faGithub}}

\newcommand{\ignore}[1]{}

\title{\boldmath Cosmological Constraints on Dark Matter Interactions with Ordinary Matter}

\author[a,b]{Manuel A. Buen-Abad,}
\author[c]{Rouven Essig,}
\author[d]{David McKeen,}
\author[e]{Yi-Ming Zhong}

\affiliation[a]{Department of Physics, Brown University, Providence, RI, 02912, USA}
\affiliation[b]{Dual CP Institute of High Energy Physics, C.P. 28045, Colima, M\'{e}xico}
\affiliation[c]{C. N. Yang Institute for Theoretical Physics, Stony Brook University, Stony Brook, NY
11794, USA}
\affiliation[d]{TRIUMF, 4004 Wesbrook Mall, Vancouver, BC V6T 2A3, Canada}
\affiliation[e]{Kavli Institute for Cosmological Physics, University of Chicago, Chicago, IL 60637, USA}

\emailAdd{manuel\_buen-abad@brown.edu}
\emailAdd{rouven.essig@stonybrook.edu}
\emailAdd{mckeen@triumf.ca}
\emailAdd{ymzhong@kicp.uchicago.edu}

\preprint{YITP-SB-2021-12}

\abstract{Dark matter interactions with electrons or protons during the early Universe  leave imprints on the cosmic microwave background and the matter power spectrum, and can be probed through cosmological and astrophysical observations. These interactions lead to momentum and heat exchange between the ordinary and dark matter components, which in turn results in a transfer of pressure from the ordinary to the dark matter. We explore these interactions using a diverse suite of data: cosmic microwave background anisotropies, baryon acoustic oscillations, the Lyman-$\alpha$ forest, and the abundance of Milky-Way subhalos. We derive constraints using model-independent parameterizations of the dark matter--electron and dark matter--proton interaction cross sections and map these constraints onto concrete dark matter models. Our constraints are complementary to other probes of dark matter interactions with ordinary matter, such as direct detection, big bang nucleosynthesis, various astrophysical systems, and accelerator-based experiments. They exclude sufficiently large cross sections for a large range of dark matter masses, which cannot be accessed by direct-detection experiments due to the overburden from the Earth’s atmosphere or crust. \githubmaster
}

\begin{document}
\maketitle
\flushbottom

\section{Introduction}
\label{sec:intro}

Dark matter makes up about 85\% of the matter in our Universe. Determining its identity and properties is one of the most important physics endeavors of our time. Although all the evidence for dark matter comes from its gravitational interactions with ordinary matter, it is plausible that dark matter also interacts with Standard Model particles with other, non-gravitational, interactions. These dark matter--baryon (DMb) interactions, where the term ``baryons'' refers to both nuclei and electrons in the cosmological parlance, have been intensively searched for in a multitude of direct detection, indirect detection, and collider experiments during the past decades. If DMb interactions are present in the early Universe, they will also leave imprints on the cosmic microwave background (CMB) and matter power spectrum, and hence can be detected through cosmological and astrophysical observations. These cosmological and astrophysical probes are an important step for identifying the nature of dark matter. They are able not only to constrain dark matter over a wide range of dark matter mass, at least for couplings to ordinary matter that are sufficiently large, but also to provide an experimental probe with unique characteristics. They explore dark matter properties on a large scale and at high redshifts, which can, in principle, be different from the current and local dark matter properties that are searched for by direct and indirect-detection experiments.

The impact of the DMb interactions can be succinctly summarized as follows. The dark matter and baryon components of the Universe tend to couple in such a way as to exchange both heat and momentum. On the one hand, the exchange of heat between both tends to cool the baryons, which can lead to recombination occurring earlier than in the absence of DMb interactions, directly impacting the CMB. On the other hand, since the baryons themselves are coupled to the photons and perceive their pressure, they can communicate this pressure to the dark matter. As a result, the formation of structure, which is driven by the clustering of dark matter, can be dampened. This generally leads to a suppression of the matter power spectrum and related observables.

Earlier studies of the effects of DMb interactions on cosmology mostly focus on the case of dark matter--proton interactions~\cite{McDermott:2010pa,Dvorkin:2013cea,Munoz:2015bca,Gluscevic:2017ywp,Xu:2018efh,Slatyer:2018aqg,Boddy:2018wzy,Nadler:2019zrb,Maamari:2020aqz,Becker:2020hzj}. The impact of dark matter--electron interactions on cosmological observables has remained mostly neglected (although see, for example, Refs.~\cite{Wadekar:2019xnf,Ali-Haimoud:2021lka}).  Within a specific particle physics model, the dark matter--proton coupling may be related to the dark matter--electron coupling, and consequently, the scattering cross sections of the two can be similarly linked.  However, there are models where dark matter is electro-philic or proton-phobic, so that constraints on dark matter--proton interactions are not applicable to those for electrons.  Moreover, since electrons are much lighter than protons, the rates for dark matter--proton and dark matter--electron scattering, and the constraints on these rates, can differ markedly, even if the fundamental coupling between the dark matter to proton and to the electron is equal. In particular, for some models (as we will see below), constraints from the dark matter--electron interactions can be stronger than those from the dark matter--proton interactions.

We will investigate the cosmological constraints on dark matter--electron interactions and dark matter--proton interactions separately in the paper. We use up-to-date data from CMB (Planck 2018) and baryon acoustic oscillations (BAO) and place constraints on dark matter--electron and dark matter--proton interactions. We also recast constraints on thermal warm dark matter (WDM) from the Lyman-$\alpha$ forest and the Milky Way subhalos on dark matter with these interactions. We then translate the bounds for two concrete models, dark matter with a dark photon mediator and dark matter with an electric dipole moment, where dark matter couples to both electrons and protons with equal couplings. During the final stages of this work, we learned of Ref.~\cite{Nguyen:2021cnb}, with which we have some overlap. Our main results are in agreement with their results.  Our work is also partially complementary, since we include BAO data in addition to CMB data, consider also constraints from the Lyman-$\alpha$ forest, and use different prescriptions for deriving constraints based on observations of Milky Way subhalos.  We also consider both dark matter scattering off protons and off electrons, mapping our results onto concrete models. In contrast, Ref.~\cite{Nguyen:2021cnb} includes also positive values for the power--velocity dependence of the momentum transfer cross section, which we do not study for reasons discussed briefly in Sec.~\ref{sec:desc}. 

The rest of the paper is organized as follows.  We give the description of the dark matter--baryon interactions in~\Sec{sec:desc} and discuss their impact on various physical observables in~\Sec{sec:obs}. We then show the implementation of the numerical analysis of the interactions in~\Sec{sec:num} and the results in~\Sec{sec:res}. We conclude in~\Sec{sec:concl}. In  \App{appA}, we show how our constraints are affected when using different cosmological or astrophysical datasets, model assumptions, and derivation methods. In \App{appB}, we show detailed derivations for the momentum-transfer cross sections for the Coulomb-like dark matter--ordinary matter interactions.

\section{Modeling Dark Matter--Baryon Interactions in Cosmology}
\label{sec:desc}

We consider a family of cosmological models, which has been dubbed ``DMb''~\cite{Becker:2020hzj}, that extend the standard cosmological model $\LC$ to include elastic scattering between dark matter particles $\chi$ and Standard Model baryons. These $\chi$ particles, with energy density $\omega_\chi$, will in general constitute a fraction $\fx \equiv {\omega_\chi}/{\omega_\dm^\tot} \le100\%$ of the total energy density in the dark matter $\omega_\dm^\tot$, the rest of it being vanilla cold dark matter (CDM) with no relevant interactions other than gravity.\footnote{We note that if the dark matter--baryon interaction is sufficiently large, the interacting component of dark matter could be tightly coupled with baryons as a single fluid through the recombination era.  The behavior of this fluid is similar to that of helium. In this case, the Planck 2015 measurement of the helium fraction restricts the DMb fraction to be $f_\chi \leq 0.6\%$ at 95\%~C.L.~\cite{dePutter:2018xte}.} In a fluid description, the scattering between dark matter particles $\chi$ and baryonic species $B$, $\chi B \to \chi B$, is often characterized by the dark matter mass $m_\chi$ and the momentum transfer cross section $\sigma_{\rm T}^{\chi B},
$\beq
\sigma_{\rm T}^{\chi B} = \int_{-1}^{+1} \dd \cos \theta_* \frac{\dd \sigma}{ \dd \cos \theta_*} (1-\cos \theta_*)\,,
\eeq
where $\theta_*$ is the scattering angle in the center-of-mass frame and ${\dd \sigma}/{ \dd \cos \theta_*}$ is the differential cross section for the scattering process. Instead of focusing on specific dark matter models, we adopt in this paper the following phenomenological description for the momentum transfer cross section,   
\beq\label{eq:sT}
\sT^{\chi B} = \sn^{\chi B} v_\text{rel}^n\, ,
\eeq 
where $v_\text{rel}$ is the magnitude of the relative velocity between the incoming dark matter and baryon (we set $\hbar = c = k_B = 1$ throughout this paper), $\sn^{\chi B}$ is the velocity-stripped momentum transfer cross section, and $n$ is a power for the relative-velocity dependence that encodes the type of interaction under consideration. Using such a phenomenological description allows us to simplify the model implementation and enlarges the applicability of the results when mapping to concrete dark matter models. In~\Sec{sec:num}, we will focus on three choices for $n$, namely $n=0$, $-2$, and $-4$, which naturally arise from contact interactions~\cite{Chen:2002yh}, electric dipole-moment-like interactions~\cite{Sigurdson:2004zp}, and Coulomb-like interactions~\cite{Melchiorri:2007sq}, respectively. In fact, both $n=0$ and $n=-4$ can be obtained from the same particle physics model, involving a dark photon mediator: they correspond to the massive and massless dark photon limits, respectively. We do not consider positive values in this work, since laboratory constraints on such models for low dark matter masses are very strong, and, as we discuss briefly in Sec.~\ref{subsec:vel_distr}, one needs to carefully treat the heat-transfer rate.  More discussions on the mapping between the phenomenological model and the concrete model can be found in \Sec{subsec:specific}.

\subsection{Evolution of cosmological perturbations}
\label{subsec:cosmo}

The presence of DMb interactions, while leaving the cosmological evolution of the background fields unchanged, modifies the equations describing the evolution of the cosmological scalar perturbations. The divergence of the interacting dark matter (baryonic) fluid velocities $\theta_\chi \,(\theta_b)$ receives additional DMb terms associated with momentum exchange~\cite{Ma:1995ey,Dvorkin:2013cea},
\beqa\label{eq:pert_x}
    \dot{\theta}_\cdm & = & - \mH \theta_\cdm + k^2 \psi \ , \\
    \dot{\theta}_\chi & = & - \mH \theta_\chi + k^2 \psi + c_\chi^2 k^2 \delta_\chi + \Rx\bl( \theta_b - \theta_\chi \br) \ , \\
    \dot{\theta}_b & = & - \mH \theta_b + k^2 \psi + c_s^2 k^2 \delta_b + R_\gamma (\theta_\gamma - \theta_b) + S \Rx \bl( \theta_\chi - \theta_b \br) \ ,\label{eq:pert_b}
\eeqa
given in the conformal Newtonian gauge. Here the $\cdm$ and $\gamma$ subindices denote the non-interacting CDM and photon fluids respectively, while the subscripts $\chi$ and $b$ denote the interacting dark matter  and baryon quantities. The dot signifies conformal time $\tau$ derivatives, $a$ is the scale factor, $\mH = \frac{\dot{a}}{a} = a H$ is the conformal Hubble expansion rate, $\delta$ is the perturbations in the energy density of the various fluids, $k$ is the comoving wavenumber of the scalar perturbation, $\psi$ is one of the metric perturbations, $c_{\chi}$ and $c_s$ are the dark matter and baryon sound speeds respectively, and $R_\gamma = \frac{4}{3} \frac{\rho_\gamma}{\rho_b} a n_e \sigma_{\rm Thomson}$ is the usual conformal Thomson scattering momentum exchange rate. In addition, $S \equiv \frac{\rho_\chi}{\rho_b}$ is ratio of dark matter-to-baryon energy densities. $\Rx$ is the conformal DMb momentum transfer rate~\cite{Dvorkin:2013cea,Gluscevic:2017ywp,Xu:2018efh,Slatyer:2018aqg,Boddy:2018wzy},
\beqa\label{eq:Rx}
    \Rx & \equiv & a \sum\limits_B \frac{Y_B \rho_b}{\mx + \mB} \sBn c_n u_B^{n+1} \ ,
\eeqa
where $c_n = \frac{2^{(5+n)/2}}{3 \sqrt{\pi}} \Gamma\bl( 3+\frac{n}{2} \br)$ is a numerical coefficient, the sum is over all the baryon species $B$ with which the dark matter interacts; $Y_B \equiv \frac{\rho_B}{\rho_b}$ is the baryon species mass fraction, $\mB$ its particle mass, $\sBn$ the velocity-stripped momentum transfer cross section between the dark matter and the $B$ species, taken from \Eq{eq:sT}. The parameter $u_B$ is the velocity dispersion, which is approximately given by
\beq
    u_B \equiv \bl( \frac{T_b}{\mB} + \frac{T_\chi}{\mx} + \frac{\langle V_{\rm bulk}^2 \rangle}{3} \br)^{\frac{1}{2}} \ , \label{eq:uB}
\eeq 
where we include contributions from the thermal velocity $\frac{T_b}{\mB} + \frac{T_\chi}{\mx}$ and the average of the squared relative bulk velocity $\langle V_{\rm bulk}^2 \rangle$. The addition of this last term approximately captures the nonlinear effects of the relative bulk velocity on the evolution equations\footnote{An improved treatment of the DMb interactions requires solving for the relative velocity between both fluids with its own evolution equation (see, for example, Refs.~\cite{Munoz:2015bca,Boddy:2018wzy}); however the simplified prescription we use in our paper works rather well as a first approximation (see \App{appA}).} \cite{Boddy:2018wzy}. For the purposes of this work, we treat $\langle V_{\rm bulk}^2 \rangle$ in two different ways. On the one hand, we conservatively take $\langle V_{\rm bulk}^2 \rangle = 0$ for $n=0$, since in this case the ratio of $\Rx$ to $\mH$ rapidly decreases with time. This means that at sufficiently early times the baryonic and dark matter fluids were tightly coupled, and thus their relative bulk velocity was vanishingly small. On the other hand, for $n=-2, -4$ we take:
\beqa\label{eq:vrms2}
    \langle V_{\rm bulk}^2 \rangle = \VRMS^2 & \simeq & 10^{-8} \quad \text{for } z>10^{3} \ , \nonumber\\
     & \simeq & 10^{-8} \bl( \frac{1+z}{1+10^3} \br)^2 \text{for } z\leq 10^3 \ ,
\eeqa
as first suggested in Refs.~\cite{Dvorkin:2013cea,Xu:2018efh}. This expression corresponds roughly to the value of $\langle V_{\rm bulk}^2 \rangle$ in pure $\LC$, where the dark matter and the baryons do not interact. This makes this prescription a good approximation to its exact value for the DMb models with $n=-2$ and $n=-4$, since these models have weak dark matter--baryon interactions at the redshifts most relevant to the datasets considered in this work.\footnote{We thank the authors of Ref.~\cite{Nguyen:2021cnb} for pointing out the impact of these two prescriptions for $\langle V_{\rm bulk}^2 \rangle$ on the various choices of $n$. We checked that our results differ by $\lesssim$20\% for $n=0$ between setting $\langle V_{\rm bulk}^2 \rangle = 0$ and setting $\langle V_{\rm bulk}^2 \rangle$ as prescribed by Eq.~\eqref{eq:vrms2}.} Finally, note that a more refined treatment of the bulk velocity, properly taking into account the non-linear regime of the thermal and cosmological equations, can be found in Refs.~\cite{Munoz:2015bca,Boddy:2018wzy}. This treatment is most relevant for $n < -3$ (for which the momentum exchange rate $\Rx$ increases with time compared to the Hubble expansion rate) and leads to generally smaller bulk velocities and therefore tighter bounds. Therefore, ``the mean-field'' treatment of \Eq{eq:vrms2} is conservative for these values of $n$, and produces looser bounds than those derived assuming the other extreme value of $\langle V_{\rm bulk}^2 \rangle = 0$.

\subsection{Thermal evolution}
\label{subsec:thermal}

The DMb interactions also impact the equations governing the evolution of the temperatures of the baryons and  the dark matter particles through an additional term quantifying the heat transfer between both fluids~\cite{Dvorkin:2013cea,Gluscevic:2017ywp,Xu:2018efh,Slatyer:2018aqg,Boddy:2018wzy}:
\beqa\label{eq:heat_x}
    \dot{T}_\chi & = & - 2 \mH T_\chi + 2 \Rxp \bl( T_b - T_\chi \br) \ , \\
    \dot{T}_b & = & - 2 \mH T_b + 2 \frac{\mu_b}{m_e} R_\gamma \bl( T_\gamma - T_b \br) + 2 S \frac{\mu_b}{\mx} \Rxp \bl( T_\chi - T_b \br) \ ,\label{eq:heat_b}
\eeqa
where the first term in both equations gives the adiabatic cooling of the fluids, $m_e$ is the electron's mass, $\mu_b$ is the mean molecular weight for the baryons, and
\beq\label{eq:Rxp}
    \Rxp \equiv a \sum\limits_B \frac{Y_B \rho_b \mx}{\bl( \mx + \mB \br)^2} \sBn c_n u_B^{n+1}
\eeq
is the DMb conformal heat exchange rate. 

Throughout the rest of this paper we will, for simplicity, assume that the dark matter couples to one baryon species $B$ at a time, either to protons ($p$) or to electrons ($e$).  Tighter bounds are expected in models in which dark matter couples to both electrons and protons, so that the constraints derived in this paper are conservative for such models. Therefore we will henceforth drop the $\chi B$ index from $\sBn$; the baryon species to which the dark matter is coupled should be clear from the context. We will restore it if there is any possibility of confusion.

\subsection{Comments on the velocity distribution of dark matter}
\label{subsec:vel_distr}

In the description above, we  assume the dark matter fluid always follows a thermal Maxwell-Boltzmann distribution characterized by a single $T_\chi$. This assumption is naturally realized when the dark matter particles have sizable elastic self-interactions, which efficiently randomize the velocity of dark matter particles to reach a Maxwell-Boltzmann distribution.\footnote{Dissipative self-interactions, in contrast, may yield a non-Maxwell-Boltzmann distribution with a suppressed high-velocity tail~\cite{Shen:2021frv}. }  If the elastic self-interaction is weak, as pointed out in~\cite{Ali-Haimoud:2018dvo,Ali-Haimoud:2021lka}, the velocity distribution of the dark matter will unavoidably depart from a Maxwell-Boltzmann distribution once the dark matter fluid decouples from the baryonic fluid.  This 
departure will affect the heat exchange between the dark matter fluid and the baryonic fluid. For $n =-2$ and $n=0$ interactions, the departure yields less than $\mathcal{O}(6\%)$ level corrections in the heat exchange~\cite{Ali-Haimoud:2018dvo}, by comparing the heat exchanges computed in the Maxwell-Boltzmann distribution and those from the Fokker-Planck equation, for baryon-to-dark matter mass ratios $\leq 10^3$. Therefore, for $n=-2$ and $n=0$ type DMb interactions, we regard treating the dark matter fluid as following a Maxwell-Boltzmann distribution as a good approximation  for most of the parameter region of interest, \ie, $m_\chi \gtrsim 1\, \Mev$ for the dark matter--proton interactions and $m_\chi \gtrsim 1 \,\keV$ for the dark matter--electron interactions. For larger $n$ values, such as $n=+2, +4$, the difference on the heat exchange between assuming a Maxwell-Boltzmann distribution and a more careful treatment can be as large as a factor of $\sim 2 - 3$~\cite{Ali-Haimoud:2018dvo,Ali-Haimoud:2021lka}. This significant difference demands a full treatment of the evolution of the velocity distribution, unless there is an additional sizable elastic self-interaction, distinct from the DMb interactions, that keeps the velocity distribution Maxwell-Boltzmann-like. For many realistic particle dark matter scenarios, such a sizable elastic dark matter self-interaction is expected if the dark matter--ordinary matter interaction is in turn sizable.

For $n=-4$ type DMb interactions, dark matter first free streams and then couples to the baryonic fluid. The initial dark matter velocity distribution, depending on the production mechanism, can be different from the Maxwell-Boltzmann distribution. For dark matter with non-thermal velocity distributions, sizable self-interactions or interactions with other thermalized particles can help bring dark matter towards thermal equilibrium. Ref.~\cite{Dvorkin:2019zdi,Dvorkin:2020xga}, for example, studied a freeze-in dark matter model where dark matter particles are non-thermally produced through in-medium photon decays. The produced dark matter particles then experience Coulomb-like interaction with baryons through its millicharge. Ref.~\cite{Dvorkin:2020xga} explicitly compares the constraints on the velocity-stripped cross section for dark matter with non-thermal velocity distribution to those with thermal velocity distribution from the CMB. Using the Planck 2018 data, the resulting constraints with the non-thermal velocity distribution are $\mathcal{O}(5-10\%)$ stronger than those with thermal velocity distribution for dark matter with mass $\mathcal{O}(10\,\keV)$. By comparing the interaction rate with the Hubble rate, Ref.~\cite{Dvorkin:2019zdi} concludes that the dark matter--baryon interaction itself is not significant enough to thermalize sub-MeV dark matter before recombination if the millicharge of dark matter is smaller than $10^{-10}$. The bounds we obtain in this work constrain millicharges above this value for most of the parameter space, and therefore the assumption of thermalization of dark matter via its interactions with baryons is warranted. In any case,  sizable self-interactions due to the millicharge or from another interaction can thermalize sub-MeV dark matter before recombination while evading constraints from cluster crossings ($\sigma_{{\rm T}}^{\chi \chi}/m_\chi\lesssim 1\,\text{cm}^2/\text{g}$~\cite{Tulin:2017ara}). Given that the thermalization condition for $n=-4$ type DMb interaction is model-dependent, we will simply assume its velocity is already in a Maxwell-Boltzmann distribution when it interacts with the baryonic fluid.

\section{Impact of Dark Matter--Baryon Interactions on Observables}
\label{sec:obs}

As described in the previous section, the DMb interactions impact the evolution of both the temperatures and the perturbations of the dark matter and the baryons. These in turn affect the various physical observables in non-trivial ways. We now turn to a brief qualitative review of these effects on some of these observables, namely the matter power spectrum, the CMB lensing spectrum, and the temperature (TT) and polarization (EE) CMB spectra. We do this by comparing the percent residuals of the DMb model predictions for these observables with respect to $\LC$ as a fiducial reference. 

For the purposes of this section, we fix the six baseline cosmological parameters, namely $\{\omega_{\rm b}, \omega_\dm^\tot, H_0, z_{\rm reio}, A_{\rm s}, n_{\rm s} \}$ (the baryon density $\omega_{\rm b}$, the total dark matter density $\omega_\dm^\tot$, the Hubble parameter $H_0$, the redshift of reionization $z_{\rm reio}$, the amplitude of scalar perturbations $A_{\rm s}$, and the spectral index $n_{\rm s}$) to their mean values within the $\LC$ model's fit to Planck 2018 TT+TE+EE+lowE+lensing + BAO data~\cite{Aghanim:2018eyx}, and take $f_\chi=100\%$ of dark matter to be interacting with the baryons. Note that the CMB+BAO analysis is also sensitive to the case for which only a sub-dominant component of dark matter interacts with baryons ($f_\chi<1$), which we will investigate in detail in~\Sec{sec:num}. For illustrative purposes of this section, we focus on two dark matter masses, $\mx = 0.01~\MeV$ and $1~\MeV$, and fix $\sn$ to the 95\% C.L. values corresponding to those choices of dark matter mass, obtained from our numerical analysis in \Sec{sec:num} and shown in \Sec{sec:res}.

\subsection{Matter and lensing power spectra}
\label{subsec:mps}

\begin{figure}[t]
  \centering
  \includegraphics[width=0.49\textwidth]{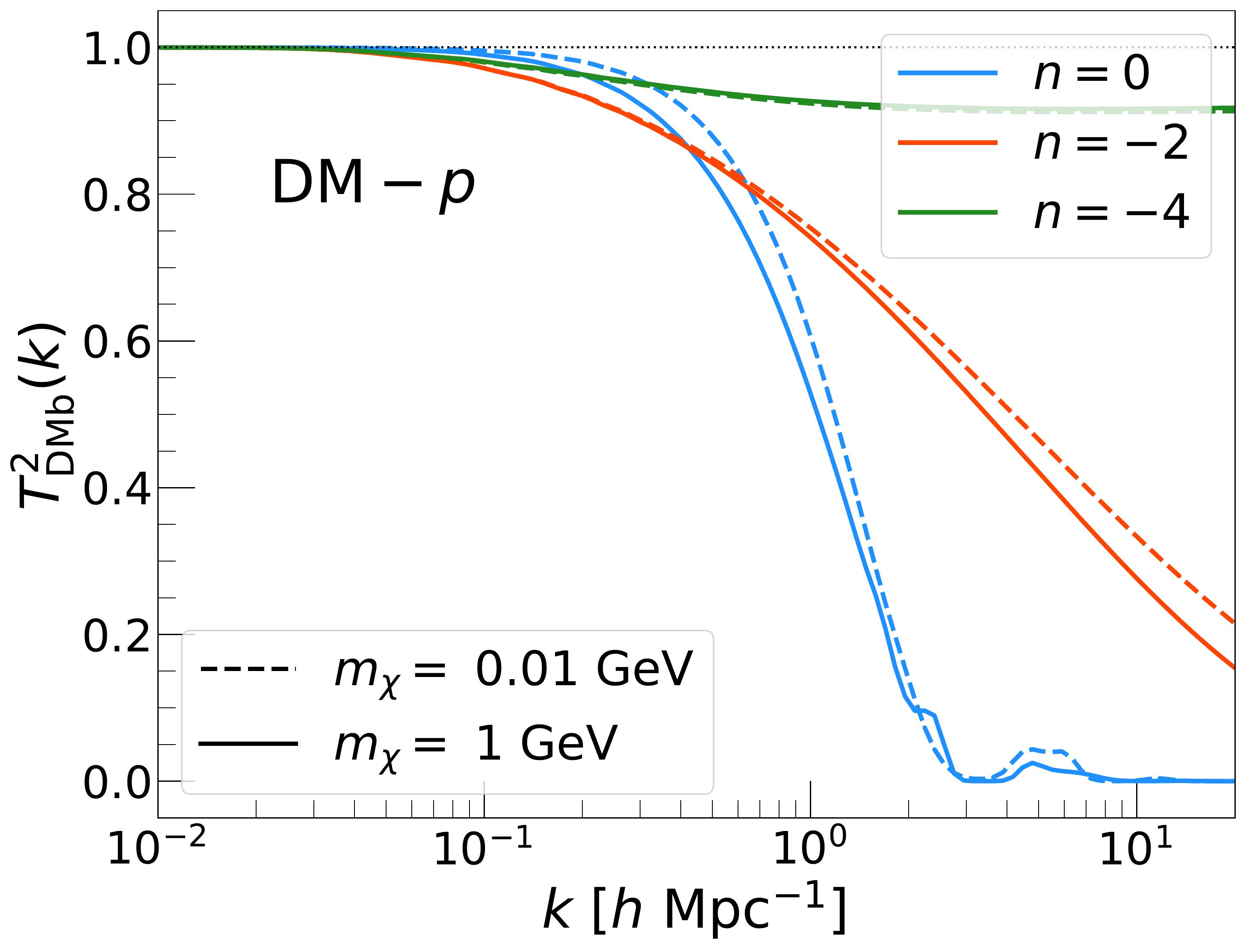}
  \includegraphics[width=0.49\textwidth]{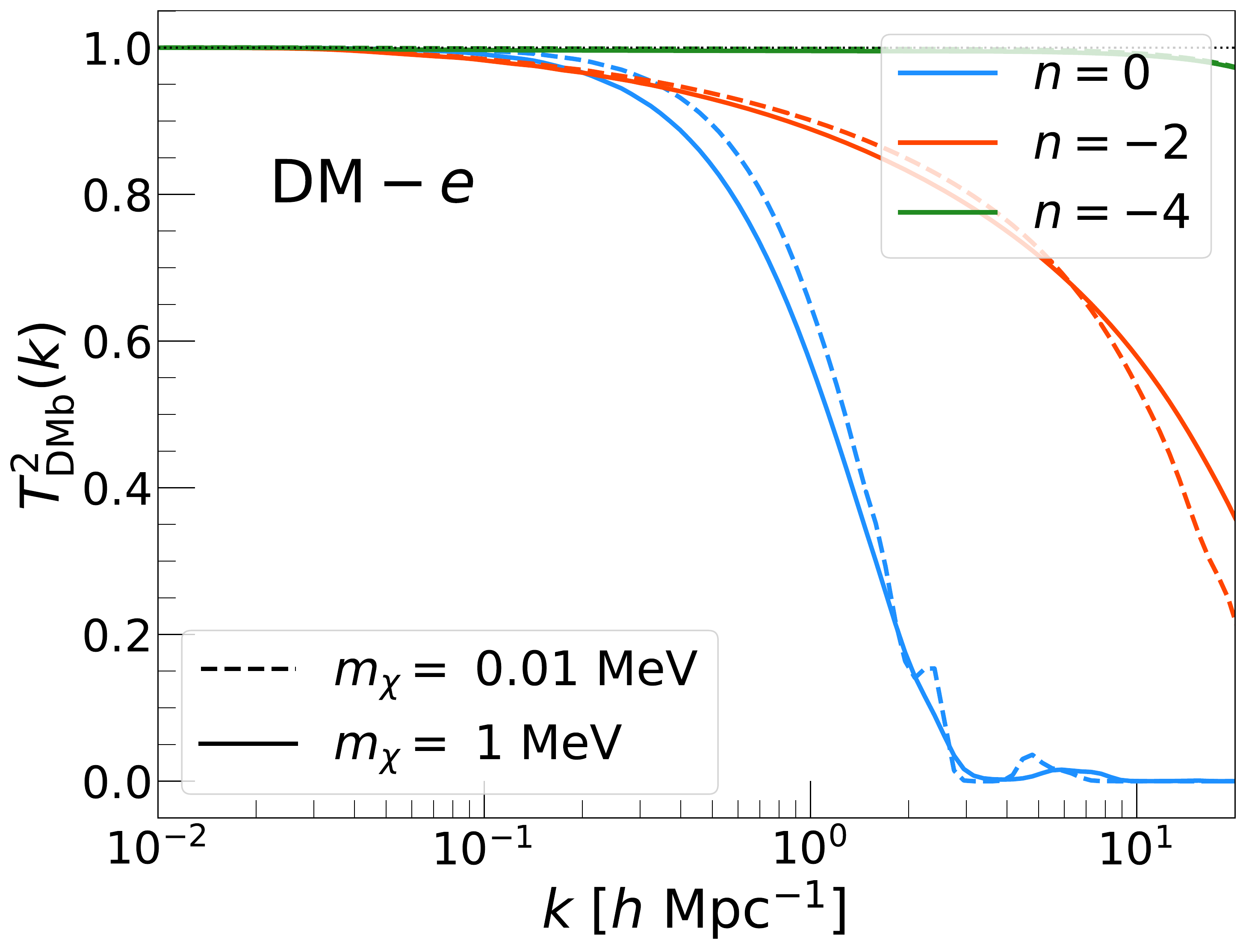}
  \caption{Squared transfer functions, $T^2_{\rm DMb}(k)$, for dark matter--proton ({\bf left}) and dark matter--electron ({\bf right}) interactions; for $n=0$ ({\bf blue}), $n=-2$ ({\bf orange}), and $n=-4$ ({\bf green}). The dark matter masses are $\mx = 0.01~\GeV$ ($0.01~\MeV$) ({\bf dashed}) and $1~\GeV$ ($1~\MeV$) ({\bf solid}) for the dark matter--proton (dark matter--electron) case. For each curve we fix $\sn$ to its corresponding 95\% C.L. value obtained with CMB+BAO data, shown in \Fig{fig:n0_bounds} (for $n=0$), \Fig{fig:n-2_bounds} (for $n=-2$), and \Fig{fig:n-4_bounds} (for $n=-4$).}
  \label{fig:Tk2}
\end{figure}

\begin{figure}[t]
  \centering
  \includegraphics[width=0.49\textwidth]{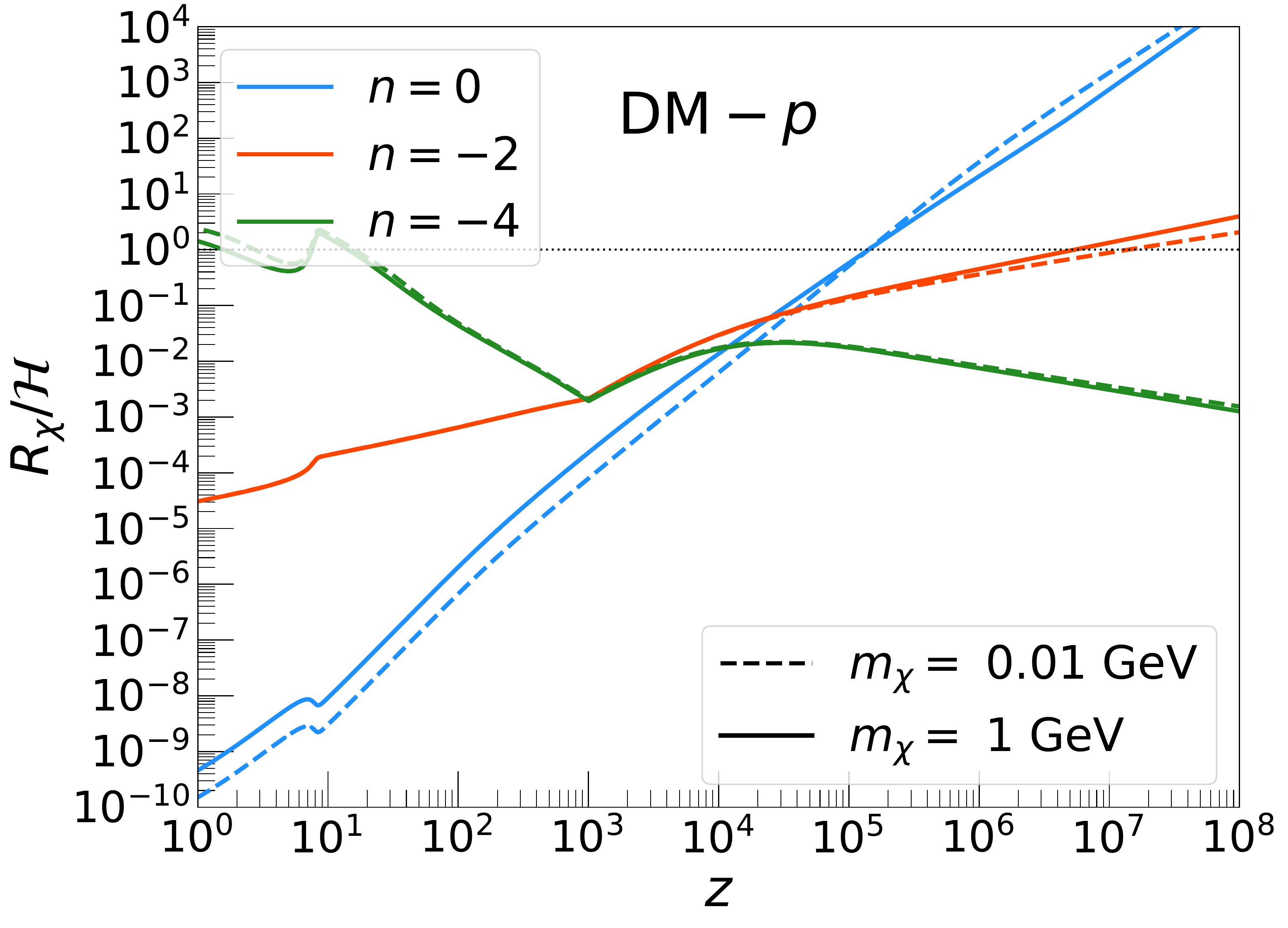}
  \includegraphics[width=0.49\textwidth]{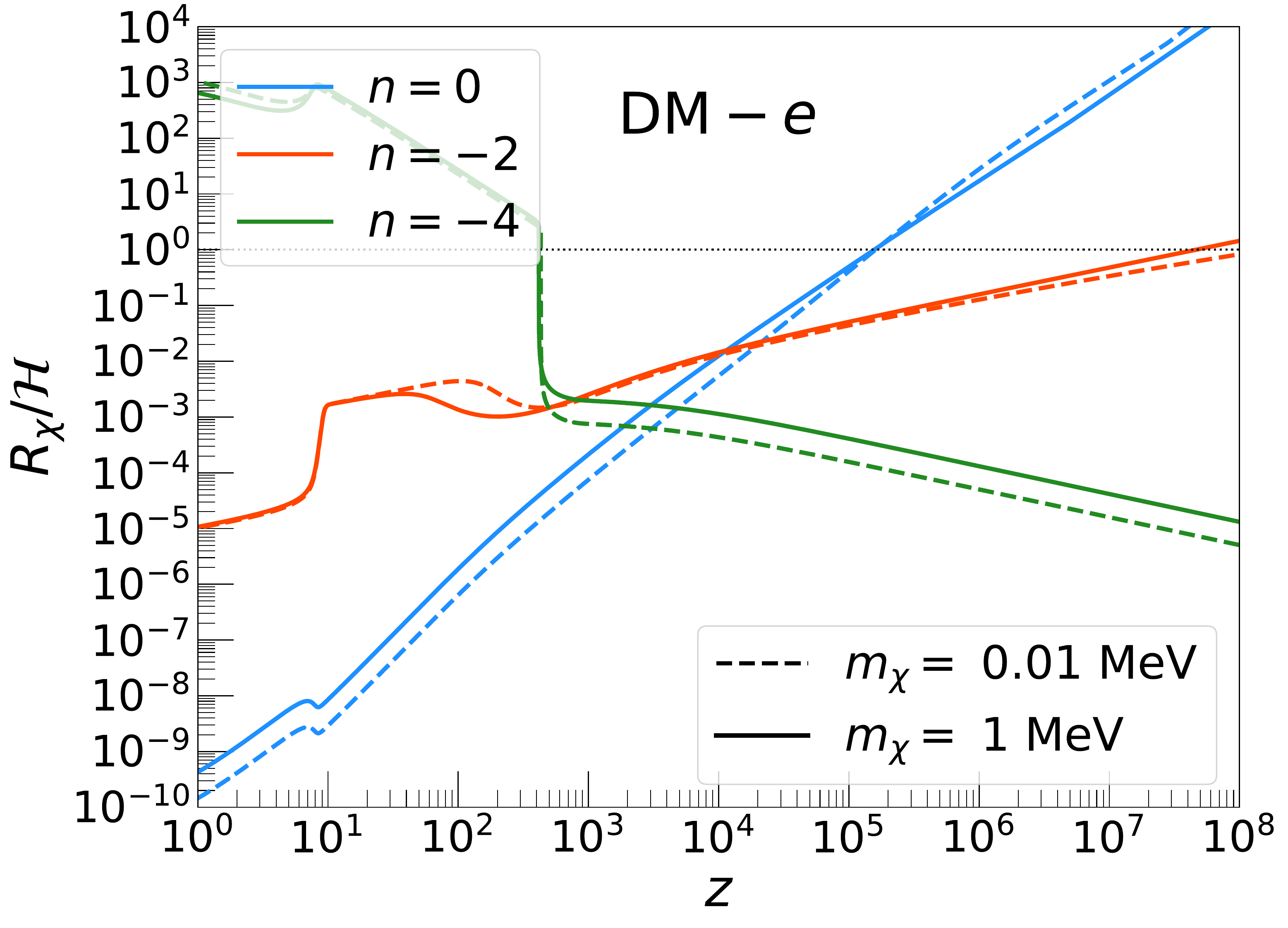}
  \caption{Momentum-exchange rates $\Rx$ normalized to the conformal Hubble expansion rate $\mH$, for dark matter--proton ({\bf left}) and dark matter--electron ({\bf right}) interactions; for $n=0$ ({\bf blue}), $n=-2$ ({\bf orange}), and $n=-4$ ({\bf green}). The dark matter masses are $\mx = 0.01~\GeV$ ($0.01~\MeV$) ({\bf dashed}) and $1~\GeV$ ($1~\MeV$) ({\bf solid}) for the dark matter--proton (dark matter--electron) case. For each curve we fix $\sn$ to its corresponding 95\% C.L. value obtained with CMB+BAO data, shown in \Fig{fig:n0_bounds} (for $n=0$), \Fig{fig:n-2_bounds} (for $n=-2$), and \Fig{fig:n-4_bounds} (for $n=-4$).}
  \label{fig:Rx}
\end{figure}

The impact of DMb interactions on the matter power spectrum is the easiest to understand. The momentum exchange rate $\Rx$ between the baryon and dark matter fluids acts as a friction on the latter. This dampens the growth of the $\delta_\chi$ perturbations, thereby suppressing structure formation. Therefore, the matter power spectrum is smaller in the DMb model than in $\LC$. The \textit{transfer function} squared, $T^2_{\rm DMb}(k)$, defined as the ratio of these spectra evaluated at a comoving wavenumber $k$,
\beq
    T^2_{\rm DMb}(k) = \frac{P_{\rm DMb}(k)}{P_{\LC}(k)} \ ,
    \label{eq:transferDMb}
\eeq
is commonly used to quantify this suppression. In \Fig{fig:Tk2}, we show the squared transfer function for dark matter--proton and electron interactions for $n=0$, $-2$ and $-4$ type interactions. The specific functional shape of $T^2_{\rm DMb}(k)$ on $k$ depends in a complicated way on the various properties of the model, and mainly on the time evolution and the size of $\Rx$. However, as shown in \Fig{fig:Rx}, since $\Rx / \mH$ decreases with time for $n=0$ and $n=-2$ type DMb interactions, we can, by comparing \Figs{fig:Tk2}{fig:Rx}, deduce the following ``rules of thumb'' to understand the qualitative features of the transfer function for these two cases:
\begin{itemize}
 \item The inflection point (or cutoff) of the transfer function, which roughly divides the $k$-modes significantly suppressed from those $k$-modes that are only moderately suppressed, occurs at the $k$ scale that enters the horizon when $\Rx \sim \mH$.
 \item The slope of the suppression is positively correlated with the rate of change of $\Rx$ as a function of time: the more quickly $\Rx$ changes, the more pronounced the suppression is for contiguous $k$ modes.
\end{itemize}

For $n=0$ a bump is present at $k$ wavenumbers above the cutoff. This corresponds to the first and most visible of a series of ``dark acoustic oscillations'' (DAO) \cite{Cyr-Racine:2015ihg}, peaks imprinted on the matter power spectrum due to the dark matter and the baryons being tightly coupled in the early Universe. This occurs because the pressure present in the baryons (due to their coupling to the photons) is being transferred to the dark matter fluid via the momentum transfer rate $\Rx$. The DAO are more visible the tighter the coupling between the dark matter and the baryons was in the early Universe, which is itself a function of $\sn$ and $n$. Refs.~\cite{Nadler:2019zrb,Maamari:2020aqz}, for example, show the transfer functions for $n \geq 0$, which display increasingly more striking DAO with larger $n$.

Since for $n=-4$ type interactions $\Rx/\mH$ actually \textit{increases} for most of the time as the Universe expands, the above qualitative understanding does not apply. However $\Rx$ in \Eq{eq:pert_x} in general acts as a friction term of the baryons on the dark matter fluid, and therefore the amount of suppression on the matter power spectrum is directly correlated with the size of $\Rx$.

\begin{figure}[t]
  \centering
  \includegraphics[width=0.49\textwidth]{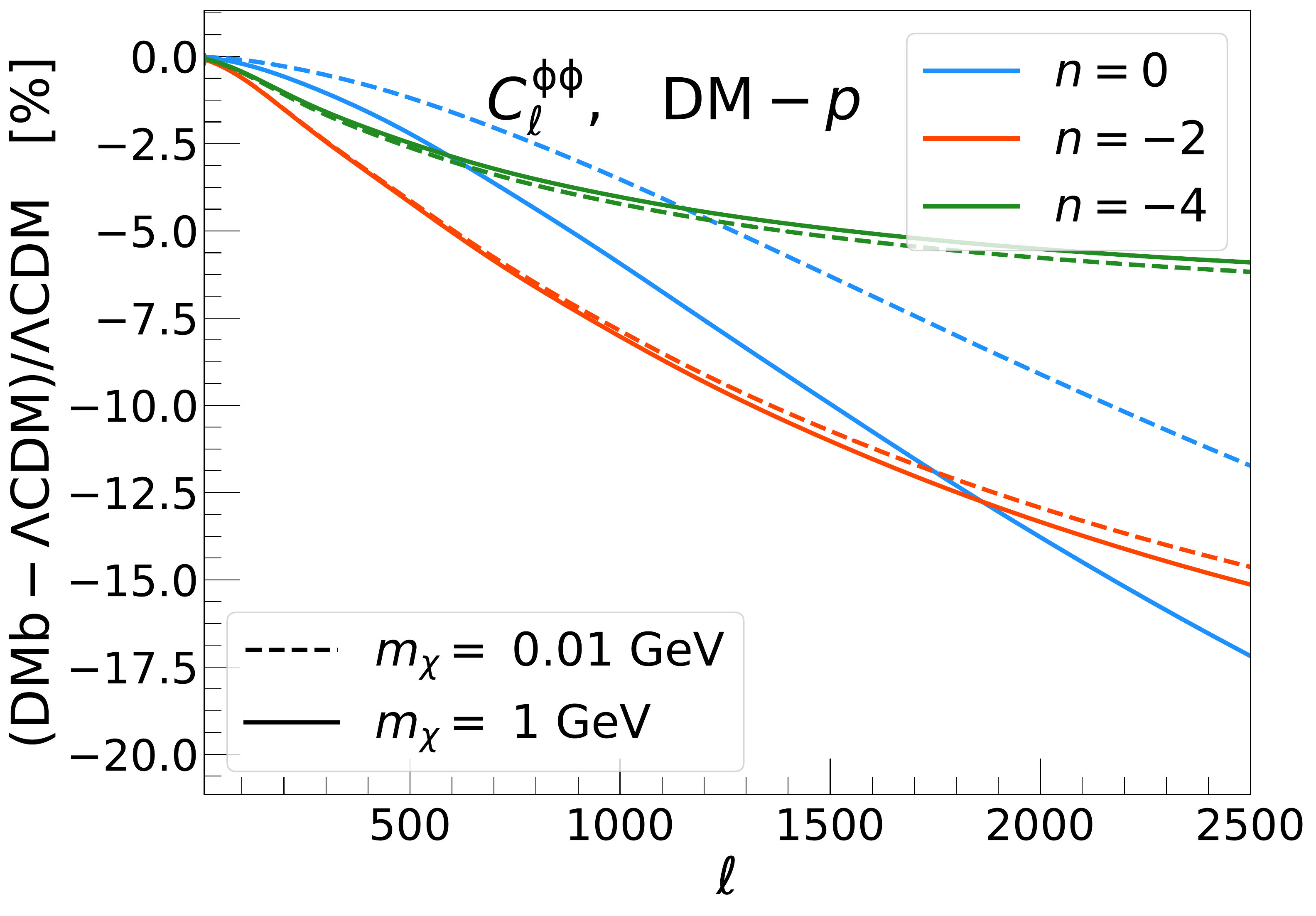}
  \includegraphics[width=0.49\textwidth]{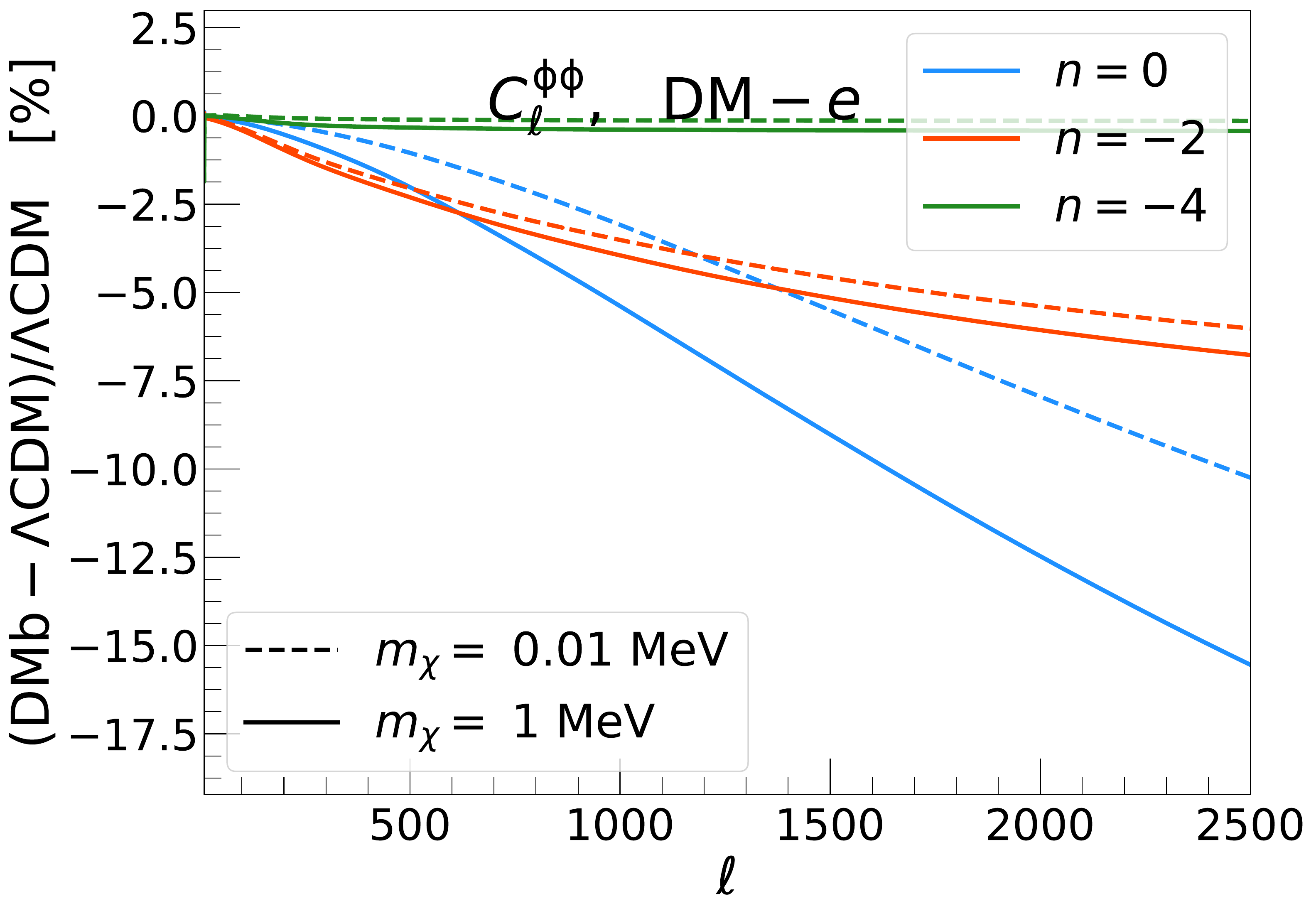}
  \caption{$C_\ell^{\phi\phi}$ percent residuals of DMb with respect to $\LC$, for dark matter--proton ({\bf left}) and dark matter--electron ({\bf right}) interactions; for $n=0$ ({\bf blue}), $n=-2$ ({\bf orange}), and $n=-4$ ({\bf green}). The dark matter masses are $\mx = 0.01~\GeV$ ($0.01~\MeV$) ({\bf dashed}) and $1~\GeV$ ($1~\MeV$) ({\bf solid}) for the dark matter--proton (dark matter--electron) case. For each curve we fix $\sn$ to its corresponding 95\% C.L. value obtained with CMB+BAO data, shown in \Fig{fig:n0_bounds} (for $n=0$), \Fig{fig:n-2_bounds} (for $n=-2$), and \Fig{fig:n-4_bounds} (for $n=-4$).}
  \label{fig:Clens}
\end{figure}

Once we have grasped the effect of DMb interactions on the matter power spectrum we can readily understand their effect on the lensing power, shown in \Fig{fig:Clens}. The lensing power is a measure of the distortion of the CMB anisotropies due to the gravitational lensing caused by the inhomogeneously distributed matter in the foreground, between the surface of last scattering and the observer. As such, it is sensitive to the matter power spectrum. Indeed, the lensing potential can be approximated using the Limber equation in the limit of large multipole index $\ell \gg 1$~\cite{1953ApJ...117..134L,Limber:1954zz,Pan:2014xua}:
\beq
    \ell^4 C_\ell^{\phi\phi} \approx 2 \int_{0}^{\chi_{\rm rec}}\!\dd\chi ~ \bl( \frac{\ell}{\chi} \br)^4 P_{\phi + \psi} \bl( k=\frac{\ell}{\chi}, z(\chi) \br) \bl( 1 - \frac{\chi}{\chi_{\rm rec}} \br)^2 \ ,
\eeq
where $\chi$ is the comoving distance as measured from the observer, $\chi_{\rm rec}$ is the distance to recombination, and $P_{\phi + \psi}(k, z)$ is the power spectrum of the sum of the $\phi$ and $\psi$ gravitational potentials (metric perturbations in conformal Newtonian gauge) at wavenumber $k$ and redshift $z$. The integrand clearly vanishes at $\chi_{\rm rec}$; on the other hand, for fixed large $\ell$, the power spectrum of the metric perturbations (which peaks at $k\approx 0.01~h/\Mpc$) grows as we probe smaller wavenumbers $k$ with increasing comoving distance \cite{Pan:2014xua}. It turns out that the integrand peaks at a comoving distance roughly halfway to recombination, with a broad support. This allows us to write the heuristic relationship $k \sim \ell/\chi_{\rm rec}$. Now, the power spectrum of the gravitational potential is related to the matter power spectrum since matter acts as a source for gravity. Therefore, the suppression of the lensing spectrum qualitatively  follows that of the matter power spectrum up to $k \sim \mO(\text{few} \times 0.1~h/\Mpc)$~\cite{Buen-Abad:2017gxg}  as can be seen by comparing \Fig{fig:Clens} to \Fig{fig:Tk2}.

\subsection{Temperature and polarization power spectra}
\label{subsec:cmb}

\begin{figure}[t]
  \centering
  \includegraphics[width=0.49\textwidth]{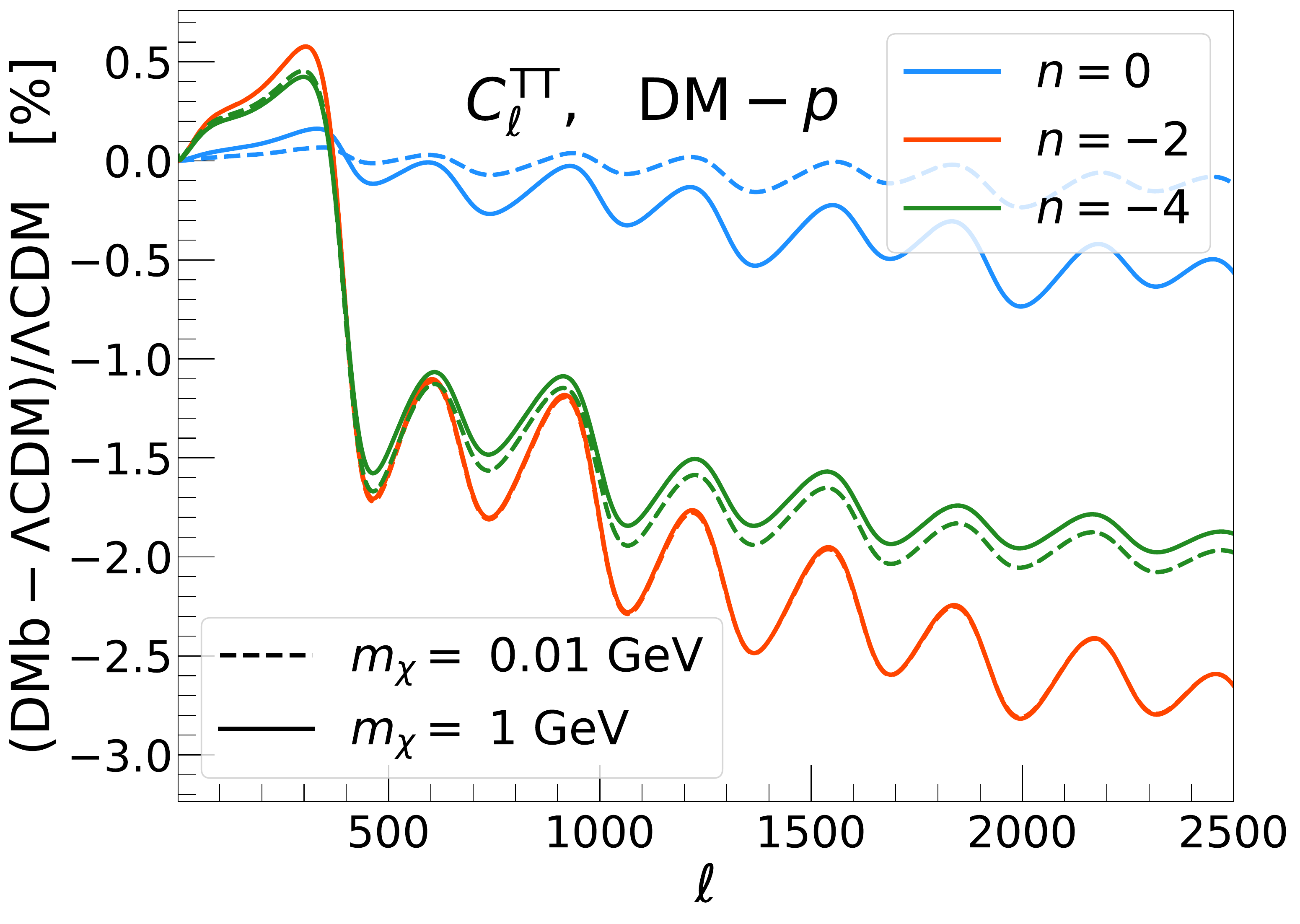}
  \includegraphics[width=0.49\textwidth]{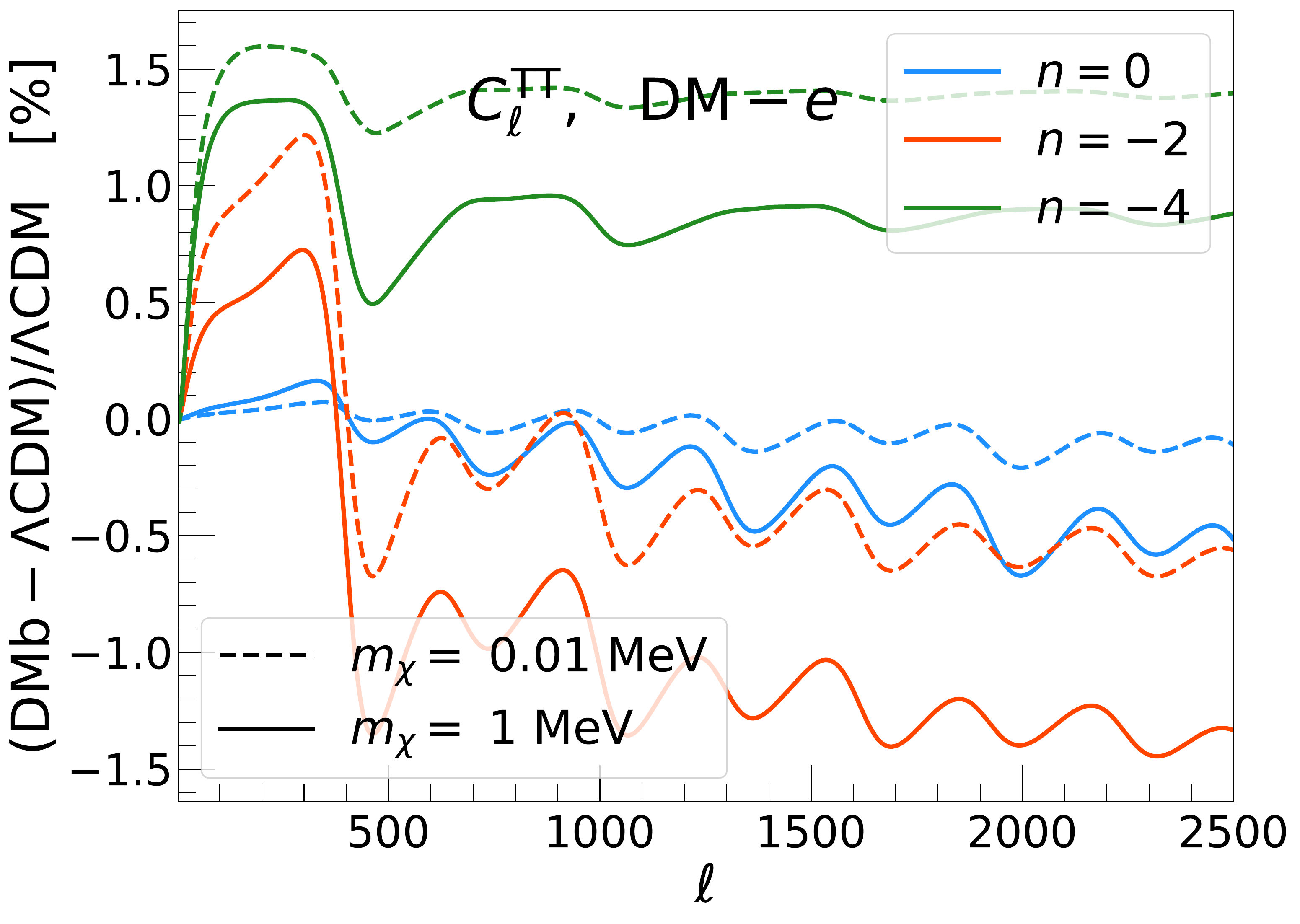}
  \caption{$C_\ell^{\rm TT}$ percent residuals of DMb with respect to $\LC$, for dark matter--proton ({\bf left}) and dark matter--electron ({\bf right}) interactions; for $n=0$ ({\bf blue}), $n=-2$ ({\bf orange}), and $n=-4$ ({\bf green}). The dark matter masses are $\mx = 0.01~\GeV$ ($0.01~\MeV$) ({\bf dashed}) and $1~\GeV$ ($1~\MeV$) ({\bf solid}) for the dark matter--proton (dark matter--electron) case. For each curve we fix $\sn$ to its corresponding 95\% C.L. value obtained with CMB+BAO data, shown in \Fig{fig:n0_bounds} (for $n=0$), \Fig{fig:n-2_bounds} (for $n=-2$), and \Fig{fig:n-4_bounds} (for $n=-4$).}
  \label{fig:CTT}
\end{figure}

The effects of DMb interactions on temperature and polarization CMB spectra are highly non-trivial (see \Figs{fig:CTT}{fig:CEE}). Nevertheless, we can repeat a qualitative analysis, similar to the previous one for the matter power spectrum, with the aid of the expressions for the sources of the CMB spectra of scalar perturbations~\cite{Zaldarriaga:1996xe,Lesgourgues:2013qba}. For the TT and EE spectra these are, respectively, given by
\beqa
    S_\text{T}(k, \tau) & = & g(\tau) ~ \bl( \frac{\delta_\gamma}{4} + \psi \br) + \frac{1}{k^2}\frac{\dd \bl( g(\tau) \theta_b \br)}{\dd \tau} + e^{-\kappa} \bl( \dot\psi + \dot\phi \br) \ , \label{eq:TT_source}\\
    S_\text{E}(k, \tau) & = & \frac{3}{4} g(\tau) ~ \bl( 2\sigma_\gamma + G_{\gamma 0} + G_{\gamma 2} \br) \ , \label{eq:EE_source}
\eeqa
where $\delta_\gamma$ is the photon density perturbation, $\sigma_\gamma$ is the photon shear stress, $G_{\gamma 0}$ ($G_{\gamma 2}$) is the 0th (2nd) polarization moment, and 
\beqa
    \dot{\kappa}(\tau) & \equiv & a n_e \sigma_{\rm Thomson} = \frac{3}{4}\frac{\rho_b}{\rho_\gamma} R_\gamma \ , \quad \text{the Thomson scattering rate,} \label{eq:thomson}\\
    \kappa(\tau) & \equiv & \int_\tau^{\tau_0} \!\dd\tau' ~ \dot{\kappa}(\tau') \ , \quad \text{the optical depth, and} \label{eq:opt_depth}\\
    g(\tau) & \equiv & -\dot{\kappa} e^{-\kappa} \ , \quad \text{the visibility function.} \label{eq:visib}
\eeqa
Note that the Thomson scattering rate $\dot{\kappa}$ depends on the ionized electron number density $n_e$, which is proportional to the electron ionization fraction, commonly denoted by $x_e$.

For the TT spectrum, the main impact of DMb interactions is on the first term of \Eq{eq:TT_source}, the so-called Sachs--Wolfe (SW) effect. The matter power spectrum suppression described in \Sec{subsec:mps} decreases the depth of the gravitational potentials. This changes both the amplitude and the zero-point of the oscillations of the SW term~\cite{Buen-Abad:2017gxg}, making it smaller and resulting in a colder effective temperature and a scale-dependent modulation of the TT spectrum. Furthermore, the change in the time-evolution of the metric perturbations modifies the third term in \Eq{eq:TT_source}, resulting in an increase of the Integrated Sachs--Wolfe (ISW) effect. A larger ISW effect enhances the first acoustic peak, whereas a smaller SW effect leads to a suppression for the rest of the peaks. This combined impact of the DMb interactions can  be clearly seen in \Fig{fig:CTT}. 

The predominant effect on the EE spectrum is a suppression at high-$\ell$ due to a reduction of the photon shear $\sigma_\gamma$, which contributes to the EE source in \Eq{eq:EE_source}. This occurs because around the time of recombination, $\sigma_\gamma$ depends on the photon velocity divergence $\theta_\gamma$, which in turn is tightly correlated to that of the baryons, $\theta_b$. Due to the DMb momentum exchange rate $\Rx$, the inertia of the baryons is increased~\cite{Boddy:2018wzy}, effectively bringing down $\theta_b$ and, with it, $\theta_\gamma$ and $\sigma_\gamma$.

\begin{figure}[t]
  \centering
  \includegraphics[width=0.49\textwidth]{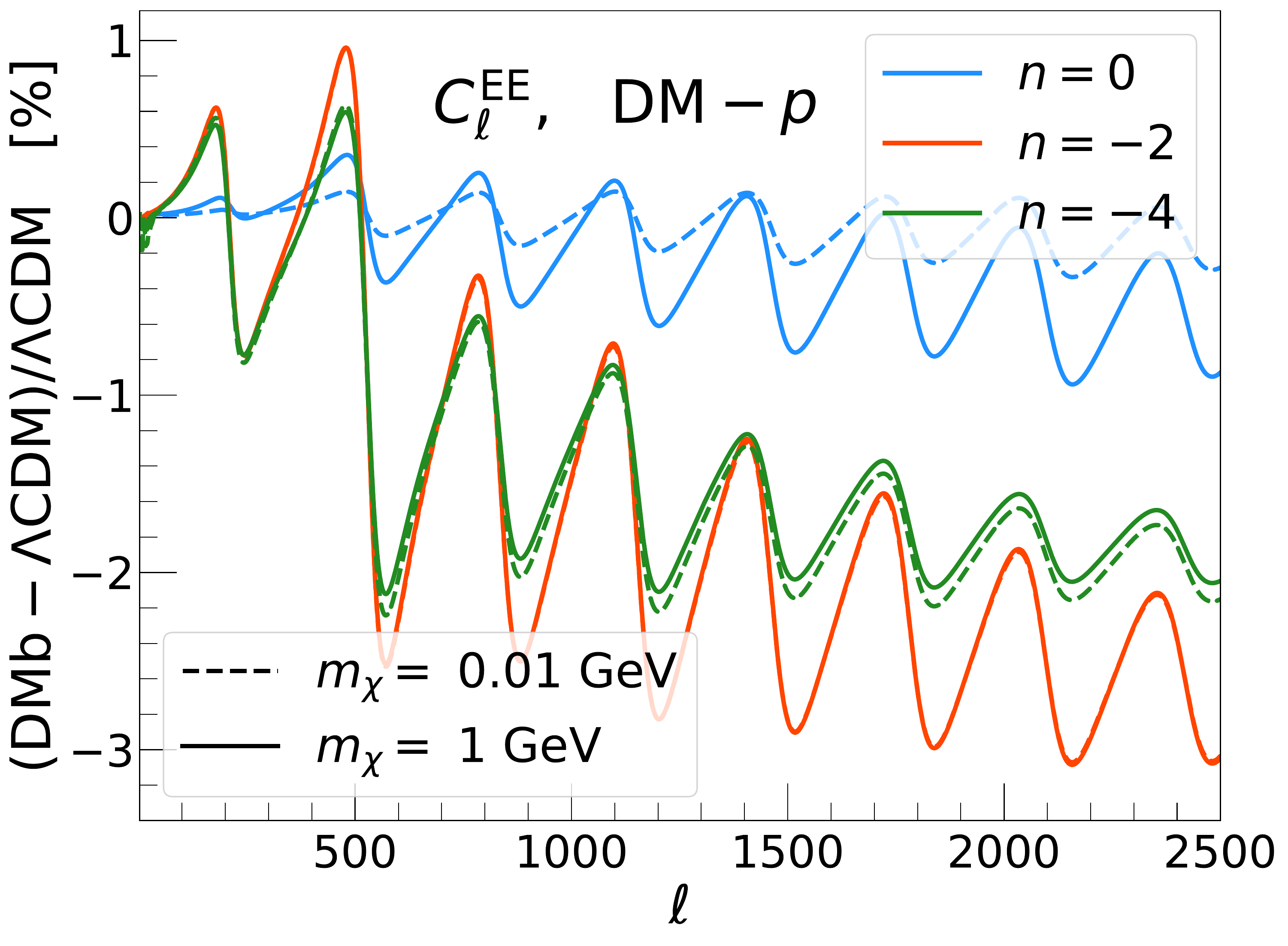}
  \includegraphics[width=0.49\textwidth]{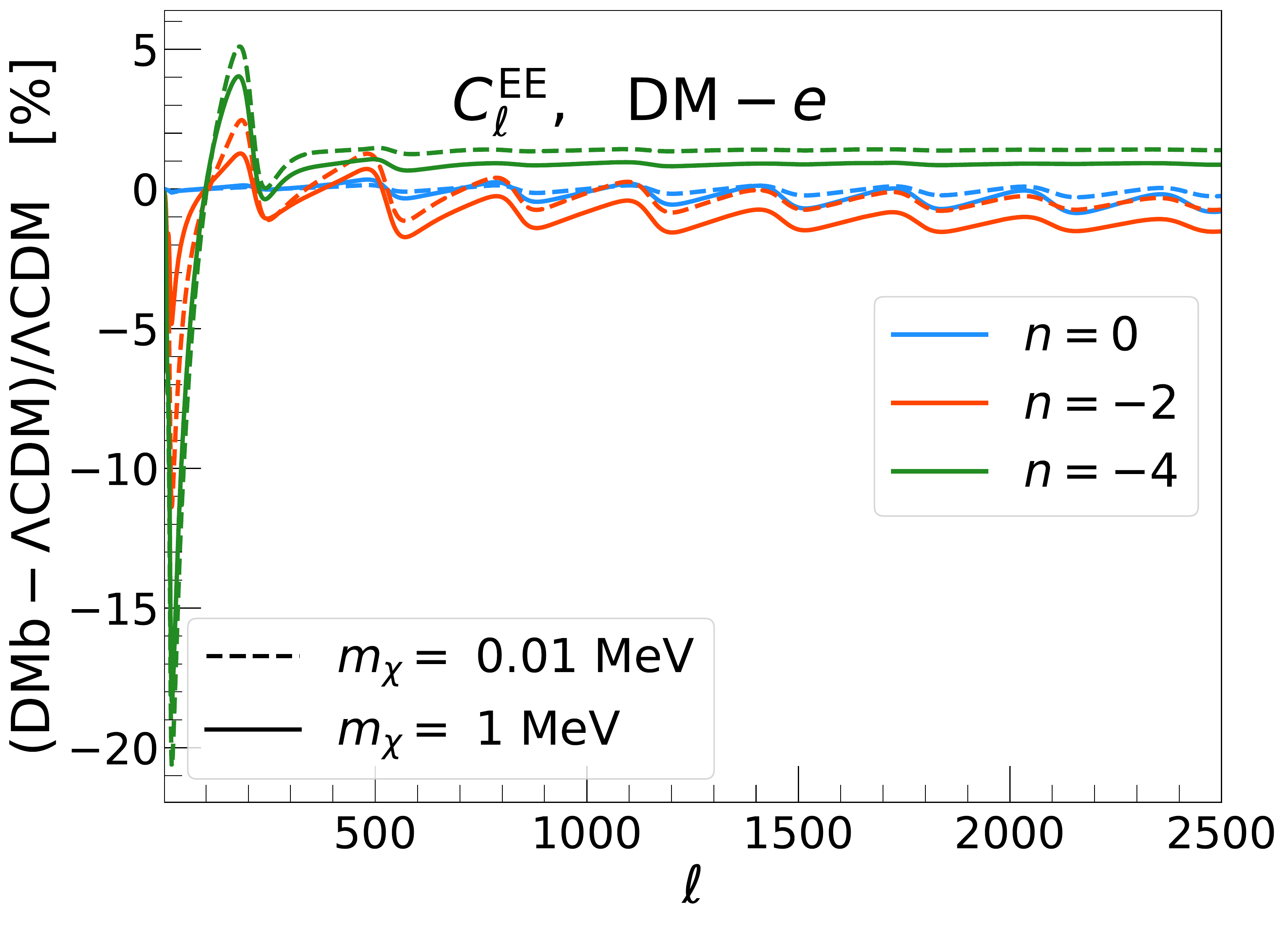}
  \caption{$C_\ell^{\rm EE}$ percent residuals of DMb with respect to $\LC$, for dark matter--proton ({\bf left}) and dark matter--electron ({\bf right}) interactions; for $n=0$ ({\bf blue}), $n=-2$ ({\bf orange}), and $n=-4$ ({\bf green}). The dark matter masses are $\mx = 0.01~\GeV$ ($0.01~\MeV$) ({\bf dashed}) and $1~\GeV$ ($1~\MeV$) ({\bf solid}) for the dark matter--proton (dark matter--electron) case. For each curve we fix $\sn$ to its corresponding 95\% C.L. value obtained with CMB+BAO data, shown in \Fig{fig:n0_bounds} (for $n=0$), \Fig{fig:n-2_bounds} (for $n=-2$), and \Fig{fig:n-4_bounds} (for $n=-4$).}
  \label{fig:CEE}
\end{figure}

The previous qualitative description of the TT and EE spectra holds for most of the DMb models considered. However it fails to accurately capture the behavior of the $n=-4$ dark matter--electron case. Indeed, the SW and ISW high-$\ell$ suppression is absent in the TT spectrum, and so is the high-$\ell$ decrease of the EE spectrum due to the baryon inertia. The previous description also does not explain the $n=-2$ and $n=-4$ dip at low-$\ell$ in \Fig{fig:CEE}. We address these issues in the paragraphs below.

\begin{figure}[t]
  \centering
  \includegraphics[width=0.49\textwidth]{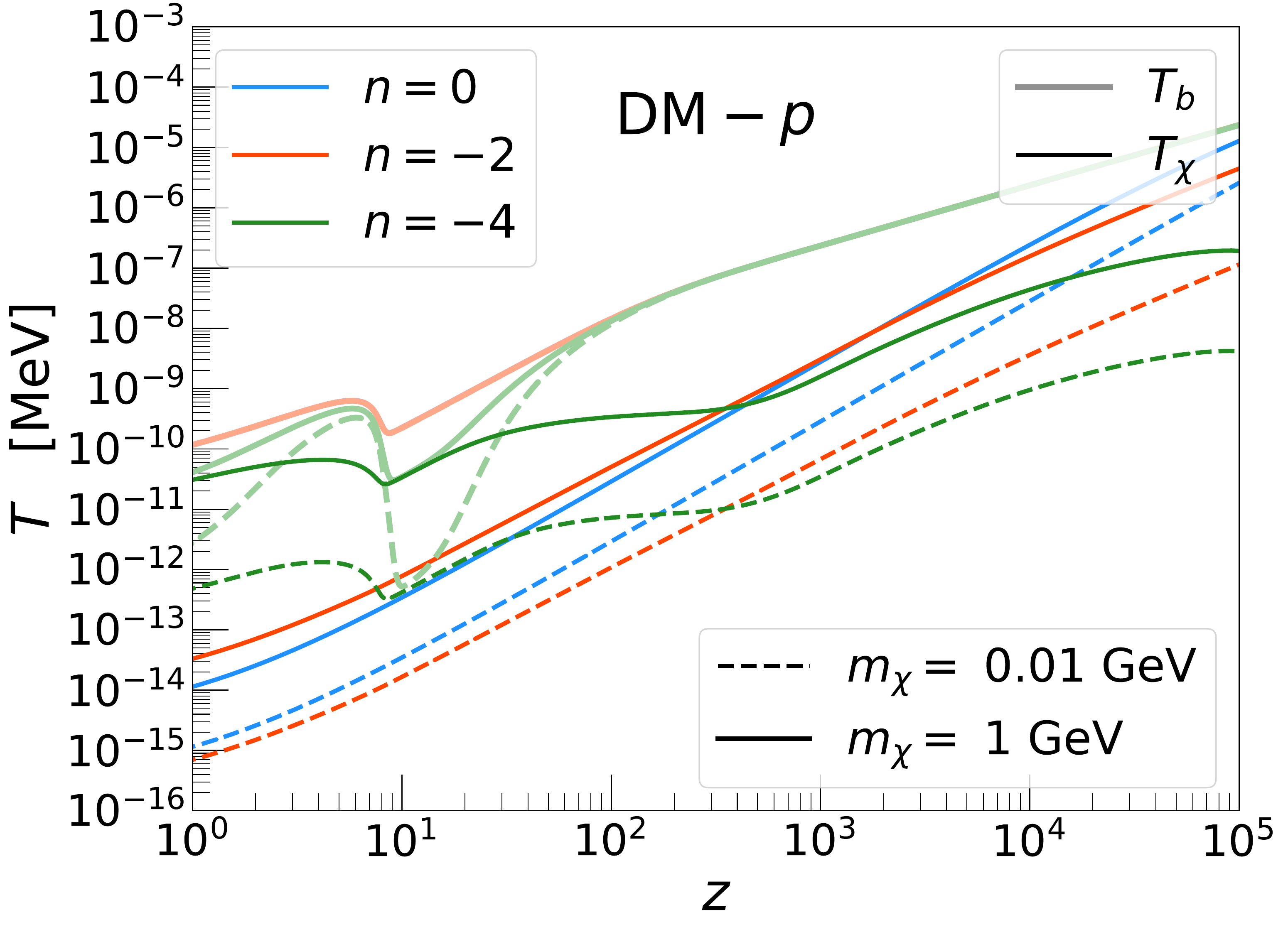}
  \includegraphics[width=0.49\textwidth]{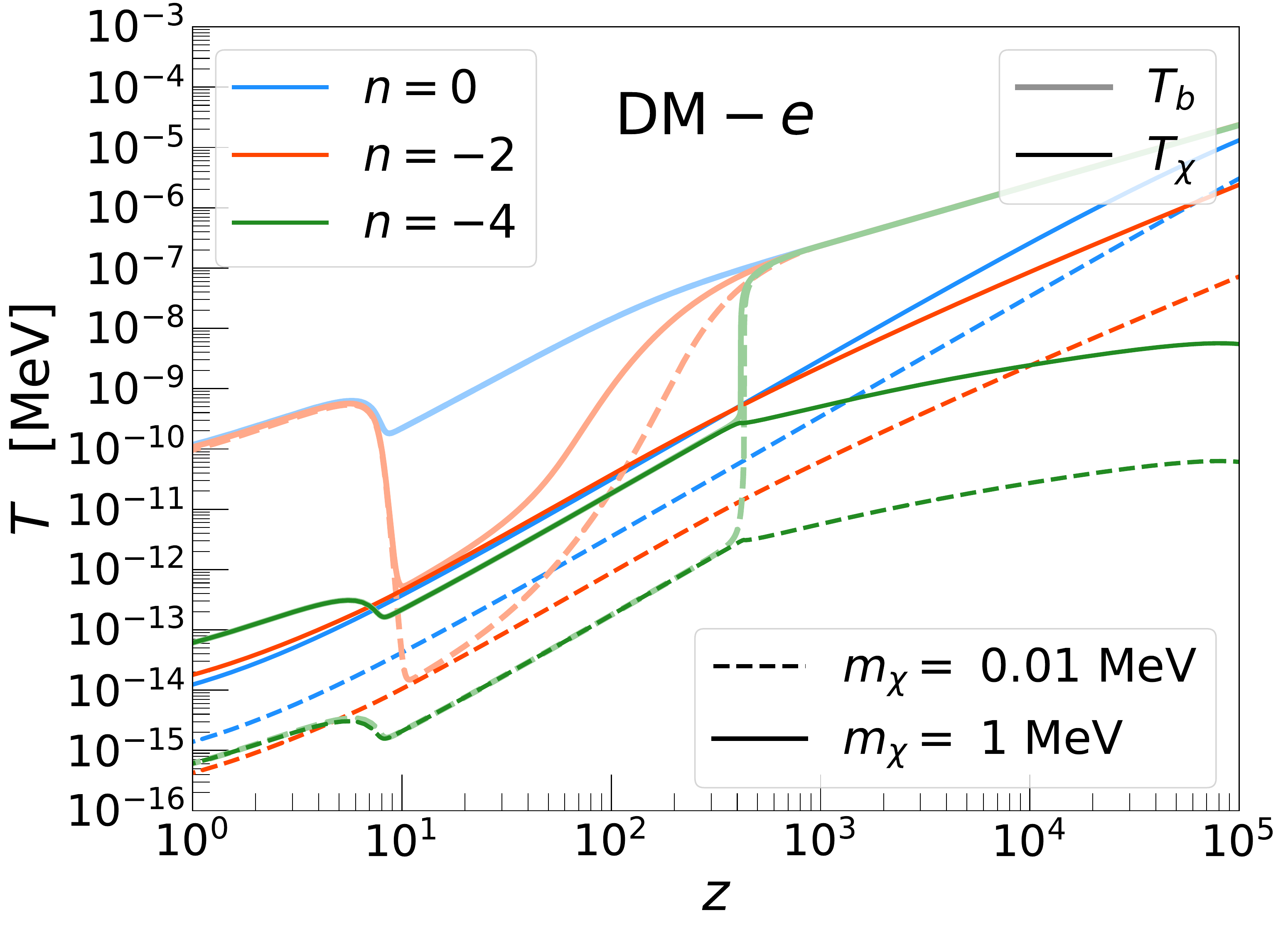}
  \caption{Dark matter ({\bf high opacity} lines) and baryon ({\bf low opacity} lines) temperatures in $\MeV$, for dark matter--proton ({\bf left}) and dark matter--electron ({\bf right}) interactions; for $n=0$ ({\bf blue}), $n=-2$ ({\bf orange}), and $n=-4$ ({\bf green}). The dark matter masses are $\mx = 0.01~\GeV$ ($0.01~\MeV$) ({\bf dashed}) and $1~\GeV$ ($1~\MeV$) ({\bf solid}) for the dark matter--proton (dark matter--electron) case. For each curve we fix $\sn$ to its corresponding 95\% C.L. value obtained with CMB+BAO data, shown in \Fig{fig:n0_bounds} (for $n=0$), \Fig{fig:n-2_bounds} (for $n=-2$), and \Fig{fig:n-4_bounds} (for $n=-4$).}
  \label{fig:temps}
\end{figure}

\begin{figure}[t]
  \centering
  \includegraphics[width=0.49\textwidth]{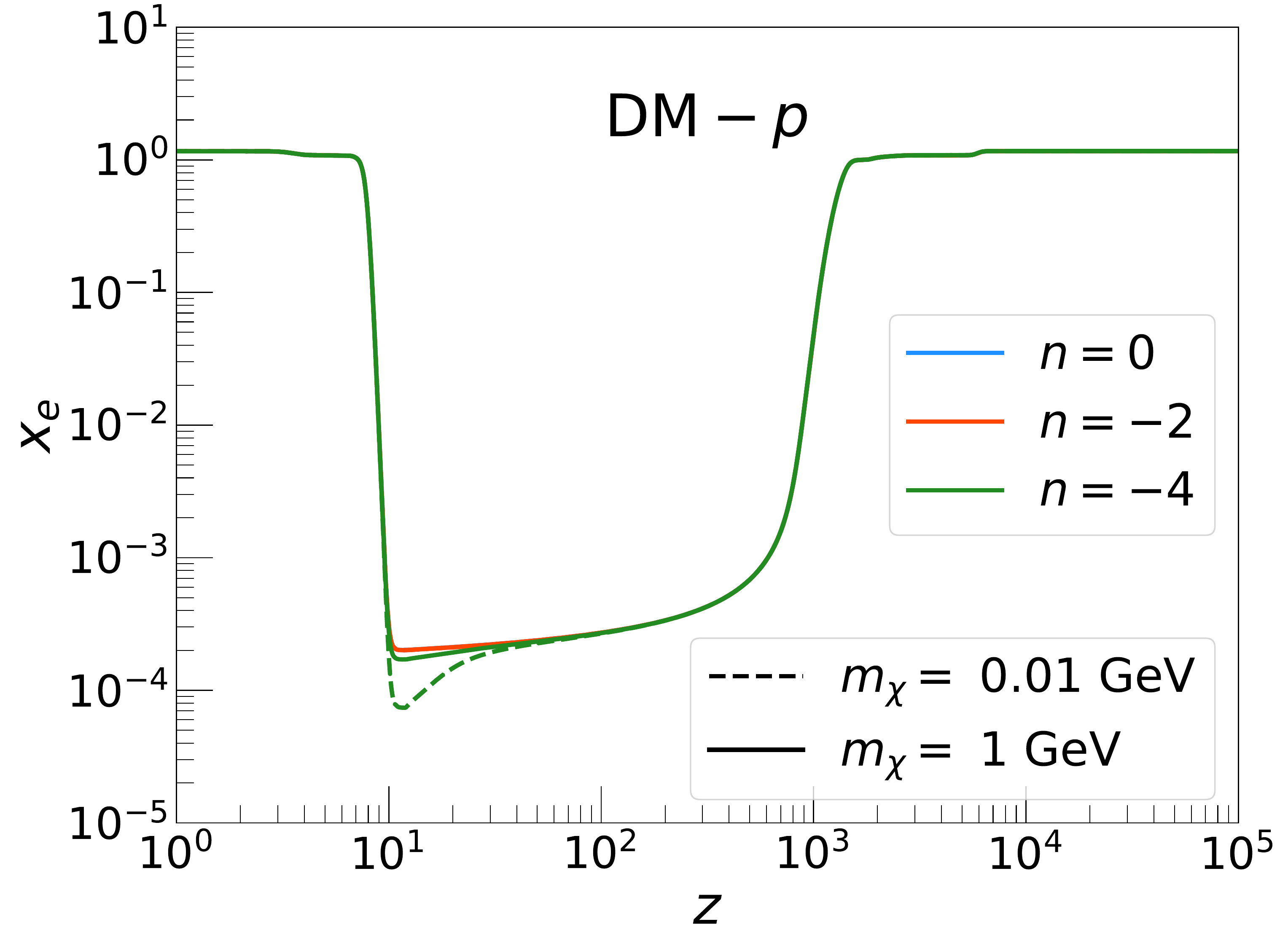}
  \includegraphics[width=0.49\textwidth]{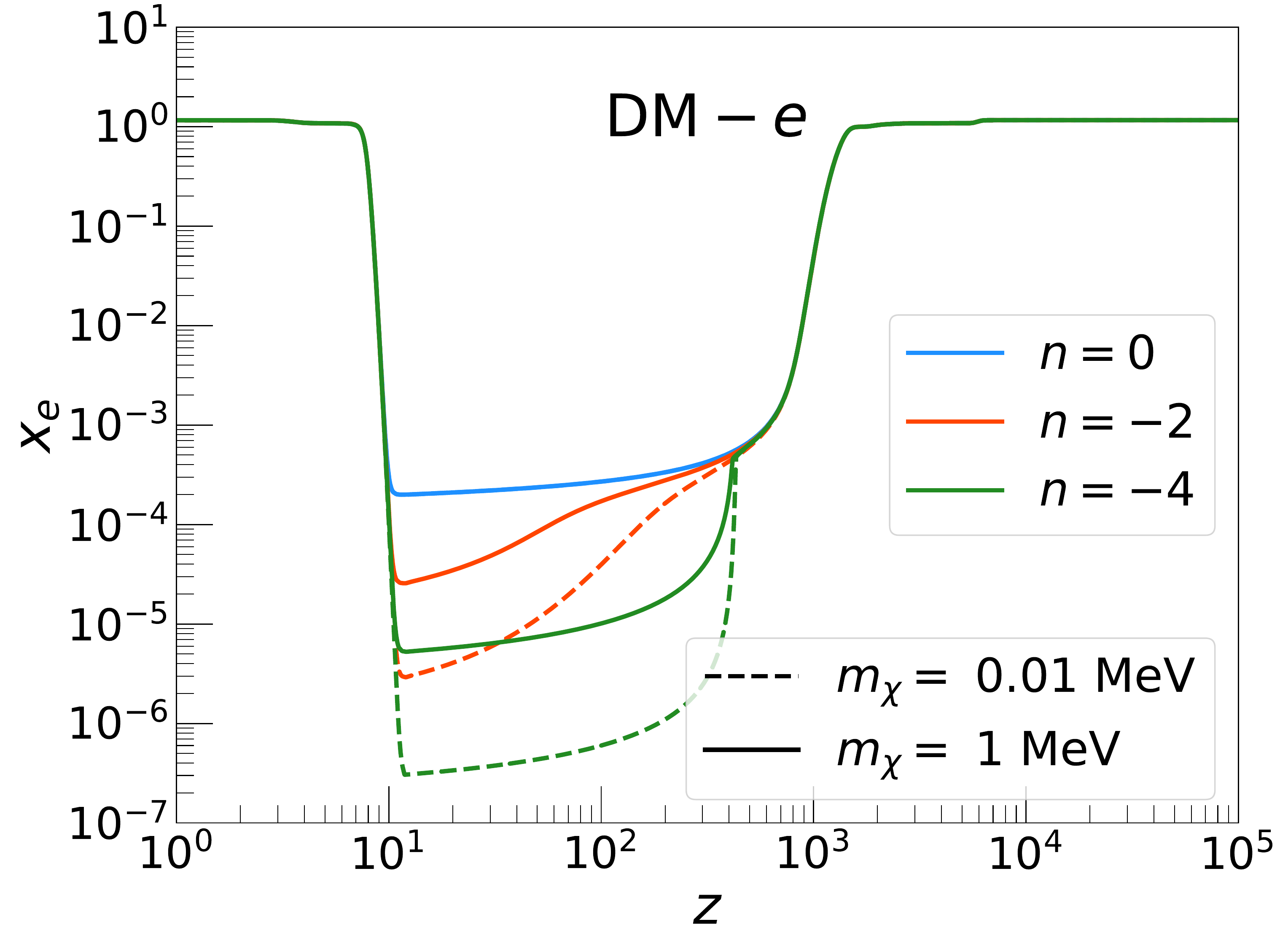}
  \caption{Electron ionization fraction $x_e$ of DMb with respect to $\LC$, for dark matter--proton ({\bf left}) and dark matter--electron ({\bf right}) interactions; for $n=0$ ({\bf blue}), $n=-2$ ({\bf orange}), and $n=-4$ ({\bf green}). The dark matter masses are $\mx = 0.01~\GeV$ ($0.01~\MeV$) ({\bf dashed}) and $1~\GeV$ ($1~\MeV$) ({\bf solid}) for the dark matter--proton (dark matter--electron) case. For each curve we fix $\sn$ to its corresponding 95\% C.L. value obtained with CMB+BAO data, shown in \Fig{fig:n0_bounds} (for $n=0$), \Fig{fig:n-2_bounds} (for $n=-2$), and \Fig{fig:n-4_bounds} (for $n=-4$).}
  \label{fig:ion}
\end{figure}

\begin{figure}[t]
  \centering
  \includegraphics[width=0.49\textwidth]{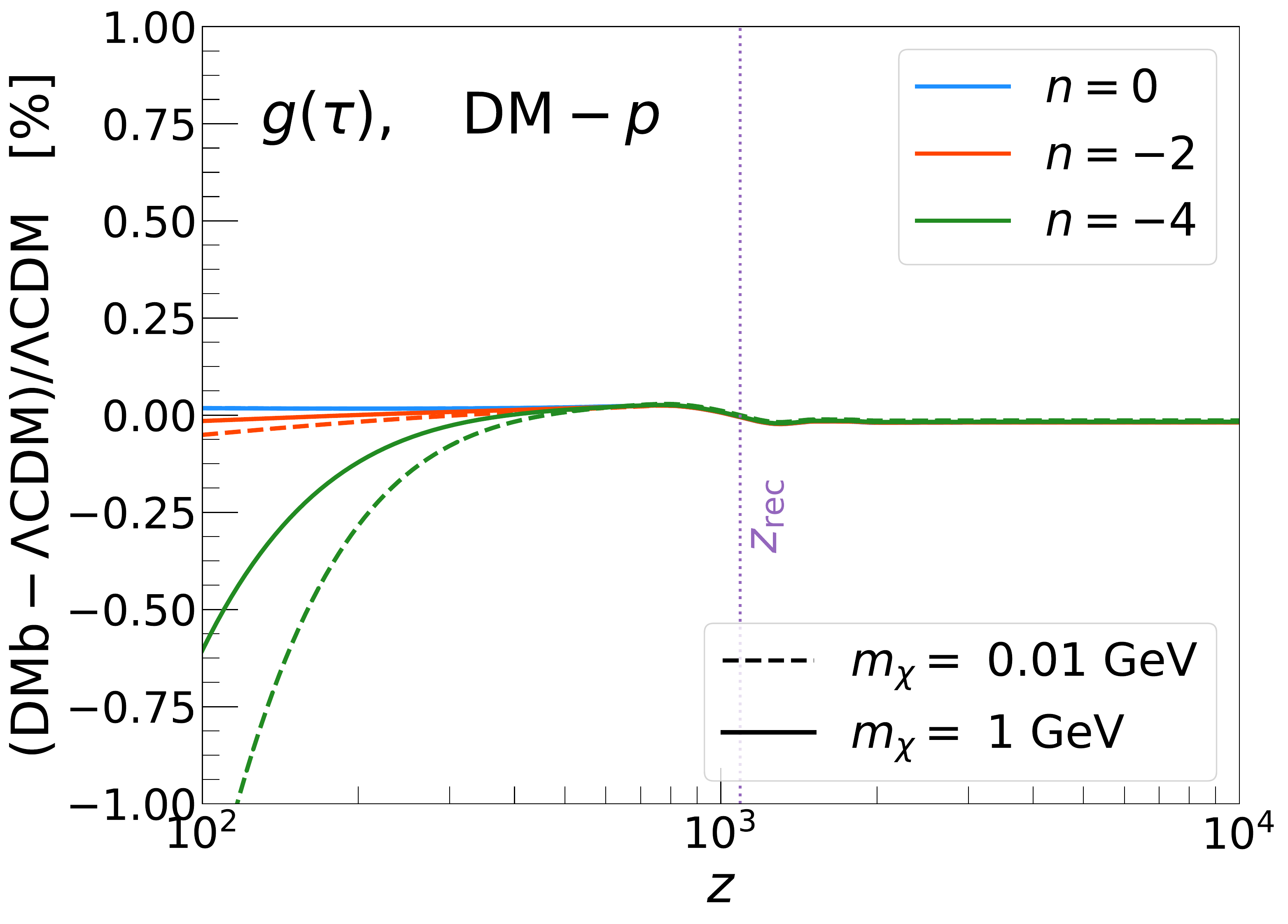}
  \includegraphics[width=0.49\textwidth]{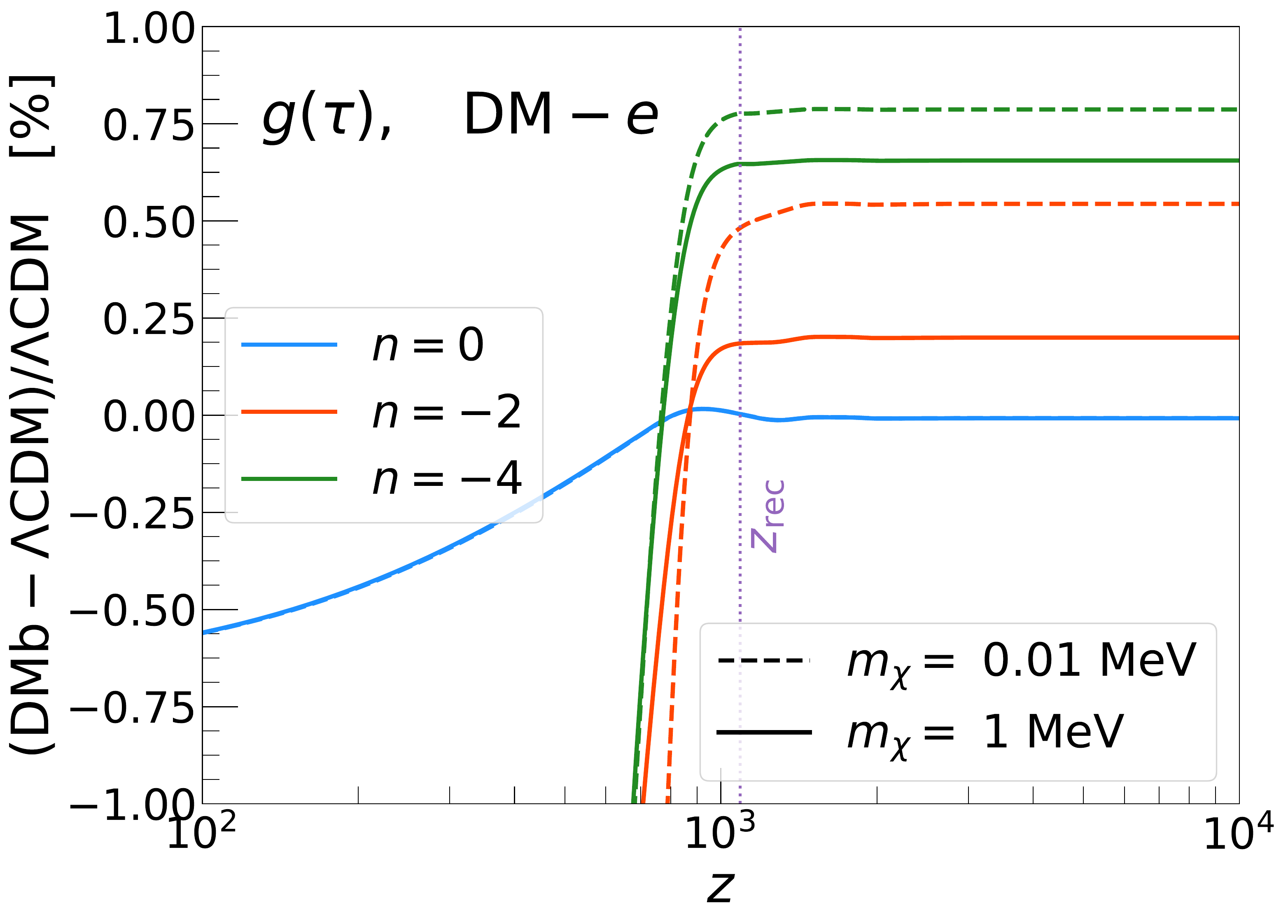}
  \caption{Visibility function $g(\tau)$ percent residuals of DMb with respect to $\LC$, for dark matter--proton ({\bf left}) and dark matter--electron ({\bf right}) interactions; for $n=0$ ({\bf blue}), $n=-2$ ({\bf orange}), and $n=-4$ ({\bf green}). The dark matter masses are $\mx = 0.01~\GeV$ ($0.01~\MeV$) ({\bf dashed}) and $1~\GeV$ ($1~\MeV$) ({\bf solid}) for the dark matter--proton (dark matter--electron) case. For each curve we fix $\sn$ to its corresponding 95\% C.L. value obtained with CMB+BAO data, shown in \Fig{fig:n0_bounds} (for $n=0$), \Fig{fig:n-2_bounds} (for $n=-2$), and \Fig{fig:n-4_bounds} (for $n=-4$).}
  \label{fig:visib}
\end{figure}

Since the baryon--dark matter interactions decrease more slowly with time for more negative $n$ (see \Fig{fig:Rx}), the $n=-2$ and $n=-4$ cases can have significant interactions between the baryons and the dark matter for longer times. Indeed, the $S\Rx$ and $S\frac{\mu_b}{m_\chi}\Rxp$ interaction terms of \Eqs{eq:pert_b}{eq:heat_b} have larger contributions to their thermal velocity dispersion $u_B$ from smaller dark matter masses (see \Eq{eq:uB}). This is enhanced for the dark matter--electron case, where the baryon mass $\mB = m_e$ is also small. For the temperature evolution, shown in \Fig{fig:temps}, this leads to a baryon cooling that is stronger than that in $\LC$. This cooling results in a lower post-recombination ionization fraction, shown in \Fig{fig:ion}, and thus a lower optical depth to recombination (which is an integral and thus sensitive to post-recombination times; see \Eqs{eq:thomson}{eq:opt_depth}). This in turn enhances the visibility function $g(\tau)$ (\Eq{eq:visib}), whose change we show in \Fig{fig:visib}. It can be seen that this change is indeed most significant for dark matter--electron interactions with $n=-2$ and $n=-4$, as we explained above, around the time of recombination ($z_{\rm rec} \approx 1100$ \cite{Aghanim:2018eyx}). Since the visibility function peaks sharply precisely at recombination \cite{Zaldarriaga:1996xe,Lesgourgues:2013qba} the CMB sources described in \Eqs{eq:TT_source}{eq:EE_source} contribute to the CMB spectra mostly at this time. The increase in the visibility function changes the SW and ISW terms in \Eq{eq:TT_source}, and the prefactor in \Eq{eq:EE_source}, increasing the TT and EE power throughout all scales. While this happens for both $n=-2$ and $n=-4$, it is largest for $n=-4$ (\Fig{fig:visib}) and, in fact, it is the dominant effect, since the SW and ISW suppressions we described in the preceding paragraphs are absent. Furthermore, since $\Rx$ actually \textit{grows} with time for $n=-4$ (see \Fig{fig:Rx}), $S\Rx$ eventually dominates in \Eq{eq:pert_b}, which means that the baryon velocity divergence $\theta_b$ starts to differ from $\theta_\gamma$ and grows to match $\theta_\chi$, thereby increasing the Doppler contribution to TT, the second term in \Eq{eq:TT_source}. These effects can be seen in the right panels of \Figs{fig:CTT}{fig:CEE}, most dramatically for $n=-4$. Finally, as Ref.~\cite{Boddy:2018wzy} noted, the lowering of the optical depth to recombination (\ie an increase in the visibility function) is opposite to what would be caused by an early reionization, which would increase the optical depth.  This suppresses the EE ``reionization bump'', which results in a decrease of power or dip at low-$\ell$ in \Fig{fig:CEE}, most evident for dark matter--electron interactions.

\section{Numerical Analysis and Cosmological and Astrophysical Datasets}
\label{sec:num}

In this section, we describe our implementation of the DMb phenomenological model in a Boltzmann solver code for its cosmological evolution. We also discuss the various datasets and methods we use to constrain the model.

\subsection{\texttt{class\_dmb:} Numerical Implementation of DMb Model}
\label{subsec:code}

We modify the publicly available {\tt CLASS} code~\cite{Blas:2011rf} to take into account DMb interactions, both between dark matter and protons and between dark matter and electrons. We achieve this by including \Eqs{eq:heat_x}{eq:heat_b} for the thermal evolution of the baryon and dark matter temperatures in the {\tt thermodynamics.c} module; and \Eqs{eq:pert_x}{eq:pert_b} for the cosmological evolution of the baryon and dark matter $\theta$ perturbations in the {\tt perturbations.c} module. In order to compute exactly the resulting non-standard recombination history we implement the {\tt ndf15} stiff integrator~\cite{Blas:2011rf} into a suitably modified version of {\tt RECFAST}~\cite{Seager:1999bc}.

The DMb model contains the usual six  parameters of $\LC$ described at the beginning of \Sec{sec:obs}, $\{\omega_{\rm b}, \omega_\dm^\tot, H_0, z_{\rm reio}, A_{\rm s}, n_{\rm s} \}$, as well as the new parameters $\{ B, n, \fx, \mx, \sn \}$; here $B$ is a binary variable that denotes whether the dark matter interacts with protons ($B=p$) or with electrons ($B=e$); $n$, $f_\chi$, $\mx$, and $\sn$ are the physical quantities introduced in \Sec{sec:desc}. Our modified {\tt CLASS} code can be found at \href{https://github.com/ManuelBuenAbad/class\_dmb}{\tt github.com/ManuelBuenAbad/class\_dmb}.

\subsection{Cosmological and Astrophysical Datasets}
\label{subsec:data}

In order to constrain the DMb parameter space, we use various methods to compare its predictions with cosmological and astrophysical observables. We perform separate analyses of three different datasets:
\begin{itemize}
    \item CMB and BAO,
    \item Lyman-$\alpha$ forest, and
    \item Abundance of Milky Way subhalos.
\end{itemize}
We devote the rest of this section to the description of these datasets and of our analysis methods. 

\subsubsection{CMB and BAO}
\label{subsubsec:cmb_bao}

We perform a full likelihood analysis on TT, TE, and EE CMB anisotropies and lensing data from Planck 2018~\cite{Aghanim:2018eyx}, as well as BAO data from the Six-degree Field Galaxy Survey (6dFGS)~\cite{Beutler:2011hx} and the Sloan Digital Sky Survey (SDSS)~\cite{Ross:2014qpa,Alam:2016hwk}. The baryon acoustic oscillations, or BAO, are the oscillations of the tightly-coupled baryon--photon plasma in the early Universe imprinted on the matter power spectrum. These oscillations are due to the pressure the photons impart on the plasma. The BAO data therefore encodes a relationship between the typical scale traveled by these acoustic ripples of matter (which would eventually become galaxies) up until the time of baryon--photon decoupling and the distance between the observer and these galaxies. It is therefore commonly used in conjunction with CMB data in order to get rid of parameter degeneracies when performing model fits to data. In order to do our likelihood analysis we use {\tt class\_dmb} code~\cite{Blas:2011rf} and sample the model parameters using the Markov chain Monte Carlo (MCMC) code {\tt MontePython}~\cite{Brinckmann:2018cvx,Audren:2012wb} with the Metropolis-Hastings algorithm. The chains are considered to have converged following the Gelman-Rubin (GR) criterion, $R < 1.01$, where $R$ is the GR statistic~\cite{10.2307/2246093}.

We then scan the parameter space, the baseline parameters $\{\omega_{\rm b}, \omega_\dm^\tot, H_0, z_{\rm reio}, A_{\rm s}, n_{\rm s} \}$ in addition to the velocity-stripped DMb cross section $\si_n$, for all combinations of fixed values for $B \in \{ e, \ p \}$ (interactions with electrons or with protons), $n \in \{ 0, -2, -4 \}$, $\fx \in \{ 1\%,\ 100\% \}$, and $\mx = 10^{N}~\MeV \ \text{for } N \in \{ -2, -1, 0, 1, 2, 3, 4, 5 \}$.  In order to focus on the effects of dark matter--baryon interactions we limit ourselves to $\mx \geq 10~\keV$. Dark matter with smaller masses will behave as WDM at early times, which will have an additional impact on some of the observables we consider. We quote the resulting 95\% C.L. upper limits on $\si_n$ as the constraint on the cross section for that choice of $\{ B, n, \fx, \mx \}$. For a comparison with CMB bounds from previous literature, based on Planck 2015 data \cite{Ade:2015xua}, see Appendix~\ref{app:cmb}.

\subsubsection{Lyman-$\alpha$ forest}
\label{subsubsec:lya}

The spectra of distant quasars (quasi-stellar objects, or QSOs) present Lyman-$\alpha$ absorption lines due to the intergalactic neutral hydrogen lying along their line of sight, the so-called \textit{Lyman-$\alpha$ forest}. Since hydrogen traces the matter distribution at intermediate redshifts of $2 \lsim z \lsim 5$ and small scales $\mathcal{O}(1)~\Mpc/h \lsim \lambda \lsim \mathcal{O}(10)~\Mpc/h$ (see~\cite{McQuinn:2015icp} and references therein), Lyman-$\alpha$ forest data can be used to probe the matter power spectrum and thereby put bounds on cosmological models. We leverage the constraining power of the HIRES/MIKE and XQ-100 data samples of QSO spectra~\cite{Viel:2013fqw,Irsic:2017ixq} by using the recently introduced \textit{area criterion}~\cite{Murgia:2017lwo,Murgia:2017cvj,Murgia:2018now}, which we summarize below.

The matter power spectrum deviation of a model $X$ with respect to $\LC$ can be parameterized by the ratio
\beq
    \xi_X(k) \equiv \frac{P_{\rm 1D}^X(k)}{P_{\rm 1D}^{\LC}(k)} \ ,
\eeq
where $P_{\rm 1D}(k)$ is the one-dimensional matter power spectrum, obtained from the usual 3D matter power spectrum as follows:
\beq
    P_{\rm 1D}(k) \equiv \frac{1}{2\pi} \int_k^\infty \!\dd k' ~ k' P(k') \ .
\eeq

Typically, an experiment probing the matter power spectrum is sensitive to an interval $[k_\minn, k_\maxx]$ of scales. In the case of the MIKE/HIRES+XQ-100 Lyman-$\alpha$ dataset, this is $0.5~h/\Mpc \lsim k \lsim 20~h/\Mpc$~\cite{Irsic:2017ixq}. Model $X$'s matter power spectrum suppression in the scales of interest can then be related to the area under $\xi(k)$ curve,\footnote{This area can be trivially related to the average suppression $\overline{\xi}_X$ of the $X$ model's 1D matter power spectrum over the relevant scales: $A_X = \overline{\xi}_X (k_\maxx - k_\minn)$.}
\beq
    A_X \equiv \int_{k_\minn}^{k_\maxx}\!\dd k ~ \xi_X(k) \ .
\eeq
Note that $A_{\LC} = k_\maxx - k_\minn$ by construction. The area criterion then consists on rejecting those models $X$ whose area $A_X$ deviates from $A_{\LC}$ below some reference value:
\beqa
    \delta A_X & \equiv & \frac{A_{\LC} - A_X}{A_{\LC}} \ ,\\
    \text{if} \quad \delta A_X > \delta A_{\rm ref} & \Rightarrow & \text{reject } X. \label{eq:area_criterion}
\eeqa

This simple and intuitive method for ruling out models has been shown to perform extremely well, if a little conservatively, for the Lyman-$\alpha$ HIRES/MIKE+XQ-100 datasets when compared to a full-fledged statistical analysis~\cite{Murgia:2018now}. Indeed, the authors of Ref.~\cite{Murgia:2018now} modeled the Lyman-$\alpha$ observables in terms of various cosmological, astrophysical, and generalized transfer function shape parameters\footnote{These shape parameters were first introduced in Refs.~\cite{Murgia:2017lwo,Murgia:2017cvj}.  In Ref.~\cite{Murgia:2018now}, this so-called $\{ \alpha, \beta, \gamma \}$ parameterization of non-CDM models was employed, generalizing the well known analytic formula for the WDM transfer function~\cite{Viel:2005qj} (see \Eqs{eq:twdm}{eq:awdm}).} with the help of a large suite of dedicated hydrodynamical $N$-body simulations, and proceeded to constrain these parameters with a likelihood analysis in an MCMC approach. They find that a model's area deviation $\delta A$ strongly and positively correlates with the $\chi^2$ of the model's fit to the Lyman-$\alpha$ data.

A previous analysis of the MIKE/HIRES+XQ-100 datasets in the context of WDM found a 95\% C.L. bound of $m_\wdm = 5.3~\keV$ and a looser one of $m_\wdm = 3.5~\keV$ under more conservative assumptions of the intergalactic medium temperature~\cite{Irsic:2017ixq}. The area deviations for WDM of these masses are $\delta A^\wdm_{5.3~\keV} = 0.31$ and $\delta A^\wdm_{3.5~\keV} = 0.46$, respectively. Taking the conservative value as $\delta A_{\rm ref}$, the authors of Ref.~\cite{Murgia:2018now} found that applying the area criterion, \Eq{eq:area_criterion}, to their generalized transfer function parameterization model accurately reproduces the 95\% C.L. constraints from their MCMC statistical analysis. We point out that the area criterion for both of these $m_\wdm$ values has recently been used to constrain feebly interacting massive particles~\cite{DEramo:2020gpr} and self-interacting dark matter~\cite{Egana-Ugrinovic:2021gnu}. Finally, it is worth mentioning that a recent study~\cite{Palanque-Delabrouille:2019iyz} of cosmological constraints on WDM, employing new high-precision Lyman-$\alpha$ measurements~\cite{Chabanier:2018rga} based on QSO spectra from the Baryon Oscillation Spectroscopic Survey (BOSS) and the Extended Baryon Oscillation Spectroscopic Survey (eBOSS) collaborations~\cite{Dawson2012,Dawson:2015wdb} in combination with XQ-100, found a 95 \% C.L. bound of $m_\wdm = 5.3~\keV$ as well.

With these encouraging results in mind we now turn to the DMb model, which presents matter power spectrum transfer function suppressions similar to those studied in Ref.~\cite{Murgia:2018now} (see \Sec{sec:obs}). We fix the six standard $\LC$ cosmological parameters to their Planck 2018 values~\cite{Aghanim:2018eyx} and apply the criterion with $\delta A_{\rm ref} = \delta A^\wdm_{5.3~\keV} = 0.31$ to a grid of points in the $( \mx, \sn )$ parameter space for $\fx = 100\%$ and all combinations of $B$ and $n$. We leave the case of the more conservative $\delta A_{\rm ref} = \delta A^\wdm_{3.5~\keV} = 0.46$ for Appendix~\ref{app:lya}.

A full statistical likelihood analysis of the MIKE/HIRES+XQ-100 or BOSS/eBOSS+XQ-100 datasets, in the spirit of those made in~\cite{Murgia:2018now,Archidiacono:2019wdp,Palanque-Delabrouille:2019iyz}, where all the DMb model parameters (including the six baseline $\LC$ parameters) are fitted to the Lyman-$\alpha$ data (either by itself or in combination with Planck's), is left to future work.\footnote{Lyman-$\alpha$ has been used previously in the literature within the context of DMb models~\cite{Dvorkin:2013cea,Xu:2018efh}. The data used was in the form of linear matter power spectrum measurements from SDSS-II low resolution, low signal-to-noise quasar spectra~\cite{McDonald:2004eu}. However, these measurements rely on modeling the matter power spectrum within $\LC$~\cite{McDonald:2004eu,McDonald:2004xn}, and it is therefore not obvious that these constraints can be used to bound models that produce non-$\LC$ matter power spectra, such as DMb.} There are two additional caveats: \textit{i.}  the WDM constraints from Lyman-$\alpha$ suffer from systematical uncertainties in the modeling of the intergalactic medium, such as its temperature fluctuations~\cite{Hui:2016ltb}; \textit{ii.} similar to the scenario of Milky Way subhalos discussed below, DMb interactions can affect Lyman-$\alpha$ forest observables through late-time effects that are different from those of WDM/non-cold DM. For example, late-time DMb interactions can change the gas temperature~\cite{Munoz:2017qpy,Slatyer:2018aqg}. We leave a dedicated study that includes those late-time effects to future work.

\subsubsection{Milky Way subhalos}
\label{subsubsec:joint}

The abundance of the Milky Way (MW) subhalos inherits the small-scale properties of the matter power spectrum. The observed abundance of MW subhalos has been  used to constrain thermal relic warm dark matter (WDM) that suppresses the small-scale structures by a transfer function squared~\cite{Bode:2000gq,Schneider:2011yu}
\beq\label{eq:twdm}
T^2_\text{WDM} (k)= {\frac{P_\text{WDM} (k)}{P_\text{CDM} (k)}}= \left[1+ (\alpha k)^{2\nu}\right]^{-\frac{10}{\nu}}\,,
\eeq
with $\nu = 1.12$ and 
\beq\label{eq:awdm}
\alpha = 0.049 \left(\frac{m_\text{WDM}}{\text{keV}}\right)^{-1.11} \left(\frac{\Omega_\text{WDM}}{0.25}\right)^{0.11} \left(\frac{h}{0.7}\right)^{1.22} \,\text{Mpc}/\text{h}\,,
\eeq
where $m_\text{WDM}$ and $\Omega_\text{WDM}$ are the WDM mass and abundance, respectively.  
For $\fx \sim 100\%$, sizable interactions between dark matter and baryons in the early Universe can suppress the matter power spectrum on small scales, allowing us to constrain the dark matter--baryon interactions from the MW subhalos data. In particular, one can translate the constraints on the transfer function of WDM to those of DMb~\cite{Nadler:2019zrb, Nadler:2020prv, Maamari:2020aqz}.\footnote{The method has also been used to constrain dark matter--photon interactions~\cite{Escudero:2018thh}.} Such translation works well if the transfer function of the DMb has a similar shape to that of the WDM. If the shapes of the two transfer functions are very different, the translation is approximate. We will address this point in more detail when discussing the matching criteria below. Note that DMb interactions during halo or galaxy formation can affect both the abundance and the properties of MW subhalos through effects such as enhanced subhalo disruptions. Addressing these late-time effects requires dedicated cosmological hydrodynamical simulations.
The enhanced subhalo disruption reduces the  abundance of MW subhalos and is therefore degenerate with the effect of a larger dark matter--baryon interactions in the early Universe.

The abundance of the MW subhalos can be inferred from various observables:
\begin{enumerate}[label=(\roman*)]
    \item \textbf{The observed luminous MW satellite galaxies.} Ref.~\cite{Nadler:2019zrb} constrained WDM masses to be $m_\text{WDM} > 3.26\,\keV$ (here and below, the constraints are quoted as 95\%~C.L.) using a population of classical and SDSS-observed MW satellite galaxies, which has been updated to $m_\text{WDM} > 6.5\,\keV$ using MW satellite galaxies from the Dark Energy Survey (DES) and the Panoramic Survey Telescope and Rapid Response System (Pan-STARRS)~\cite{Nadler:2020prv}. Ref.~\cite{Newton:2020cog}  found a weaker constraint of $m_\text{WDM} > 2.02\,\keV$ using MW satellite galaxies from SDSS and DES. The discrepancy between the two groups is most likely driven by differences in their accounting of the number of satellite galaxies formed in MW subhalos~\cite{Enzi:2020ieg,Newton:2020cog}: the analysis of Refs.~\cite{Nadler:2019zrb, Nadler:2020prv} is based on an abundance matching technique extrapolated to low magnitudes of the luminosity function of the MW satellite galaxies~\cite{Wechsler:2018pic}, whereas that of Ref.~\cite{Newton:2020cog} is based on semi-analytical models of galaxy formation.
    \item \textbf{Stellar streams.} The MW subhalos can perturb stars in the stellar streams and leave imprints~\cite{Banik:2018pjp, Bonaca:2018fek, Banik:2019smi, Dalal:2020mjw}. Using GD-1 and Pal-5 streams, together with classical MW satellite galaxies data, Ref.~\cite{Banik:2019smi} found $m_\text{WDM} >  6.3\,\keV$.
    \item \textbf{Strong gravitational lensing.} The MW subhalos also affect  flux ratios and  image distortions of lensed  compact sources~\cite{Birrer:2017rpp,Vegetti:2018dly, Ritondale:2018cvp,gilman2019probing,Hsueh:2019ynk,gilman2020warm}. Ref.~\cite{Nadler:2021dft} constrained WDM to require $m_\text{WDM} > 4.9\,\keV$ using flux ratios and image positions of the lensed image of eight quasars. A joint constraint of $m_\text{WDM} > 9.7\,\keV$ is reached after combining the strong lensing constraint with the MW satellite population from Ref.~\cite{Nadler:2020prv}. Ref.~\cite{Enzi:2020ieg} found $m_\text{WDM} > 2.0\,\keV$ ($m_\text{WDM} > 0.12\,\keV$) by re-analyzing  low-(high-)redshift galaxy--galaxy lens systems from Refs.~\cite{Vegetti:2018dly, Ritondale:2018cvp} using the surface brightness data of magnified arcs or Einstein rings of lensed galaxies. Combining the results with the MW satellite population of Ref.~\cite{Newton:2020cog} and the Lyman-$\alpha$ forest leads to a joint constraint of $m_\text{WDM} > 6.733\,\keV$.
\end{enumerate}

\begin{figure}[t]
  \centering
  \includegraphics[width=0.49\textwidth]{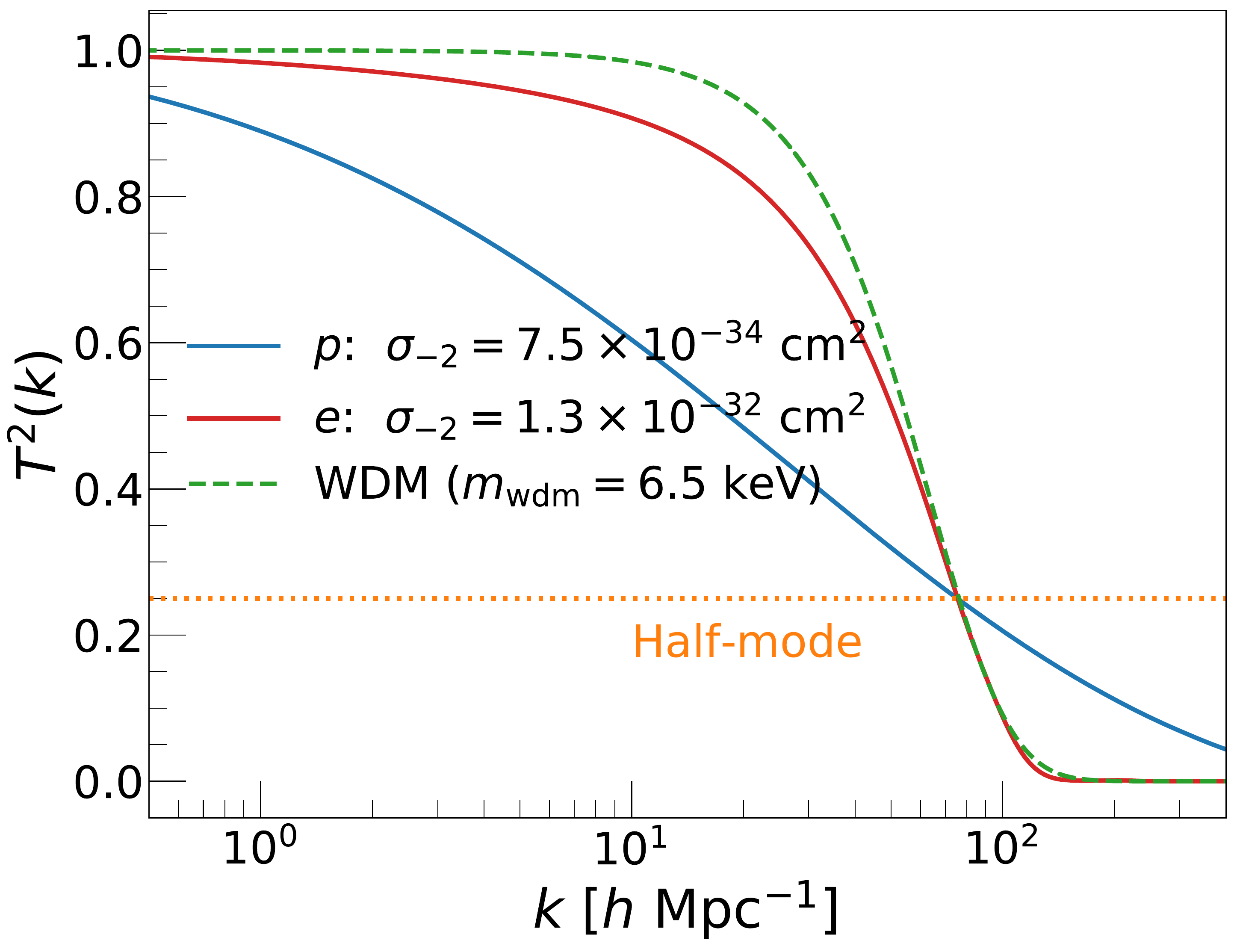}
  \includegraphics[width=0.49\textwidth]{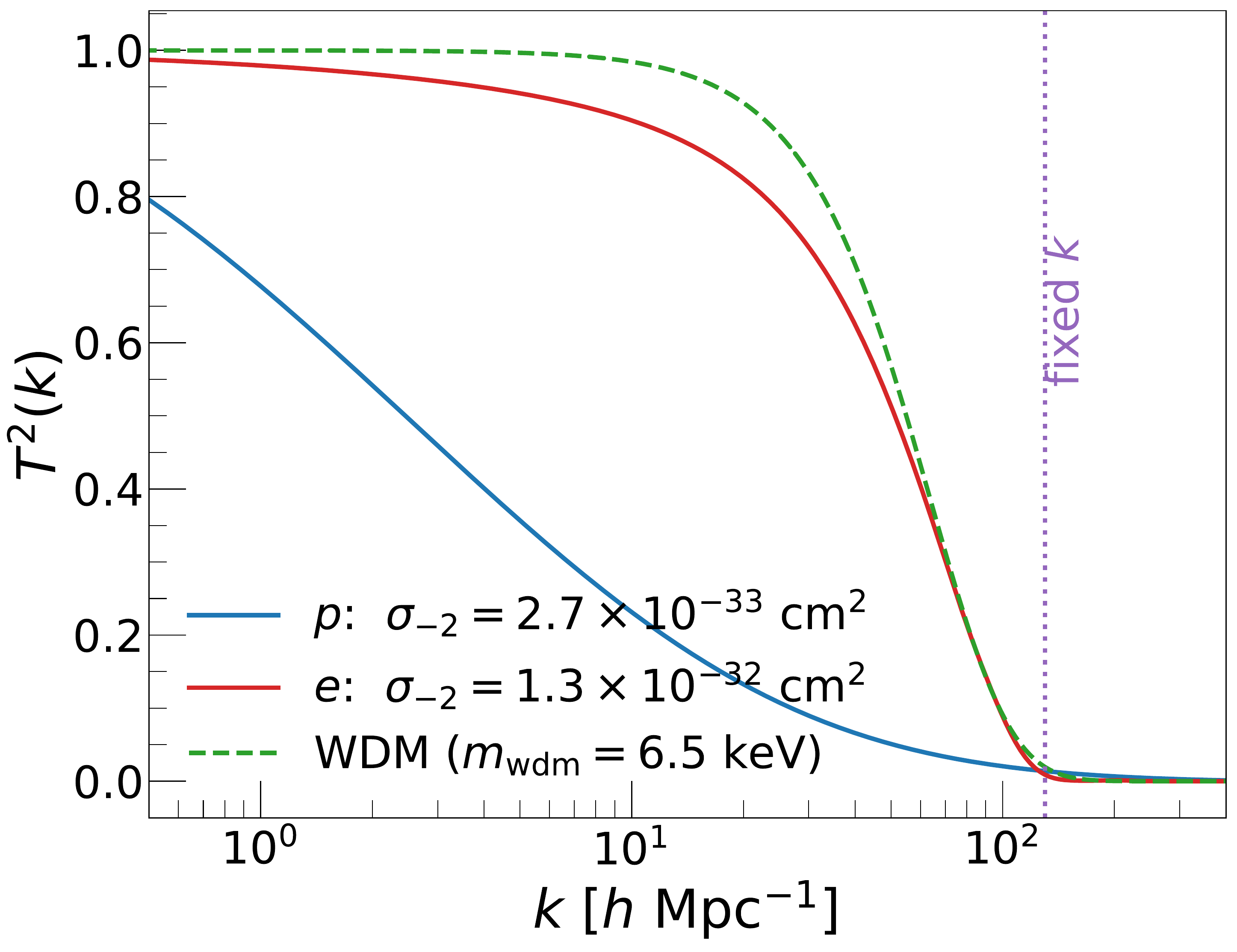}
  \caption{Examples of the squared transfer functions for $n=-2$ dark matter--protons ({\bf blue}) and dark matter--electron ({\bf red}) interactions constrained by bounds from MW subhalo data. In the {\bf left} panel, the {\bf blue} ({\bf red}) line corresponds to dark matter--protons (dark matter--electron) interactions with $m_\chi = 1\, \MeV$ and $\sigma^{\chi p}_{-2} = 7.5\times 10^{-34}\,\text{cm}^2$ ($\sigma^{\chi e}_{-2} = 1.3\times 10^{-32}\,\text{cm}^2$). In the {\bf right} panel, the {\bf blue} ({\bf red}) line corresponds to dark matter--proton (dark matter--electron) interactions with $m_\chi = 1\, \MeV$ and $\sigma^{\chi p}_{-2} = 2.7\times 10^{-33}\,\text{cm}^2$ ($\sigma^{\chi e}_{-2} = 1.3\times 10^{-33}\,\text{cm}^2$). The {\bf dashed green} line is the transfer function for WDM with $m_{\rm wdm} = 6.5~\keV$. The {\bf left} ({\bf right}) panel uses the \textit{half-mode} (\textit{fixed $k$}) matching scheme, represented by the {\bf horizontal orange} ({\bf vertical purple}) {\bf dotted} line at $T^2(k)=0.25$ ($k=k_\text{match}=130~\Mpc/h$). Note that in the dark matter--electron case, when the DMb transfer function has a very similar shape to that of WDM, both schemes give approximately the same results. }
  \label{fig:Tk_mws}
\end{figure}

\begin{table}[t]
   \centering
   \begin{tabular}{@{} llr|rr @{}}
      \hline
      Ref. & Data type & $m_\text{WDM}^\text{95\%C.L.}$ [keV] & $h$ & $\Omega_\text{dm}$ \\
      \hline
     \cite{Nadler:2019zrb}       & luminous satellite galaxies & $3.26$ &  0.67 & 0.265\\
     \cite{Nadler:2020prv} & luminous satellite galaxies & $6.5$ & 0.7 & 0.24\\
     \cite{Newton:2020cog} & luminous satellite galaxies & $2.02$ & 0.7 & 0.23\\
     \cite{Banik:2019smi} & stellar stream, luminous satellite galaxies & $6.3$ & 0.67 & 0.26 \\ 
     \cite{Nadler:2021dft} & strong lensing, luminous satellite galaxies & $9.7$ & 0.7 & 0.24 \\
     \cite{Enzi:2020ieg} & strong lensing, luminous satellite galaxies, Lyman-$\alpha$ & 6.733 & 0.7 & 0.25\\
      \hline
   \end{tabular}
   \caption{95\% C.L. lower limits on the WDM mass (3rd column) from MW subhalo data reported in the literature. The 2nd, 4th, and 5th columns respectively show the data type, the value of $h$, and the value of $\Omega_\text{dm}$ that was used in the derivation of the limits.}
   \label{tab:mw-limit}
\end{table}
Although the most stringent constraint on $m_\text{WDM}$ comes from Ref.~\cite{Nadler:2021dft}, which combines strong gravitational lensing with MW satellite population, the strong lensing analysis based on the flux ratios is highly sensitive to the central densities of the subhalos (see \eg~\cite{Nadler:2021dft} and references therein).\footnote{We thank D.~Gilman and E.~Nadler for drawing our attention to this issue.} The concentration--mass relation has been properly studied for WDM and CDM, but not for interacting dark matter.\footnote{Besides, self-interactions of dark matter can also change the central densities of subhalos and thus affect strong lensing observables (see \eg~\cite{Gilman:2021sdr}).} We thus adopt the most stringent constraint that relies on luminous MW satellite galaxies without strong lensing input, namely $m_\text{WDM} > 6.5\, \keV$~\cite{Nadler:2020prv}, for the main text of this work. We show the constraints on the DMb parameter space under $m_\text{WDM} > 2.02\,\keV$, $6.73 \, \keV$, and $9.7\,\keV$ in Appendix~\ref{app:mwsg}. The corresponding cosmological parameters that used in $T_\text{WDM}$ are $h=0.7$ and $\Omega_\text{dm} =0.24$~\cite{hinshaw2013nine}. To translate bounds on thermal WDM to bounds on the DMb model, we scan over dark matter mass $m_\chi$, its velocity stripped momentum transfer cross section with protons or electrons $\sigma^{\chi p/\chi e}_n$, and the velocity power $n \in \{ 0, -2, -4 \}$. For each parameter set, we generate the linear matter power spectra using \texttt{class\_dmb}. The parameter set is allowed if the resulting transfer function of DMb, given by~\eqref{eq:transferDMb}, is less than or equal to the transfer function of WDM with the critical mass $T_\text{WDM,crit} = T_\text{WDM} (k; m_\text{WDM} = 6.5\,\keV)$. More precisely, we use two matching criteria for the comparison process:
\begin{enumerate}[label=(\alph*)]
    \item {\bf Half-mode:} Adopted in Refs.~\cite{Nadler:2019zrb, Nadler:2020prv}. $T_\text{DMb}$ matches $T_\text{WDM,crit}$ at $T_\text{WDM,crit} = 50\%$ (\ie, $T^2_\text{WDM,crit} = 25\%$). This
 corresponds to matching the two transfer functions   at the comoving wavenumber of $k_\text{match}=k_\text{half-mode}\simeq 74\,h/\Mpc$ for our choice of critical WDM mass (\ie, $m_\text{WDM} > 6.5\, \keV$). 
    \item {\bf Fixed $k$:} Adopted in Ref.~\cite{Maamari:2020aqz}. $T_\text{DMb}$ matches $T_\text{WDM,crit}$ at the comoving wavenumber $k_\text{match} = 130\,h/\Mpc$, which corresponds to $T_\text{WDM,crit} \approx 2\%$, and guarantees  $T_\text{WDM,crit}$ always exceeds $T_\text{DMb}$ at a smaller $k$.
\end{enumerate}
To illustrate the differences in the constraints from the two matching criteria, we show four squared transfer functions  with four $(m_\chi, \sigma_{-2})$ benchmarks for $n=-2$ type DMb interactions in \Fig{fig:Tk_mws} as an example. The two transfer functions in the left (right) panel are matched to the squared transfer function of the WDM under the half-mode (fixed $k$) criterion. As \Fig{fig:Tk_mws} shows, if the two transfer functions have a similar shape (\eg, as for dark matter--electron interactions with $m_\chi = 1\, \MeV$), criteria (a) and (b) yield similar constraints. If the DMb transfer function is significantly different from that of WDM (\eg, as for dark matter--proton interactions with $m_\chi = 1\, \MeV$), criterion (a) can easily disfavor a given set of parameters even if these parameters are still allowed by criterion (b). In Appendix~\ref{app:mwsg}, we also derive  constraints under the criteria of the smallest halo mass~\cite{Nadler:2019zrb}, which has been used to derive constraints for various types of DMb interactions in the literature~\cite{ Bringmann:2006mu,kasahara2009neutralino, Binder:2016pnr, Gondolo:2012vh, Bertoni:2014mva, Boehm:2014vja, Gondolo:2016mrz, Chu:2018qrm}.

\section{Results}
\label{sec:res}

In this section, we present our constraints derived from the three datasets discussed in the previous section. These bounds were obtained in the context of the DMb phenomenological models described in \Sec{sec:desc} and used in our Boltzmann code, where the dark matter particles couple only to a single baryon species $B$ (either electrons or protons) with a momentum-transfer cross section that depends on the relative velocity to some power $n$. In \Sec{subsec:pheno}, we show these bounds in the $(\mx, \sn)$ parameter space, for each independent combination of $B$ and $n$ (Figs.~\ref{fig:n0_bounds}, \ref{fig:n-2_bounds}, and \ref{fig:n-4_bounds}). In \Sec{subsec:specific}, we embed these phenomenological models in a specific particle physics model that naturally gives rise to the velocity power $n$. These models generally lead to dark matter couplings with both protons \textit{and} electrons simultaneously. We therefore combine our bounds on $\si^{\chi p}_n$ and $\si^{\chi e}_n$ in the context of these specific particle models, and describe the bounds in terms of the (model-dependent) form factor-stripped DMb cross section $\sbar$ commonly used in direct-detection experiments (Figs.~\ref{fig:n0_dp_bounds}, \ref{fig:n-2_dp_bounds}, and \ref{fig:n-4_dp_bounds}).

\subsection{Constraints on phenomenological models}
\label{subsec:pheno}

\begin{figure}[t]
  \centering
  \includegraphics[width=0.49\textwidth]{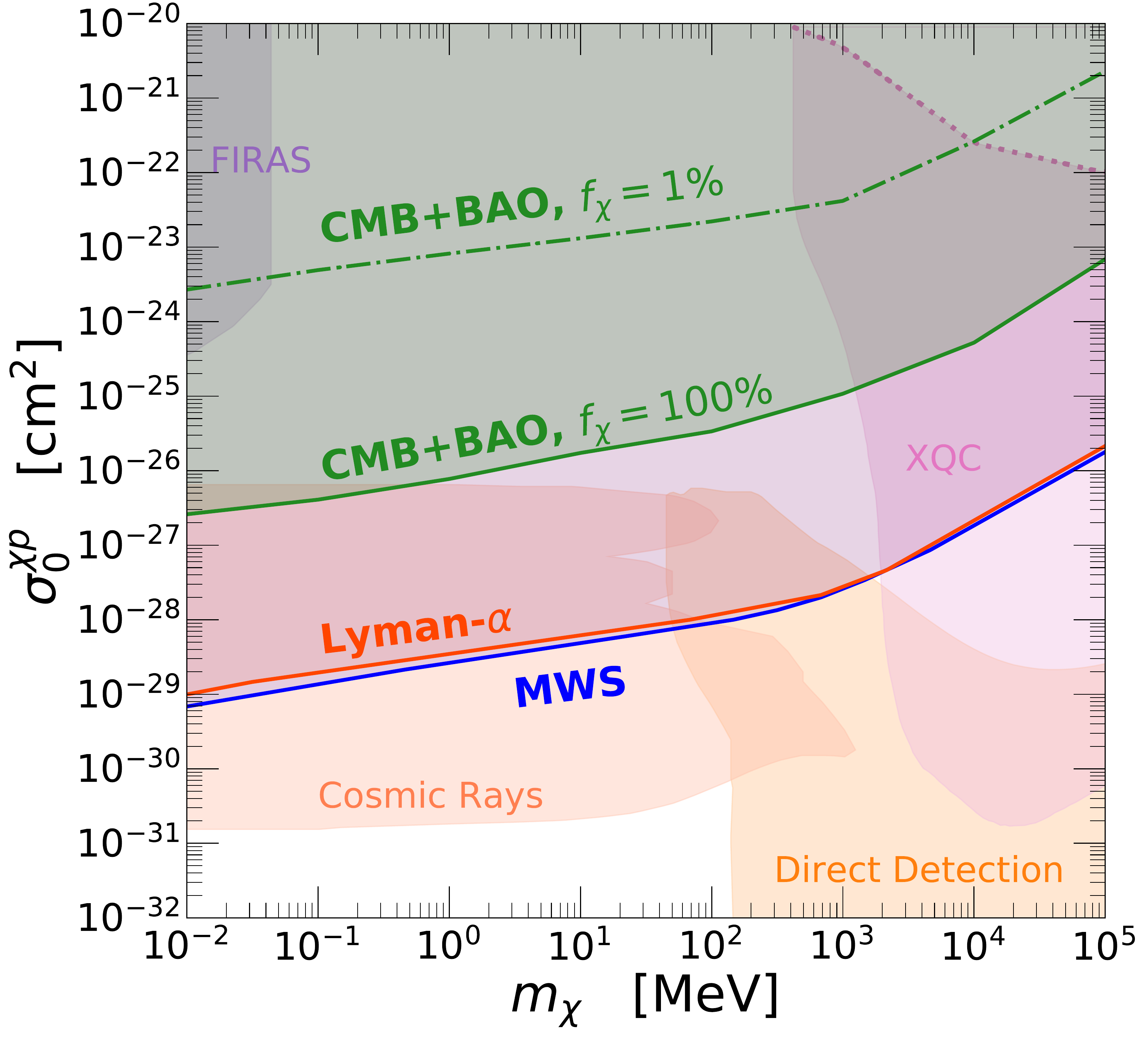}
  \includegraphics[width=0.49\textwidth]{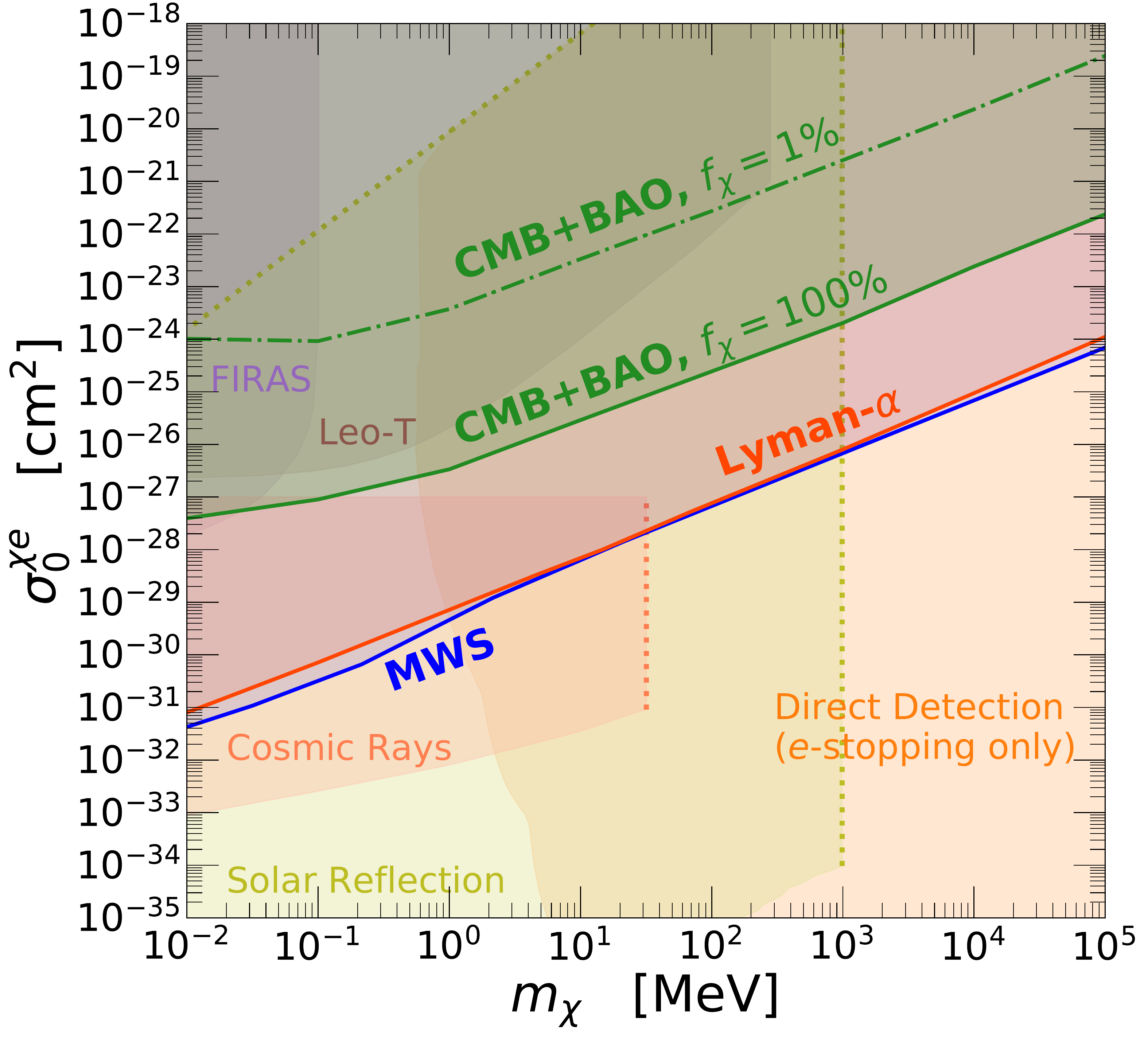}
  \caption{95\% C.L. bounds from CMB+BAO ({\bf solid green}), MW subhalos ({\bf solid blue}; both the half-mode and fixed $k$ matching schemes yield the same result), and Lyman-$\alpha$ forest ({\bf solid red}) datasets on the $(\mx, \sn)$ parameter space, for $n=0, f_\chi = 100\%$ and ({\bf left}) dark matter--protons and ({\bf right}) dark matter--electron interactions. The {\bf dash-dotted green} line shows the constraint for $f_\chi = 1\%$ based on CMB+BAO data. Bounds from MW subhalos and Lyman-$\alpha$ forest are recast from bounds for thermal WDM/non-cold DM. See text for details of the recast and the limitations. We also include constraints from previous literature for $f_\chi=100\%$ dark matter--baryon interactions. For dark matter--proton interactions ({\bf left}): XQC rocket ({\bf pink})~\cite{McCammon:2002gb,Erickcek:2007jv,Mahdawi:2018euy}, various direct-detection experiments ({\bf orange}: CRESST-III~\cite{Abdelhameed:2019hmk,Abdelhameed:2019mac}, CRESST surface run~\cite{Angloher:2017sxg}, XENON1T~\cite{Aprile:2017iyp}, and Migdal effect-based EDELWEISS~\cite{Armengaud:2019kfj}), and cosmic-ray accelerated dark matter \cite{Yin:2018yjn} ({\bf coral red}: MiniBOONE and XENON1T~\cite{Bringmann:2018cvk}; and Daya Bay and KamLAND~\cite{Cappiello:2019qsw}). For dark matter--electron interactions ({\bf right}): gas cooling of Leo-T dwarf galaxy ({\bf brown}) ~\cite{Wadekar:2019xnf}, various direct-detection experiments as considered in~\cite{Emken:2019tni} ({\bf orange}: XENON10~\cite{Essig:2012yx,Essig:2017kqs,Angle:2011th}, XENON100~\cite{Essig:2017kqs,Aprile:2016wwo}, DarkSide-50~\cite{Agnes:2018oej}, CDMS-HVeV~\cite{Agnese:2018col}, protoSENSEI~\cite{Crisler:2018gci,Abramoff:2019dfb}, and SENSEI at MINOS~\cite{Barak:2020fql}), all with their upper boundaries obtained with electronic stopping only; cosmic-ray accelerated dark matter ({\bf coral red}: Super-K and MiniBooNE~\cite{Ema:2018bih, Cappiello:2019qsw}), and bounds from solar reflection ({\bf yellow}) \cite{An:2017ojc,Emken:2021lgc}. The {\bf dotted} lines represent our conservative assumptions about the limits of the XQC ({\bf pink}, with the ceiling from~\cite{Erickcek:2007jv}); cosmic-ray ({\bf coral red}, taken from the plotting range of~\cite{Ema:2018bih}), and solar reflection ({\bf yellow}) bounds. For the latter  we take its upper boundary to be given by the same electron-only stopping as in the case of direct-detection experiments. In both panels: FIRAS ({\bf purple})~\cite{Fixsen:1996nj,Ali-Haimoud:2015pwa,Ali-Haimoud:2021lka}. 
  }
  \label{fig:n0_bounds}
\end{figure}

\begin{figure}[t]
  \centering
  \includegraphics[width=0.49\textwidth]{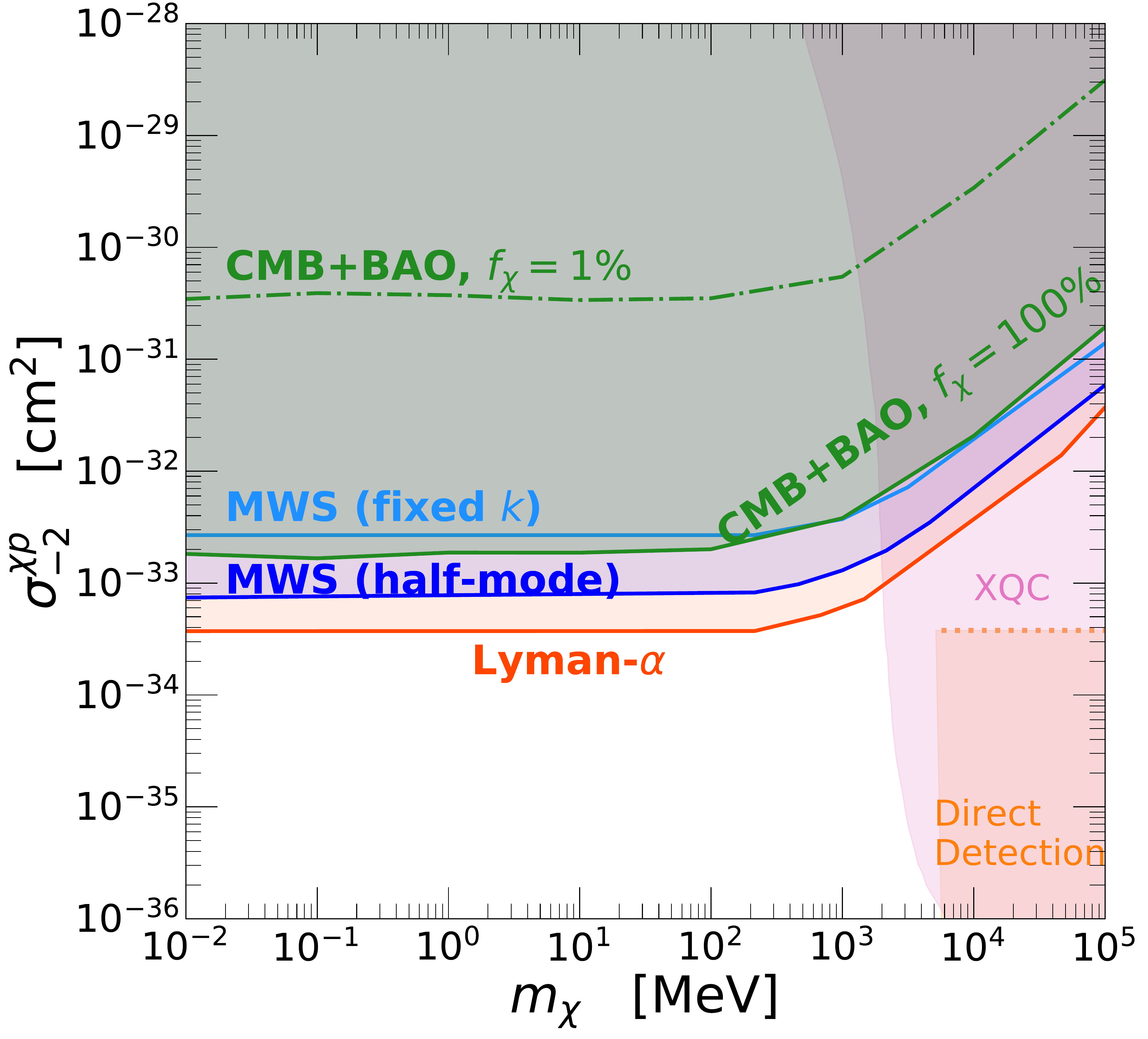}
  \includegraphics[width=0.49\textwidth]{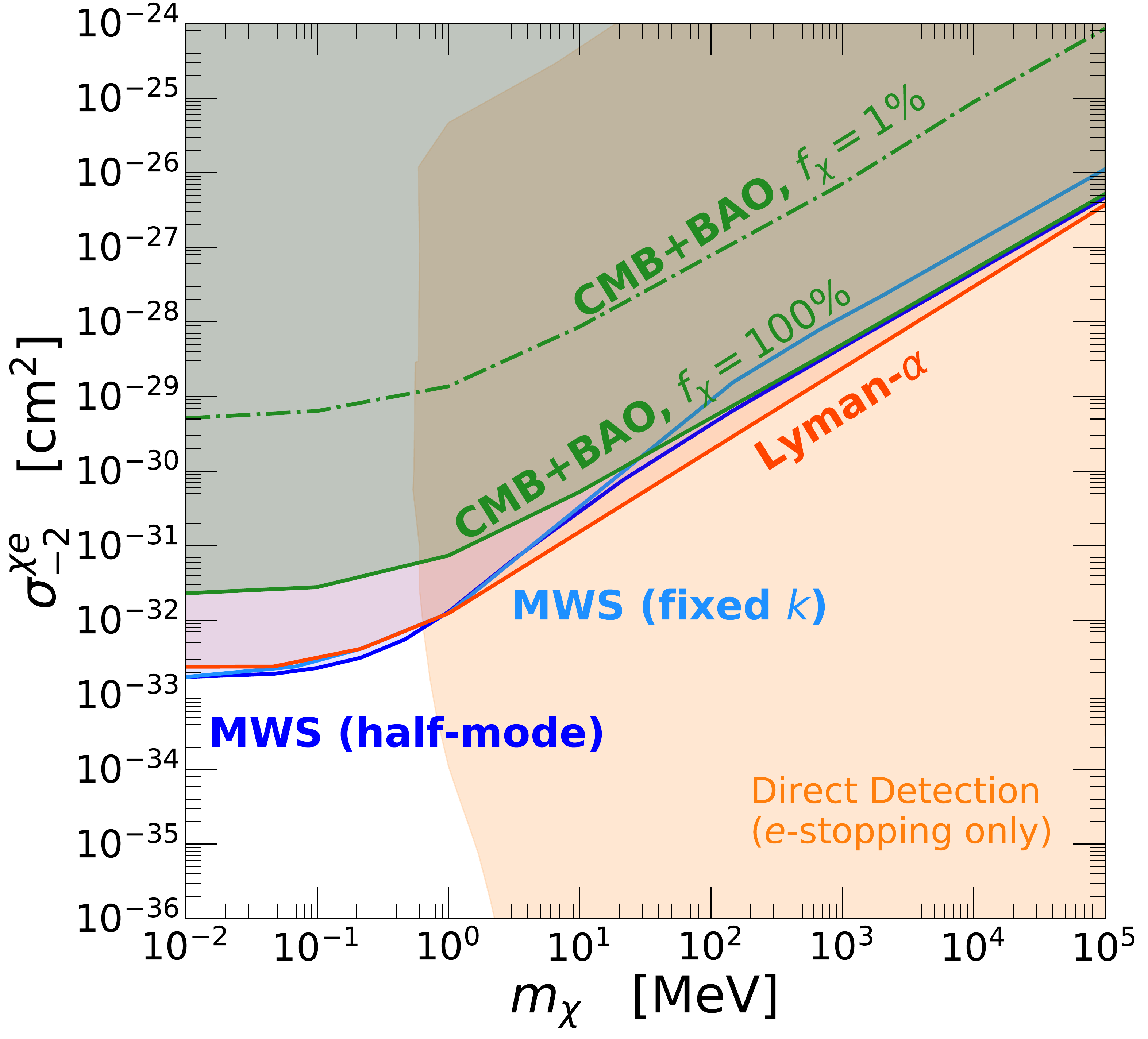}
  \caption{95\% C.L. bounds from CMB+BAO ({\bf solid green}), MW subhalos ({\bf solid blue}), and Lyman-$\alpha$ forest ({\bf solid red}) datasets on the $(\mx, \sn)$ parameter space, for $n=-2, f_\chi = 100\%$ and ({\bf left}) dark matter--protons and ({\bf right}) dark matter--electron interactions. The {\bf dash-dotted green} line shows the constraint for $f_\chi = 1\%$ based on CMB+BAO data. Bounds from MW subhalos and Lyman-$\alpha$ forest are recast from bounds for thermal WDM/non-cold DM. For the MW subhalos constraints, we consider both the half-mode ({\bf dark blue}) and the fixed $k$ ({\bf light blue}) matching criteria for the recast procedure. See text for more details of the recast and the limitations. We also include constraints from previous literature for $f_\chi=100\%$ dark matter--baryon interactions. For dark matter--proton interactions ({\bf left}): XENON1T direct detection bounds ({\bf orange})~\cite{Arina:2020mxo}, and XQC rocket ({\bf pink}) \cite{McCammon:2002gb,Mahdawi:2018euy}. For dark matter--electron interactions ({\bf right}): various direct-detection experiments as considered in~\cite{Emken:2019tni} ({\bf orange}: XENON10~\cite{Essig:2012yx,Essig:2017kqs,Angle:2011th}, XENON100~\cite{Essig:2017kqs,Aprile:2016wwo}, DarkSide-50~\cite{Agnes:2018oej}, CDMS-HVeV~\cite{Agnese:2018col}, protoSENSEI~\cite{Crisler:2018gci,Abramoff:2019dfb}, and SENSEI at MINOS~\cite{Barak:2020fql}), all with their upper boundaries obtained with electronic stopping only. In both panels: CRESST~\cite{Angloher:2002in}, and CDMS ({\bf orange})~\cite{Akerib:2003px}, as compiled in~\cite{Sigurdson:2004zp}. Also present but not shown are constraints from RRS balloon ({\bf purple})~\cite{Rich:1987st}, which overlap with those from XQC and direct detection.}
  \label{fig:n-2_bounds}
\end{figure}

\begin{figure}[t]
  \centering
  \includegraphics[width=0.49\textwidth]{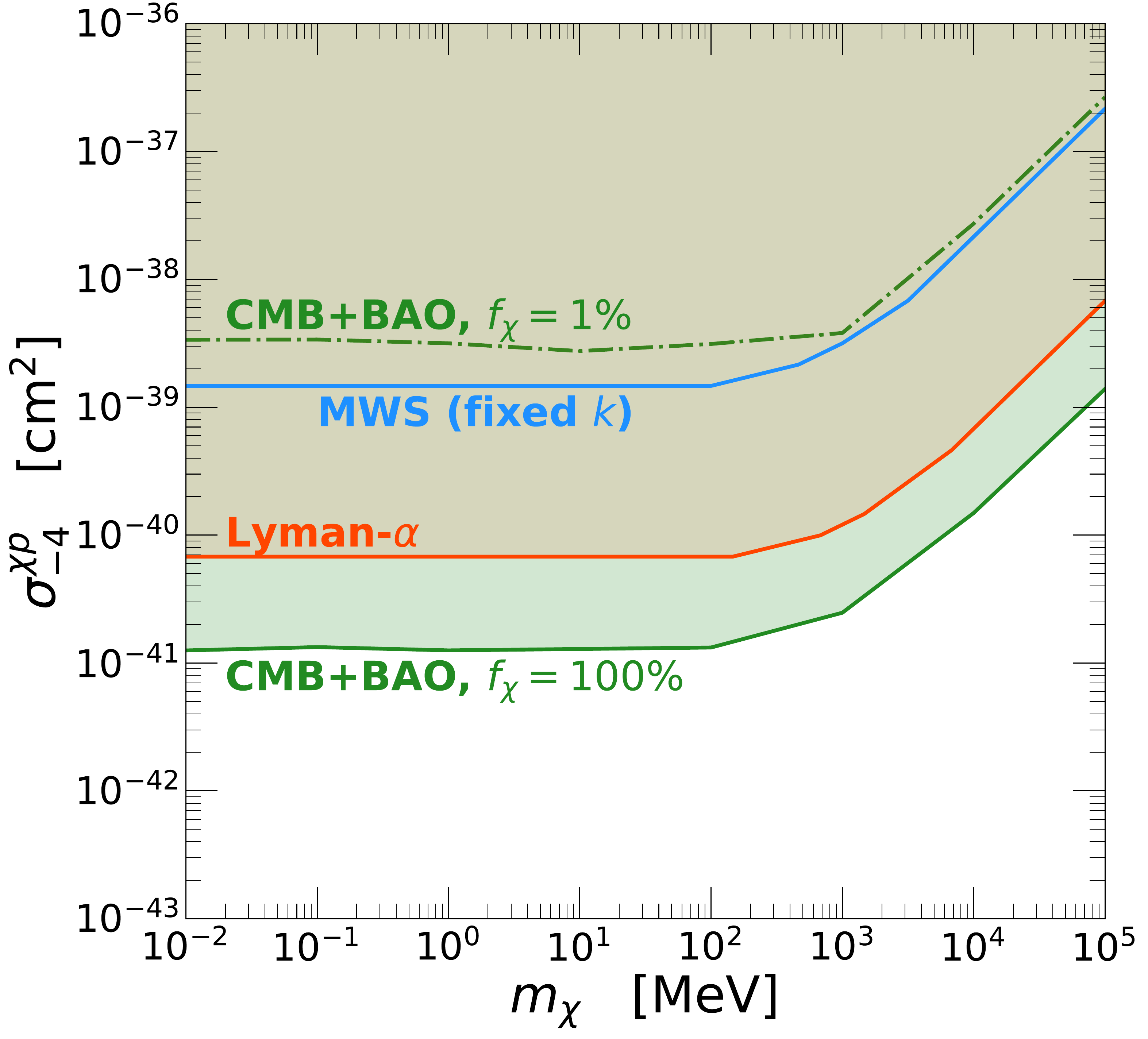}
  \includegraphics[width=0.49\textwidth]{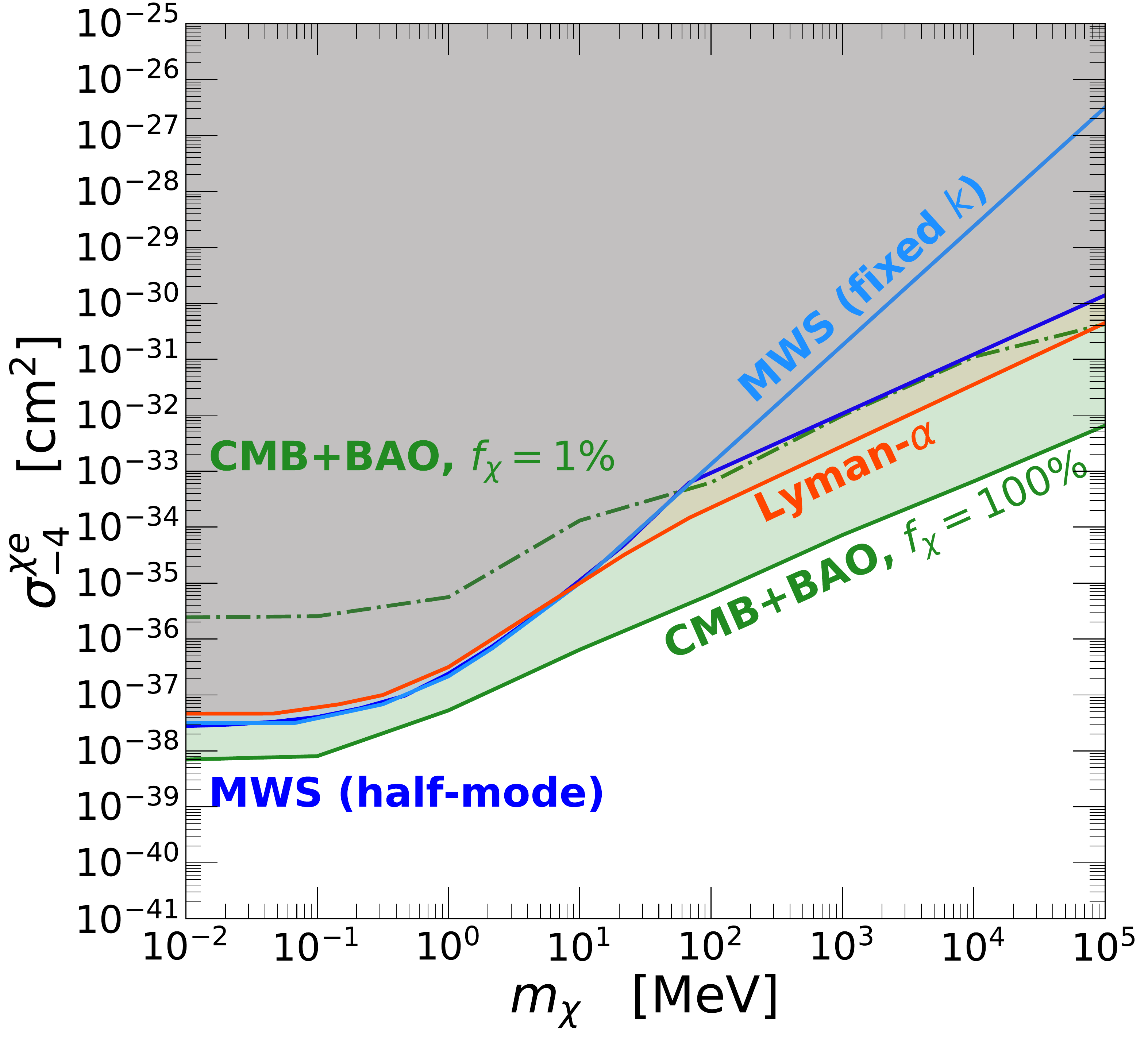}
  \caption{95\% C.L. bounds from CMB+BAO ({\bf solid green}), MW subhalos ({\bf solid blue}), and Lyman-$\alpha$ forest ({\bf solid red}) datasets on the $(\mx, \sn)$ parameter space, for $n=-4, f_\chi = 100\%$ and ({\bf left}) dark matter--protons and ({\bf right}) dark matter--electron interactions. The {\bf dash-dotted green} line shows the constraint for $f_\chi = 1\%$ based on CMB+BAO data. Bounds from MW subhalos and Lyman-$\alpha$ forest are recast from bounds for thermal WDM/non-cold DM. For the MW subhalos constraints we consider the fixed $k$ matching criterion ({\bf light blue}) for both dark matter interactions with protons and electrons when recasting thermal WDM bounds to DMb, while we use the half-mode matching criterion ({\bf dark blue}) only for interactions with electrons. The reason is that the DMb transfer function for the dark matter--proton case is of a very different functional shape when compared to that for WDM. Note that most experimental bounds for $n=-4$ found in the literature are written in terms of the direct detection cross section $\sbar^{\chi B}_{-4}$ (defined in \Sec{subsec:specific}). See text for more details of the recast and the limitations. Since the mapping between $\sbar^{\chi B}_{-4}$ and $\si^{\chi B}_{-4}$ is not straightforward (see text for details), we do not include them in these plots. In \Sec{subsec:specific} we consider a specific model with ultra-light dark photons that can give rise to $n=-4$, and we defer showing the applicable constraints from the literature until then. 
  }
  \label{fig:n-4_bounds}
\end{figure}

The methods and datasets discussed in \Sec{sec:num} can constrain the family of DMb phenomenological models described in \Sec{sec:desc} in the $(\mx, \sn)$ parameter space for each pairwise choice of $B$ and $n$. Figs.~\ref{fig:n0_bounds}, \ref{fig:n-2_bounds}, and \ref{fig:n-4_bounds} show these for $n=0$, $n=-2$, and $n=-4$, respectively, and separately for dark matter--proton and dark matter--electron interactions, all for $f_\chi = 100\%$.\footnote{We found our bound on $\sigma_{0}^{\chi p}$ from MW satellite galaxy data (with $m_\text{WDM} > 6.5\, \keV$) agrees with that shown in the right panel of Fig.~2 of~\cite{Nadler:2020prv}.} In addition, we show CMB+BAO constraints for $f_\chi = 1\%$ dark matter that interacts with protons or electrons. In general, we find that bounds from MW subhalos and Lyman-$\alpha$ are respectively stronger, comparable, and weaker than those from CMB+BAO for $n=0$, $-2$, and $-4$ type interactions.  This is expected given the different shapes of the transfer functions of the three types of DMb interactions as explained in~\Sec{subsec:mps}: while the suppression of the transfer function for $n=0$ type interactions is most significant at small scales, which can be effectively probed by Lyman-$\alpha$ or MW subhalo data, the suppression for $n=-4$ type interactions mainly occurs at larger scales, which are better probed by CMB and BAO data. As pointed out in~\Sec{subsec:data}, these constraints are susceptible to input datasets, model assumptions (\eg, the temperature evolution of the intergalactic medium for Lyman-$\alpha$ forest data; galaxy formation for the MW subhalos), and derivation methods (\eg, whether to match the WDM transfer function or to use the smallest halo mass when setting the limits with MW subhalos data). We address those variations and compare the resulting limits in the appendices.

It is interesting to compare the cosmological constraints with other constraints on dark matter--proton or dark matter--electron interactions.  
Direct-detection experiments constrain the same $t$-channel elastic scattering process as the cosmological probes.   
Assuming both dark matter and baryons are non-relativistic, the differential dark matter--baryon scattering cross section is
\beq
\frac{\dd \sigma}{ \dd \cos \theta_*} = \frac{ \overline{|\mathcal M|^2}}{32 \pi (m_B +m_\chi)^2} \ ,
\eeq
where $ \overline{|\mathcal M|^2}$ is the spin-averaged amplitude squared of dark matter--baryon scattering. The form-factor-stripped direct detection cross section is given by
\beq\label{eq:form_factor_xsec}
\sbarBn \equiv \frac{\left. \overline{|\mathcal M|^2}\right|_{q = q_\text{ref}}  }{16\pi (m_B+m_\chi)^2}=\frac{ \mu_{\chi B}^2 \left. \overline{|\mathcal M|^2}\right|_{q = q_\text{ref}} }{16 \pi m_\chi^2 m_B^2} \ ,
\eeq
where $\left. \overline{|\mathcal M|^2}\right|_{q = q_\text{ref}}$ is the spin-averaged amplitude squared evaluated at a fixed magnitude for the three-momentum transfer $q = q_\text{ref}$ and $\mu_{\chi B}=\frac{m_\chi m_B}{m_\chi +m_B}$ is the dark matter--baryon reduced mass. A common choice for the reference transfer momentum for dark matter--electron scattering is $q_\text{ref} =\alpha m_e$. For $n=0$ ($n=-2$) type interactions, a simple translation, between the velocity-stripped momentum transfer cross section $\sigma_n^{\chi B}$ and the form-factor-stripped direct detection cross section $\bar \sigma^{\chi B}$ is possible if dark matter interacts with baryons through contact interactions (electric dipole interactions), according to \Eq{eq:sbar-s0_n0_2} (\Eq{eq:sbar-s0_n-2_2}) that we will derive in~\Sec{subsec:specific}. In~\Figs{fig:n0_bounds}{fig:n-2_bounds}, we show the translated constraints for $n=0$ and $n=-2$ type interactions from the direct detection constraints from surface or underground direct-detection experiments (denoted as ``Direct Detection'')~\cite{Angloher:2002in,Akerib:2003px,Abdelhameed:2019hmk,Abdelhameed:2019mac,Angloher:2017sxg,Aprile:2017iyp,Armengaud:2019kfj,Essig:2012yx,Essig:2017kqs,Angle:2011th,Aprile:2016wwo,Agnes:2018oej,Agnese:2018col,Crisler:2018gci,Abramoff:2019dfb,Barak:2020fql,Arina:2020mxo} available in the literature, for $f_\chi = 100\%$. We also include bounds from the XQC rocket experiment~\cite{McCammon:2002gb,Erickcek:2007jv,Mahdawi:2018euy}, also for $f_\chi = 100\%$, conservatively assuming a thermalization efficiency factor of 2\% \cite{Mahdawi:2018euy}.\footnote{We take the constraints on dark matter--nucleon interaction as the constraints on dark matter--proton interaction by assuming equal couplings for dark matter--proton and neutron interactions.}  In contrast, there is no straightforward translation between $\sigma_n^{\chi B}$ and $\bar \sigma^{\chi B}$ for $n=-4$ type interactions, as we discuss further in Sec.~\ref{subsubsec:darkphoton}. Therefore we do not show the direct detection bounds in~\Fig{fig:n-4_bounds}. 
Besides bounds from direct-detection experiments for the halo dark matter, we also include bounds for dark matter being accelerated by cosmic rays (denoted as ``Cosmic Rays'')~\cite{Yin:2018yjn,Bringmann:2018cvk, Ema:2018bih, Cappiello:2019qsw}\footnote{See Ref.~\cite{Ema:2020ulo} for constraints on cosmic-ray accelerated dark matter whose interactions with quarks are mediated by (pseudo-)scalars .} or by the Sun (denoted as ``Solar Reflection'')~\cite{An:2017ojc,Emken:2021lgc}. 
We also show bounds from CMB spectral distortions with FIRAS data (dubbed ``FIRAS'')~\cite{Fixsen:1996nj,Ali-Haimoud:2015pwa, Ali-Haimoud:2021lka} and gas cooling of Leo-T dwarf galaxy (dubbed ``Leo-T'')~\cite{Wadekar:2019xnf}, all for $f_\chi = 100\%$.

Note that due to the attenuation of dark matter in the atmosphere, earth, or shields on rockets, there are lower limits (\ie ceilings or upper boundaries for the excluded region) for dark matter--baryon interaction for ground- and rocket-based direct-detection experiments. In~\Figs{fig:n0_bounds}{fig:n-2_bounds}, we quote the ceilings that are available in the literature, such as the exclusion regions for the surface/underground direct-detection experiments (both panels of ~\Fig{fig:n0_bounds} and right panel of ~\Fig{fig:n-2_bounds}) and those for the cosmic-ray accelerated dark matter that intact with nucleons (left panel of ~\Fig{fig:n0_bounds}). Such ceilings are not commonly shown for the exclusions of $n=0$ type dark matter--electron interactions for  solar-reflected dark matter~\cite{An:2017ojc,Emken:2021lgc} and cosmic-ray accelerated dark matter~\cite{Cappiello:2019qsw}. In order to include the ceilings for the solar-reflected dark matter--electron interactions, we assume the same electron-only stopping as in the case of the surface/underground direct-detection experiments and present the resulting ceilings as the yellow dotted line in the right panel of~\Fig{fig:n0_bounds}. Such assumption is conservative because the solar-reflected dark matter (and the cosmic-ray accelerated dark matter) is generically more energetic than the halo dark matter and attenuated only at higher cross sections. The ceiling for the cosmic-ray accelerated dark matter--electron interactions is simply set by the plotting range of the exclusion region of the MiniBooNE experiment from~\cite{Ema:2018bih}. The attenuation argument should also apply to the rocket shielding. The ceiling of the XQC exclusion in the left panel of~\Fig{fig:n0_bounds} is taken from Ref.~\cite{Erickcek:2007jv} and applied to the mass region constrained by Ref.~\cite{Mahdawi:2018euy}. This ceiling may need to be recomputed according to the more recent treatment in Ref.~\cite{Mahdawi:2018euy}; its uncertain nature is therefore represented by a dotted line. Finally, the ceiling for the XENON1T experiment in the left panel of \Fig{fig:n-2_bounds} is also set simply by the plotting range of the exclusion plots from Ref.~\cite{Arina:2020mxo}. These limits on the available bounds should be treated as conservative estimates, and we  again emphasize their uncertainty by using dotted lines.

There is one more complication for probing dark matter--nucleon interactions at large cross sections.
As pointed out in~\cite{Digman:2019wdm,Cappiello:2020lbk}, when dark matter interacts with a nuclear target with atomic mass number $A > 1$ through a spin-independent elastic scattering, the  scaling relation for coherent scattering, $\sigma^{\chi A} = A^2 (\mu_{\chi A}/\mu_{\chi N})^2 \sigma^{\chi N}$, which translates the  dark matter--nucleus cross section $\sigma^{\chi A}$ into the corresponding dark matter--nucleon cross section $\sigma^{\chi N}$,  is no longer applicable for $\sigma^{\chi N} \gtrsim 10^{-31} \, \text{cm}^2$ and $m_\chi \gtrsim 1\, \text{GeV}$. The breakdown of the scaling relation is due to the failure of the first Born approximation for the parameter region. Consequently this may invalidate the derivation of the exclusions based on the XQC rocket data, the RRS balloon data~\cite{Rich:1987st}, and some of the surface and underground direct-detection experiments. To obtain the correct limits, one needs to specify a dark matter model that could bypass the unitarity bound and then compute the relevant form factors. We leave this task for other work and display the original constraints in the relevant figures. Note that this complication does not affect our constraints on dark matter--proton interactions from CMB+BAO, MW subhalos, and Lyman-$\alpha$ data, since here $A=1$.

As \Figs{fig:n0_bounds}{fig:n-2_bounds} show, the cosmological probes do not suffer from the same detection threshold effects that limit the sensitivity of direct-detection experiments for dark matter with small recoil momentum. They are also free from the attenuation effects that limit the sensitivity of direct-detection experiments for dark matter that has a large interaction with ordinary matter. In addition, they do not depend on current and local distributions of the dark matter density or the dark matter velocity, which do affect direct-detection experiments. Therefore cosmological probes provide a unique window into the nature of dark matter.

\subsection{Constraints on concrete models}
\label{subsec:specific}

Here we translate the phenomenological DMb bounds on the velocity-stripped momentum-transfer cross section $\sBn$ ($B = e, p$) into constraints on the form-factor stripped dark matter--baryon direct-detection cross section, $\sbarBn$. It is important to note that this translation is model-dependent. We will therefore choose two concrete dark matter models to which constraints for $n=0,-2,-4$ type DMb can be naturally mapped.

We display our results along with published constraints from other observables, \eg the effective number of light degrees of freedom, $N_\text{eff}$, during big bang nucleosynthesis (BBN) and cooling of astrophysical objects, applicable to each DMb model realization. These other observables can be quite constraining, particularly for low mass dark matter. However, we note that these constraints can often be weakened in extensions of the models; such model building is beyond the scope of our work.

The two dark matter models we consider are dark matter interacting with a dark photon (for both heavy and light dark photon masses, which yield the $n=0$ and $n=-4$ cases, respectively,) and dark matter interacting with an electric dipole moment.\\

\subsubsection{Dark matter with dark photon mediator}
\label{subsubsec:darkphoton}

In this model, the dark matter $\chi$ is a Dirac fermion and it interacts with the Standard Model electron and protons through a dark photon $A'$. The relevant interactions are
\beq
\mathcal L \supset  \epsilon e A'_\mu e\gamma^\mu e- \epsilon e A'_\mu p\gamma^\mu p + g_\chi A'_\mu \bar{\chi} \gamma^\mu \chi \ ,
\label{eq:darkphotonlag}
\eeq
where $\epsilon$ is the kinematic mixing between the dark photon and the ordinary photon, and $g_\chi$ is the coupling between dark photon and dark matter.  The spin-averaged amplitude squared in the small transfer momentum limit is 
\beq
\overline{|\mathcal M|^2}\approx \frac{16 \epsilon^2 e^2 g_\chi^2 m_B^2 m_\chi^2}{ (t-m_{A'}^2)^2} \ ,
\eeq
where the Mandelstam variable $t = q^2 = - |\vec q|^2 =2 p_\text{CM}^2 (\cos \theta_\text{CM} -1)$ and $p_\text{CM} = \mu_{\chi B} v_\text{rel}$ is the magnitude of particle momentum in the center-of-mass frame. 
We now consider two limits for the dark photon mass to which the $n=0$ and $n=-4$ type DMb models can be naturally mapped. 

\noindent{\bf  Heavy dark photon mediation.} In the heavy mediator limit, the averaged squared amplitude can be approximated by 
\beq
\overline{|\mathcal M|^2} \approx \frac{16 \epsilon^2 e^2 g_\chi^2 m_B^2 m_\chi^2}{ m_{A'}^4} \ ,
\eeq
which does not depend on the scattering angle or transfer momentum. Therefore, we have
\beq\label{eq:sbar-s0_n0_1}
\sigma_{\rm T}^{\chi B} = \frac{16 \pi \alpha \alpha_\chi \epsilon^2 \mu_{\chi B}^2}{m_{A'}^4} = \sbarB \ ,
\eeq
where $\alpha_\chi \equiv g_\chi^2/4\pi$. This is also a velocity-independent momentum-transfer cross section. Therefore the form factor-stripped direct-detection cross section is simply 
\beq\label{eq:sbar-s0_n0_2}
 \sbarB_0 = \si^{\chi B}_0 \ .
\eeq

\noindent{\bf Ultra-light dark photon mediation.} Here  the interaction is Coulomb-like, and its spin-averaged amplitude squared from \Eq{eq:darkphotonlag} is given by
\beq
\overline{|\mathcal M|^2} \approx \frac{16 \epsilon^2 e^2 g_\chi^2 m_B^2 m_\chi^2}{t^2},
\label{eq:massless}
\eeq
where $\epsilon e$ is coupling of dark photon to electron/proton and $g_\chi$ is the coupling of dark photon to dark matter.\footnote{Constraints on DMb interactions can be also set by combining the constraints on $\epsilon$ from dark photon searches and the constraints on $g_\chi$ from observations of self-interacting dark matter halos. Such combined constraints are more model-dependent and we do not explore them here.} By setting $g_\chi = e (\alpha_\chi =\alpha)$, we recover the scenario of millicharge dark matter. 
The resulting direct detection cross section is therefore given by
\beq
\sbarB \equiv \frac{1 }{16\pi (m_B+m_\chi)^2}\left. \overline{|\mathcal M|^2}\right|_{q = q_\text{ref}} = \frac{16 \pi \epsilon^2 \alpha \alpha_\chi \mu_{\chi B}^2}{q_\text{ref}^4} \ .
\label{eq:ddultralight}
\eeq

The computation of the momentum-transfer cross section between dark matter and charged particles (electrons and protons) in the baryonic fluid is more complicated. A naive integration over \Eq{eq:massless} leads to a logarithmic divergence due to forward scattering, which is well known for Coulomb-like interactions. To regulate the forward scattering singularity, Ref.~\cite{McDermott:2010pa} suggested setting the maximal impact parameter to be the Debye screening length, which we dub ``Debye-length screening''. Under this prescription, the momentum-transfer cross section is given by:
\beq
\begin{aligned}
\sigma_{\rm T}^{\chi B} 
&\approx \frac{2\pi \epsilon^2 \alpha \alpha_\chi}{\mu_{\chi B}^2 v_\text{rel}^4} \ln \left(\frac{9 T_b^3}{8\pi \epsilon^2 \alpha^2 \alpha_\chi x_e n_b}\right) \quad \text{(Debye-length screening)} \,
\label{eq:ultralight}
\end{aligned}
\eeq
where  $n_b$, $T_b$, and $x_e = n_e/n_b$ are the baryon number density, baryon temperature, and ionization fraction, respectively. See  \App{appB} for the derivation of  \Eq{eq:ultralight}. Before recombination, the ionization fraction $x_e \approx 1$ and the ratio of $T_b^3/n_b = T_0^3/n_0$ is fixed. Therefore the logarithmic factor $\ln(9 T_b^3/8\pi\epsilon^2 \alpha^2 \epsilon_\chi x_e n_b)$ is a redshift-independent constant. The relative velocity dependence in \Eq{eq:ultralight} only appears as $v_\text{rel}^{-4}$ in the pre-factor. Taking the constraints on the $v_\text{rel}^{-4}$-stripped cross section $\si^{\chi B}_{-4}$, we can first numerically solve for the constraint on $\epsilon^2 \alpha_\chi$ using \Eq{eq:ultralight}, and then substitute it into \Eq{eq:ddultralight} to get the constraint on the form factor-stripped direct detection cross section $\sbarB_{-4}$.

In the above computation for the momentum-transfer cross section based on the Debye-length screening prescription, the in-medium thermal effect of the photon has been considered at the level of the phase space integral. But as Ref.~\cite{Dvorkin:2019zdi} explicitly suggested, such a consideration is inadequate because in-medium thermal effects change the dispersion relation of the photon and excites its longitudinal polarization. Therefore in-medium effects should be considered at the level of the matrix element. Including the in-medium effects, the averaged amplitude squared in the non-relativistic limit is given by,\footnote{The result is similar to Eq.~(D2) of Ref.~\cite{Dvorkin:2019zdi} when setting $\alpha_\chi=\alpha$, except for a factor of 2 difference in the definition of the Debye mass square~\eqref{eq:static30}.} 
\beq
\overline{|\mathcal M|^2} \approx \frac{16 \epsilon^2 e^2 g_\chi^2  m_\chi^2 m_B^2}{\left[2p_\text{CM}^2(1-\cos \theta_\text{CM}) + m_\text{D}^2\right]^2} \ .
\label{eq:ultralight2me}
\eeq
where the Debye mass is 
\beq
m_\text{D}^2 =  \frac{e^2 (n_e + n_p)}{ T} = \frac{2 e^2 x_e n_b}{T_b} \ . 
\label{eq:static30}
\eeq
The derivation of \Eq{eq:ultralight2me} can be found in  \App{appB}. With this matrix element regularization prescription, which we dub ``Debye-mass screening'', the momentum-transfer cross section between dark matter and baryons, mediated by the longitudinal mode of the in-medium photon, is given by:
\beq
\begin{aligned}
\sigma_{\rm T}^{\chi B} &\approx \frac{2\pi \epsilon^2 \alpha \alpha_\chi \mu_{\chi B}^2}{ p_\text{CM}^4 }\ln \left(\frac{4 p_\text{CM}^2}{m_\text{D}^2} \right) 
\\
&= \frac{2\pi \epsilon^2 \alpha \alpha_\chi}{\mu_{\chi B}^2 v_\text{rel}^4 }\ln \left(\frac{4 \mu_{\chi B}^2 v_\text{rel}^2}{m_\text{D}^2} \right) \quad \text{(Debye-mass screening)} \ .
\label{eq:ultralight2}
\end{aligned}
\eeq
 Note that since $m_\text{D}^2 \propto x_e n_b/ T_b$, we only recover the redshift-independent combination (before recombination) $T_b^3/x_e n_b$ if $p_\text{CM} \sim T_b$ inside the logarithmic factor $\ln(4p_\text{CM}^2/m_\text{D}^2)$. More generally, the logarithmic factor will carry a redshift dependence. This prevents an unambiguous translation of our cosmological constraints on the velocity-stripped cross section $\si^{\chi B}_{-4}\equiv \sigma_\text{T}^{\chi B} v_\text{rel}^4$ (defined in \Eq{eq:sT}) into constraints on $\sigma_\text{T}^{\chi B}$ (which includes the Debye logarithm) or $\sbarB_{-4}$ (the form-factor-stripped direct detection cross section, defined in \Eq{eq:form_factor_xsec}), given the fact that we assumed $\sigma_{-4}^{\chi B}$ to be independent of the redshift when deriving our constraints.

To translate our constraints on $\si^{\chi B}_{-4}$ from \Sec{subsec:pheno} into those on $\sbar^{\chi B}_{-4}$ under the Debye-mass screening prescription, we take $v_{\rm rel}$ and $m_\text{D}$ in the argument of the logarithm of \Eq{eq:ultralight2} to be given \Eqs{eq:uB}{eq:static30} respectively. In order to do this, we compute the thermal evolution of the DMb model at each point in the $(\mx, \sn)$ parameter space with our \texttt{class\_dmb} implementation, having fixed all the other cosmological parameters to their $\LC$ Planck 2018 mean values~\cite{Aghanim:2018eyx}. We expect that a more accurate treatment including their marginalization would lead only to small deviations from our results. Note that since this argument is redshift-dependent, we have to evaluate it at a specific  $z$ value, which we take to be $z=10^5$. This redshift approximately corresponds to the time at which the largest CMB multipoles probed by the Planck satellite enter the horizon. Before recombination, $m_\text{D}^2 \sim (1+z)^2$, while $v_{\rm rel}^2 = u_B^2 \sim (1+z)$, which in turn means that the argument of the logarithm scales like $(1+z)^{-1}$. Therefore, increasing the $z$ benchmark we used from $z=10^5$ to a larger value translates into a smaller logarithm. Thus, for a given constraint on $\si^{\chi B}_{-4}$ from our analysis, larger $z$ values are degenerate with larger values of $\sbar^{\chi B}_{-4}$ and hence weaker constraints. All these subtleties stem from the fact that within the ultra-light mediator model, $\sT \sim v_{\rm rel}^{-4} \ln(v_{\rm rel}^2/m_\text{D}^2)$ instead of $\sT \sim v_{\rm rel}^{-4}$ exactly. A precise constraint on dark matter with ultra-light dark photon mediators can be solved with a dedicated implementation of this model into a Boltzmann solver, which we leave for future work. If instead we use the Debye-length screening prescription (\Eq{eq:ultralight} and Ref.~\cite{McDermott:2010pa}), the logarithm is redshift-independent, and thus the ambiguity in the translation between $\si^{\chi B}_{-4}$ and $ \sbar^{\chi B}_{-4}$ does not exist.

Finally we would like to point out that while we have described two different methods to translate between $\si^{\chi B}_{-4}$ and $\sbar^{\chi B}_{-4}$, the Debye-mass screening prescription (which includes in-medium effects at the matrix element level) seems to be on firmer footing than the Debye-length screening prescription \cite{Dvorkin:2019zdi}. We have nevertheless included in this work a description and derivation of the latter in order to facilitate comparisons with the previous literature. Computing the $\si^{\chi B}_{-4} \rightarrow \sbar^{\chi B}_{-4}$ translation using the Debye-length screening method leads to constraints stronger than those from Debye-mass screening by a factor of $\mO(1)$.

\subsubsection{Dark matter with an electric dipole moment (EDM)} 
For dark matter with an EDM~\cite{Sigurdson:2004zp}, the interaction between a Dirac fermion dark matter $\chi$ and SM electromagnetic field strength $F^{\mu\nu}$ is
\beq
\mathcal{L}\supset -\frac{i D}{2}\bar \chi \sigma_{\mu \nu} \gamma_5 \chi F^{\mu\nu} \ ,
\eeq
where $D$ is the electric dipole moment. The averaged amplitude square for DMb interactions is thus
\beq
\overline{|\mathcal M|^2}\approx\frac{64 \pi \alpha D^2 m_B^2 m_\chi^2}{-t} \ .
\eeq
The form factor-striped direct detection cross section is then 
\beq\label{eq:sbar-s0_n-2_1}
\sbarB = \frac{4 \alpha D^2 \mu_{\chi B}^2}{q_\text{ref}^2} \ ,
\eeq
and the momentum-transfer cross section for the dark matter--baryon elastic scattering in vacuum is 
\beq
\sigma_{\rm T}^{\chi B} = \frac{2\alpha D^2}{v_\text{rel}^2} \ .
\eeq

In the baryonic plasma, the interaction also receives modifications due to the thermal effect of the in-medium photons. Like in the previous case of DMb interactions mediated by the ultra-light dark photons, in the non-relativistic limit, we expect the matrix element is dominated by the time components of the baryonic current and the dark current. The corresponding spin-averaged amplitude square is given by
\beq
\overline{|\mathcal M|^2}\approx\frac{64 \pi \alpha D^2 m_B^2 m_\chi^2 \times 2 p_\text{CM}^2 (1-\cos \theta_\text{CM})}{(2 p_\text{CM}^2 (1-\cos \theta_\text{CM}) + m_\text{D}^2)^2} \ ,
\eeq
with the Debye mass $m_\text{D}$ given by~\Eq{eq:static3}. This in turn leads to a momentum-transfer cross section of
\beq
\sigma_\text{T}^{\chi B} \approx \frac{2\alpha D^2}{v_\text{rel}^2} \left[1 + \mathcal{O} \left( \frac{m_\text{D}^2}{4p_\text{CM}^2}\right)\right] \ .
\eeq

Given that $m_\text{D}^2/p_\text{CM}^2 \ll 1$, we neglect the thermal correction. Hence the velocity stripped cross section is simply given by $\sigma^{\chi B}_{-2} = 2 \alpha D^2$, and it is related to the direct detection cross section by the relation
\beq\label{eq:sbar-s0_n-2_2}
\sbarB_{-2} = \frac{2 \mu_{\chi B}^2}{q_\text{ref}^2} \si^{\chi B}_{-2} \ .
\eeq

\subsubsection{Constraints}

\begin{figure}[t]
  \centering
  \includegraphics[width=0.9\textwidth]{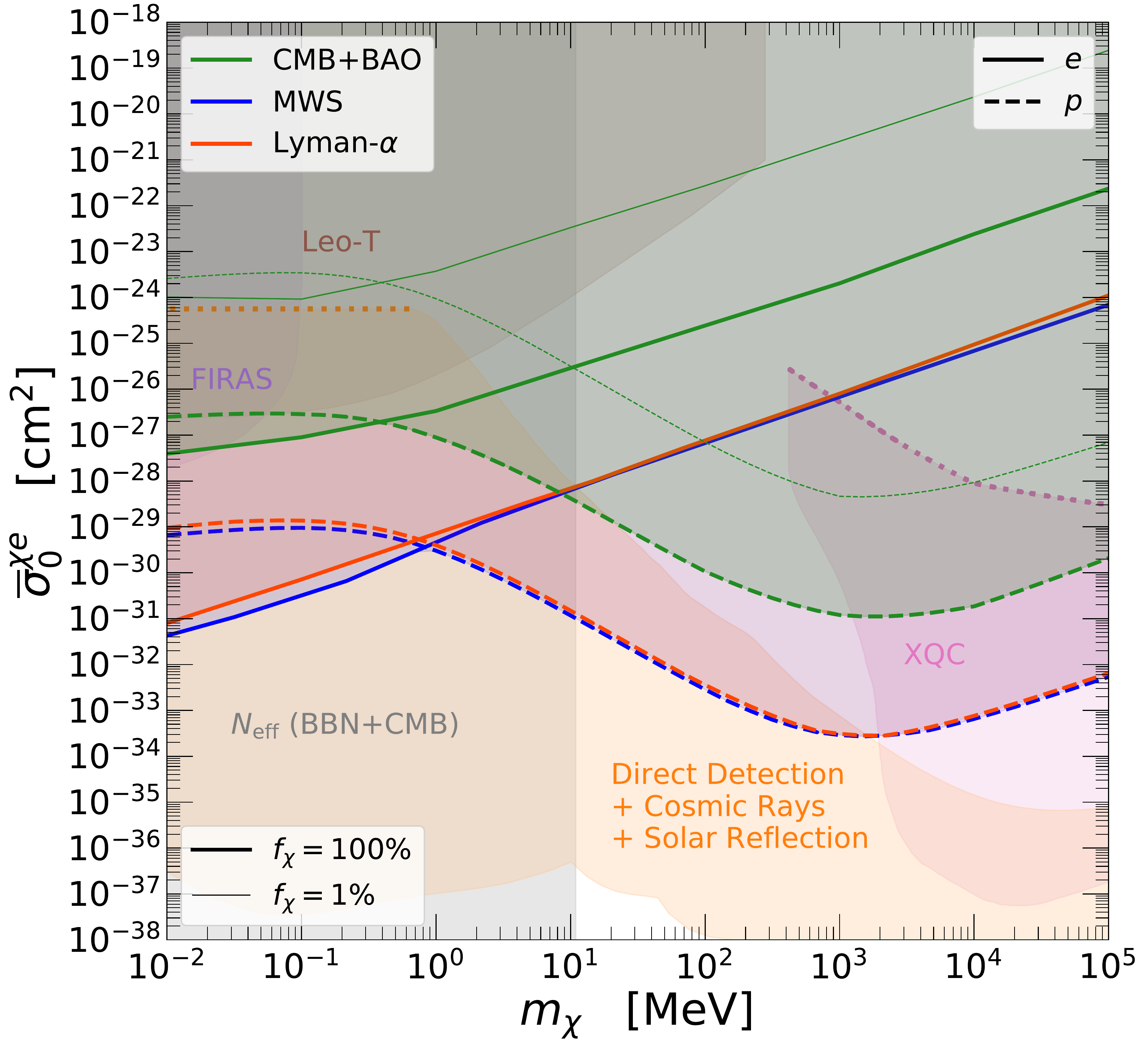}
  \caption{95\% C.L. bounds on the $(\mx, \sbar^{\chi e}_n)$ parameter space for $n=0$, assuming dark matter interacts with a heavy dark photon, and $f_\chi = 100\%$. We translate our $\si^{\chi B}_0$ constraints from dark matter--electron interactions ({\bf thick solid} lines) and dark matter--protons interactions ({\bf thick dashed} lines) based on CMB+BAO ({\bf green}), MW subhalos ({\bf blue}), and Lyman-$\alpha$ forest ({\bf red}) to the direct detection cross section $\sbar^{\chi e}_0$, using \Eqs{eq:sbar-s0_n0_1}{eq:sbar-s0_n0_2}.   The {\bf thin green} lines show the corresponding  constraints for $f_\chi = 1\%$ from dark matter--electron interactions ({\bf solid} line) and dark matter--protons interactions ({\bf dashed} line) based on CMB+BAO data. Bounds from MW subhalos and Lyman-$\alpha$ forest are recast from bounds for thermal WDM/non-cold DM.  Both the fixed $k$ and half-mode matching schemes give the same results for the MW subhalos bounds. See text for more details of the recast and the limitations. The {\bf orange} region shows the combined constraints from direct-detection experiments of the halo dark matter, some compiled in~\cite{Emken:2019tni} (\cite{Abdelhameed:2019hmk,Abdelhameed:2019mac,Angloher:2017sxg,Aprile:2017iyp,Armengaud:2019kfj,Essig:2017kqs,Angle:2011th,Aprile:2016wwo,Agnes:2018oej,Agnese:2018col,Crisler:2018gci,Abramoff:2019dfb,Barak:2020fql,Arina:2020mxo}), the cosmic-ray accelerated dark matter~\cite{Yin:2018yjn,Bringmann:2018cvk,Ema:2018bih, Cappiello:2019qsw}, and the solar-reflected dark matter~\cite{Emken:2021lgc,An:2017ojc} with total (electronic+ionic) stopping.  The {\bf pink} region shows the constraint from the XQC rocket experiment \cite{Erickcek:2007jv,Mahdawi:2018euy}. The {\bf pink dotted} line represents its ceiling, taken from~\cite{Erickcek:2007jv}. The {\bf purple} region shows constraints from CMB spectral distortions with FIRAS data~\cite{Fixsen:1996nj,Ali-Haimoud:2015pwa, Ali-Haimoud:2021lka}. The {\bf brown} region shows constraints from the gas cooling rate of Leo-T dwarf galaxy~\cite{Wadekar:2019xnf}. The {\bf gray} region is constrained by $N_{\rm eff}$ from BBN+CMB data~\cite{Sabti:2019mhn}.  The {\bf yellow dotted} line represents our conservative assumptions about the limits of the solar reflection bounds. For that we take the upper boundary to be given by the same total (electronic+ionic) stopping as in the case of direct-detection experiments. In the interest of clarity we omit freeze-out thermal relic abundance lines, which can be found in \eg Ref.~\cite{Battaglieri:2017aum}. The thermal relic line for asymmetric fermion dark matter is model-dependent; other models are not studied this work. For a review on dark matter production through dark photons see \eg Ref.~\cite{Hambye:2019dwd}. Constraints translated from the dark matter–nucleus interaction may suffer complications due to the failure of the first Born approximation as discussed in~Sec.~\ref{subsec:pheno}.}
  \label{fig:n0_dp_bounds}
\end{figure}

\begin{figure}[t]
  \centering
  \includegraphics[width=0.9\textwidth]{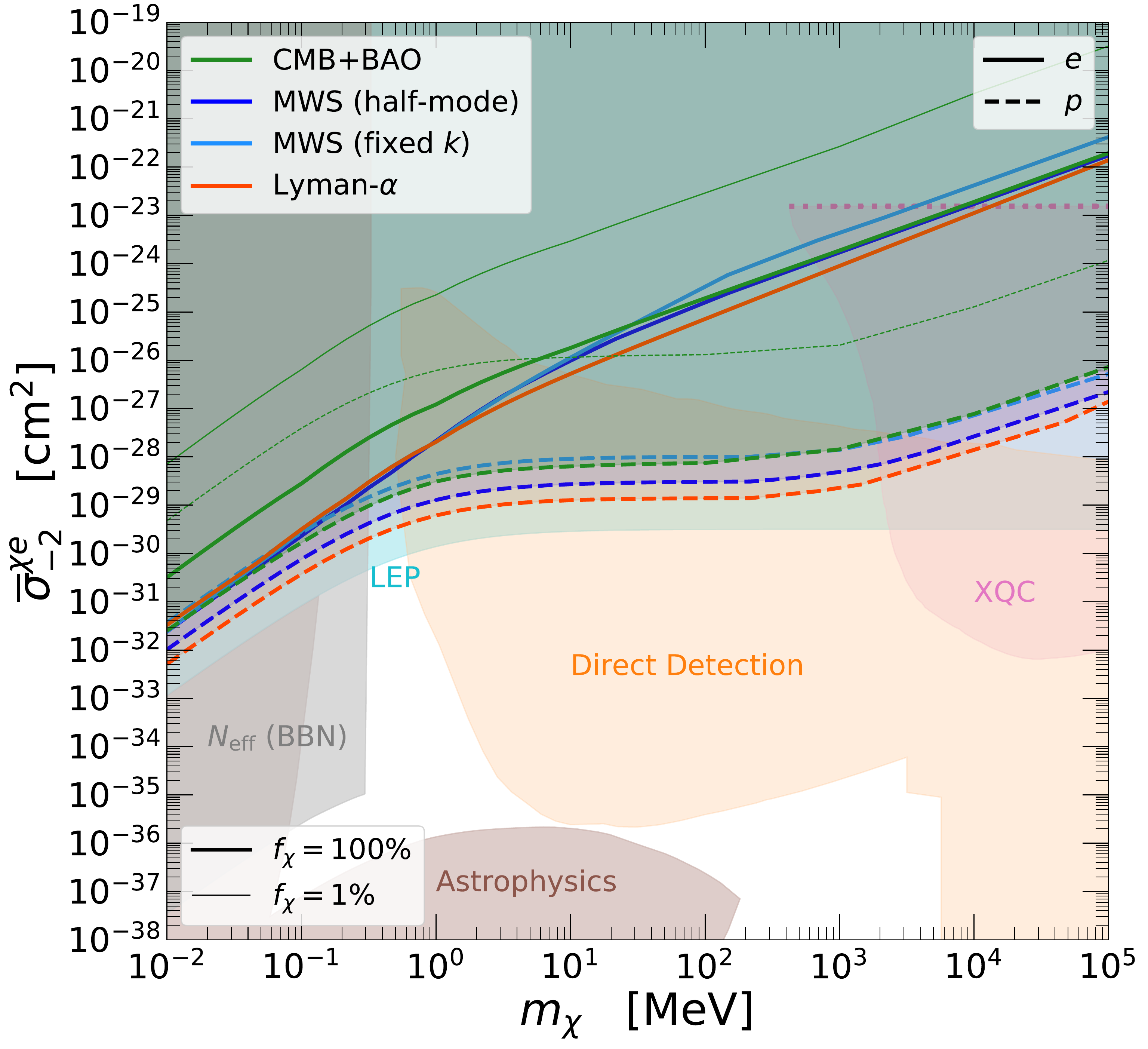}
  \caption{95\% C.L. bounds on the $(\mx, \sbar^{\chi e}_n)$ parameter space for $n=-2$, assuming dark matter interacts with an electric dipole moment, and $f_\chi = 100\%$. We translate our $\si^{\chi B}_{-2}$ constraints from dark matter--electron interactions ({\bf thick solid} lines) and dark matter--protons interactions ({\bf thick dashed} lines) based on CMB+BAO ({\bf green}), MW subhalos ({\bf darker/lighter blue}), and Lyman-$\alpha$ forest ({\bf red}) to the direct detection cross section $\sbar^{\chi e}_{-2}$, using \Eqs{eq:sbar-s0_n-2_1}{eq:sbar-s0_n-2_2}. The {\bf thin green} lines show the corresponding  constraints for $f_\chi = 1\%$ from dark matter--electron interactions ({\bf solid} line) and dark matter--protons interactions ({\bf dashed} line) based on CMB+BAO data. Bounds from MW subhalos and Lyman-$\alpha$ forest are recast from bounds for thermal WDM/non-cold DM. For the MW-subhalos based constraints, we consider both the half-mode ({\bf darker blue}) and the fixed $k$ ({\bf lighter blue}) matching criteria when recasting the bounds for the recast procedure. See text for more details of the recast and the limitations. The {\bf orange} region shows the combined constraints from the direct-detection experiments compiled by~\cite{Sigurdson:2004zp} (\cite{Angloher:2002in,Akerib:2003px}) and those  compiled in~\cite{Emken:2019tni} (\cite{Essig:2017kqs,Angle:2011th,Aprile:2016wwo,Agnes:2018oej,Agnese:2018col,Crisler:2018gci,Abramoff:2019dfb,Barak:2020fql}) with total (electronic+ionic) stopping. The {\bf pink} region shows the constraint from the XQC rocket experiment \cite{McCammon:2002gb,Mahdawi:2018euy}. The {\bf pink dotted} line represents our conservative choice for its ceiling, taken from the plotting range of~\cite{Mahdawi:2018euy}. The {\bf gray} and {\bf brown} regions correspond to constraints from cosmology ($N_{\rm eff}$ at the time of BBN) and astrophysics (cooling of horizontal branch and red giant stars, the Sun, and supernova 1987A), respectively~\cite{Chu:2019rok} (see also~\cite{Chang:2019xva}). The {\bf cyan} region corresponds to bounds from LEP~\cite{Fortin:2011hv}.
  Also present but not shown are constraints from RSS balloon~\cite{Rich:1987st,Sigurdson:2004zp}, which overlap with those from direct detection and XQC. Finally, we omit weaker bounds from CMB distortions by energy injections~\cite{Lambiase:2021xcj} for the sake of clarity. For the same reason we omit relic abundance lines, which can be found in Refs.~\cite{Chu:2018qrm,Chu:2019rok}. Constraints translated from the dark matter–nucleus interaction may suffer complications due to the failure of the first Born approximation as discussed in~Sec.~\ref{subsec:pheno}.}
  \label{fig:n-2_dp_bounds}
\end{figure}

\begin{figure}[t]
  \centering
  \includegraphics[width=0.9\textwidth]{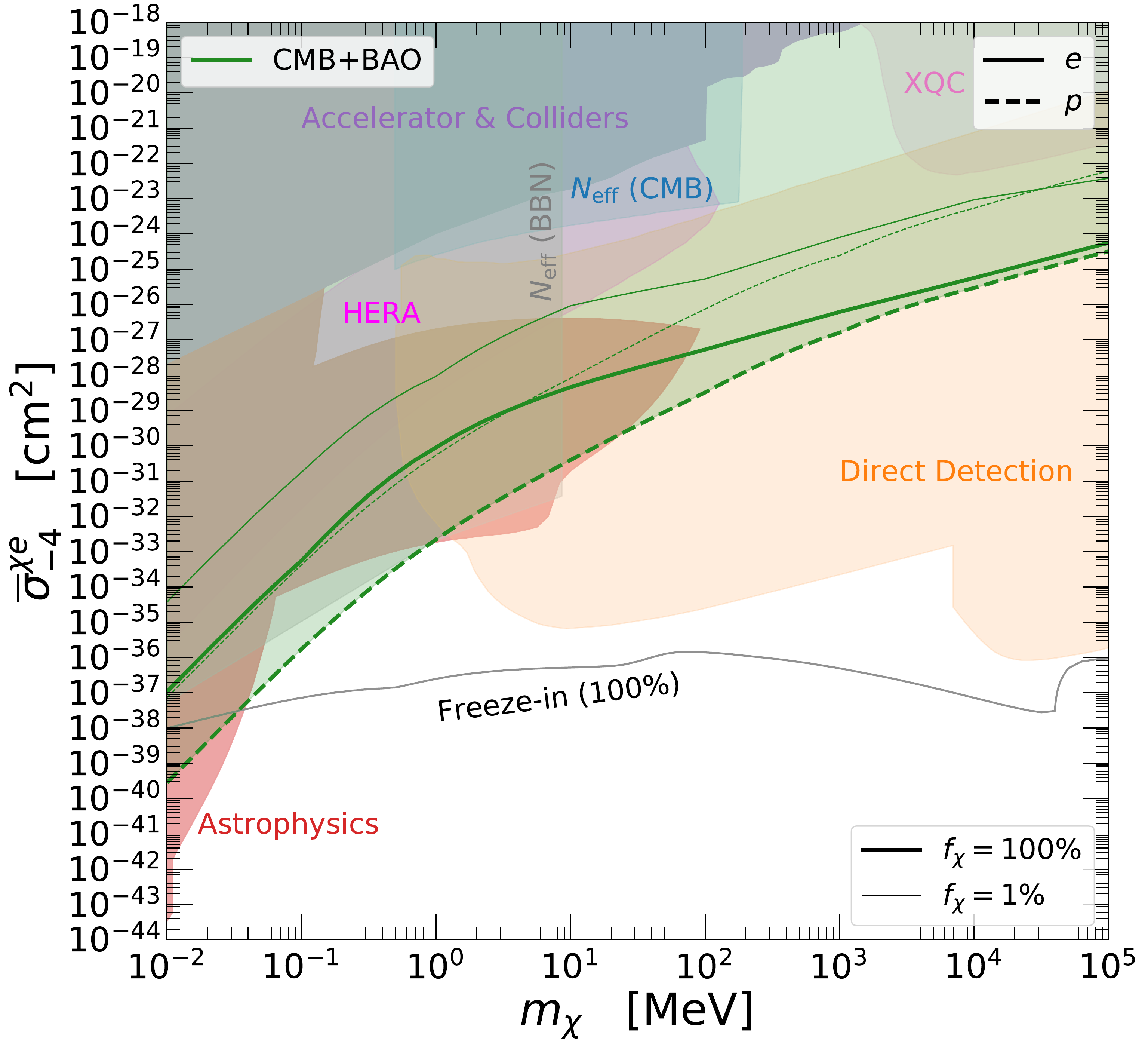}
  \caption{95\% C.L. bounds on the $(\mx, \sbar^{\chi e}_n)$ parameter space for $n=-4$, for millicharged dark matter or dark matter with an ultra-light dark photon mediator. We translate our $\si^{\chi B}_{-4}$ constraints from dark matter--electron interactions ({\bf thick green solid} line) and dark matter--protons interactions ({\bf thick green dashed} line) based on CMB+BAO to the direct detection cross section $\sbar^{\chi e}_{-4}$ To perform this translation we have used the Debye-mass screening prescription described in \Eq{eq:ultralight2}, as well as \Eq{eq:ddultralight}. We take $v_{\rm rel}$ to be given by $u_B$ in \Eq{eq:uB}, and evaluated it at $z=10^5$, which is approximately the redshift at which the largest CMB multipoles probed by the Planck satellite enter the horizon. The redshift-independent Debye-length screening prescription~\cite{McDermott:2010pa} gives bounds that are stronger by an $\mO(1)$ factor. The {\bf thin green} lines show the corresponding constraints for $f_\chi = 1\%$. Note that we omit the weaker bounds derived from the Milky Way subhalos and Lyman-$\alpha$ datasets. The {\bf orange} region shows constraints from the direct-detection experiments~\cite{Essig:2017kqs,Angle:2011th,Aprile:2016wwo,Agnes:2018oej,Agnese:2018col,Crisler:2018gci,Abramoff:2019dfb,Barak:2020fql,Hambye:2018dpi} considering total (electronic+ionic) stopping.  The {\bf red} region shows cooling constraints from astrophysics (horizontal branch and red giant stars~\cite{Vogel:2013raa}, as well as supernova 1987A~\cite{Chang:2018rso}. The {\bf blue} and {\bf gray} regions show constraints from $N_\text{eff}$ from CMB and BBN respectively~\cite{Creque-Sarbinowski:2019mcm,Munoz:2018pzp}. The {\bf pink} region shows the constraint from XQC rocket experiment \cite{Mahdawi:2018euy}. The {\bf magenta} region shows the constraint from the 21 cm experiment HERA band 1, assuming a dark matter fraction of $0.5\%$ \cite{HERA:2021noe}. The {\bf purple} region shows the combined constraints for the milicharged dark matter searches from SLAC~\cite{Prinz:1998ua}, ArgoNeuT~\cite{Acciarri:2019jly}, BaBar~\cite{Liang:2019zkb}, SuperK~\cite{Plestid:2020kdm}, neutrino experiments~\cite{Magill:2018tbb} and collider experiments~\cite{Vogel:2013raa,Davidson:1991si}. Constraints from perturbativity lie above the uppermost limit in our vertical axis. Finally, the {\bf black} line corresponds to those points that yield 100\% of the dark matter relic abundance through freeze-in \cite{Essig:2011nj,Chu:2011be}. For a review on dark matter production through dark photons see \eg Ref.~\cite{Hambye:2019dwd}. Constraints translated from the dark matter–nucleus interaction may suffer complications due to the failure of the first Born approximation as discussed in~Sec.~\ref{subsec:pheno}.
  }
  \label{fig:n-4_dp_bounds}
\end{figure}

Based on the particle physics realizations of the $n=0,-2$ and $-4$ DMb models we described above, we can now translate our cosmological constraints on the velocity-stripped momentum-transfer cross section $\sBn$ to bounds on the form factor-stripped direct detection cross section, $\sbarBn$, commonly used in the literature for direct-detection experiments. For $n=0$ this translation is trivial and given by \Eq{eq:sbar-s0_n0_2}. For $n=-2$ the translation is \Eq{eq:sbar-s0_n-2_2}. The translation for $n=-4$ is more involved as described in~\Sec{subsubsec:darkphoton}. Our constraints for concrete dark matter models are translated from those for the phenomenological model that are shown in \Fig{fig:n0_dp_bounds} ($n=0$, heavy dark photon mediator), \Fig{fig:n-2_dp_bounds} ($n=-2$, dark matter with and electric dipole moment), and \Fig{fig:n-4_dp_bounds} ($n=-4$, ultra-light dark photon mediator). Under the assumptions that the couplings for dark matter--electron interactions equal those for dark matter--proton interactions, we combine bounds on dark matter--proton interactions with those on dark matter--electron interactions for the same value of $n$ by using the relation
\beq
    \sbar^{\chi e}_n = \frac{\mu_{\chi e}^2}{\mu_{\chi p}^2} ~ \sbar^{\chi p}_n \ . \label{eq:p_to_e}
\eeq
As mentioned in~\Sec{subsec:thermal}, we expect that the translated constraints agree with the exact constraints (\ie those arising from a direct implementation of the concrete dark matter models into the Boltzmann solver) when either the dark matter--electron or dark matter--proton interaction is dominant; when the two types of interactions are comparable our constraints are conservative.

 We also collect relevant constraints previously published in the literature for the three model scenarios for $f_\chi = 100\%$ dark matter interacts with ordinary matter. Those include constraints from the direct-detection experiments (underground/surface/rocket) for halo/solar-reflected/cosmic ray-accelerated dark matter, CMB spectral distortion, and gas cooling of dwarf galaxies that are mentioned in~\Sec{subsec:pheno}. Both electronic stopping and ionic stopping are included for the attenuation limits of the direct-detection experiments. In addition, we include constraints from measurements of $N_\text{eff}$~\cite{Sabti:2019mhn,Chu:2019rok,Creque-Sarbinowski:2019mcm,Munoz:2018pzp}, stellar cooling~\cite{Chu:2019rok,Vogel:2013raa,Chang:2018rso}, and accelerator and collider searches~\cite{Fortin:2011hv,Prinz:1998ua,Acciarri:2019jly,Liang:2019zkb,Plestid:2020kdm,Magill:2018tbb,Vogel:2013raa,Davidson:1991si} for the three model scenarios where applicable and when available in the literature.

As shown in \Fig{fig:n0_dp_bounds}, for the heavy mediator case ($n=0$), most of the $(\mx, \sbar^{\chi e}_0)$ parameter space under consideration is already excluded by various efforts (\eg, direct detection searches and dark matter acceleration through solar reflection). The cosmological bounds from Milky Way subhalos, Lyman-$\alpha$, and CMB+BAO (in order of increasing strength) are complementary to these searches, closing dark matter--baryon cross sections above the attenuation ceilings. It also does not suffer from the uncertainty on the cross sections when translating those for  spin-independent elastic dark matter--nucleus scattering to those for dark matter--nucleon scattering~\cite{Digman:2019wdm,Cappiello:2020lbk}. Finally, through the comparison of constraints for dark matter--electron scattering and dark matter--proton scattering, we find that for masses below $\sim 0.5~\MeV$,  the oft-neglected dark matter interactions with electrons dominate over those with protons and hence are more sensitive probes for the dark matter model. Note that the dark matter can come into chemical equilibrium with baryonic fluid through $\chi \bar \chi \leftrightarrow A'\leftrightarrow e^+ e^-$, which will affect the number of light degrees of freedom during BBN and CMB~\cite{Boehm:2013jpa,Nollett:2014lwa,Sabti:2019mhn}. In \Fig{fig:n0_dp_bounds}, we adopt $m_\chi >10.9\, \MeV$ from a combined analysis on $N_\text{eff}$ from BBN and Planck 2018 data~\cite{Sabti:2019mhn}.

In the case where dark matter has an electric dipole moment ($n=-2$), \Fig{fig:n-2_dp_bounds}, the constraints on $(\mx, \sbar^{\chi e}_{-2})$ derived from the three cosmological datasets we considered are of similar strength, lying roughly within one order of magnitude. These bounds are dominated by dark matter--proton interactions above dark matter masses of $\sim 0.5~\MeV$. For $m_\chi \lesssim 0.5\, \MeV$,  dark matter interactions with electrons yield similar results as those from dark matter--proton interactions. Similar to the dark matter with heavy dark photon mediator case, the cosmological probes are complementary to dark matter direct detection searches~\cite{Angloher:2002in,Akerib:2003px,Essig:2017kqs,Angle:2011th,Aprile:2016wwo,Agnes:2018oej,Agnese:2018col,Crisler:2018gci,Abramoff:2019dfb,Barak:2020fql}, as well as observations for stellar cooling~\cite{Chu:2019rok,Chang:2019xva}, ruling out cross sections which are beyond the reach of these experiments due to attenuation or observations due to trapping. Constraints from the mono-photon searches at the Large Electron--Positron Collider (LEP) give an even stronger bound across the mass range~\cite{Fortin:2011hv}. The exclusion may also have a ceiling if dark matter has QCD-level interactions with detector material and shows the behavior of neutrons or $K_L^0$ instead of missing momentum~\cite{Daci:2015hca}. The upper (lower) limit from LEP searches can be surpassed by searches at the Large Hadron Collider if a dedicated mono-photon (trackless-jet) search for dark matter is performed for the high-luminosity run. A similar bound from $N_\text{eff}$ during BBN as described for the $n=0$ case that arises due to dark matter coming into chemical equilibrium with the baryonic plasma is also shown~\cite{Chu:2019rok}.

For the case with an ultra-light dark photon mediator ($n=-4$), shown in \Fig{fig:n-4_dp_bounds}, we present our bounds on $(\mx, \sbar^{\chi e}_{-4})$ derived from CMB+BAO. As mentioned above, we employ the redshift-dependent Debye-mass screening prescription (\Eq{eq:ultralight2}, Ref.~\cite{Dvorkin:2019zdi}). We have omitted the bounds derived from the Milky Way subhalos and Lyman-$\alpha$ datasets, since \textit{i.} they are much weaker than those from CMB+BAO (see \Fig{fig:n-4_bounds}), and \textit{ii.} the computation of their translated $\sbar^{\chi e}_{-4}$ bounds would require solving for the DMb thermal evolution at each point in the mass-cross section parameter space in order to compute the Debye logarithm, which would be computationally expensive. 
Unlike the previous two dark matter model scenarios, we can see that the bounds derived from dark matter--proton interactions dominate for dark matter masses below $1~\GeV$, while for larger masses both dark matter--proton and dark matter--electron bounds become comparable. The same complementary with direct-detection experiments ~\cite{Essig:2017kqs,Angle:2011th,Aprile:2016wwo,Agnes:2018oej,Agnese:2018col,Crisler:2018gci,Abramoff:2019dfb,Barak:2020fql, Hambye:2018dpi} and stellar cooling observations~\cite{Vogel:2013raa,Chang:2018rso} mentioned in the previous two scenarios is observed, stemming from the inherent limitation of these experiments or observations at large cross section due to attenuation or trapping. As in the previous two cases, for large enough couplings, dark matter can come into chemical equilibrium with baryons at early times. Bounds from this effect on $N_\text{eff}$ during BBN and CMB were computed in Refs.~\cite{Creque-Sarbinowski:2019mcm,Munoz:2018pzp} and are shown in \Fig{fig:n-4_dp_bounds}.

\section{Conclusions}
\label{sec:concl}

We have derived new constraints for dark matter--electron and dark matter--proton interactions from up to date cosmological observations, including CMB (Plank 2018) and BAO data. The constraints are placed on a phenomenological model of the dark matter--baryon transfer cross section that scales with the relative velocity as $\sigma_\text{T}\propto v^n$ for $n=0$, $-2$, and $-4$. These constraints are translated to concrete dark matter models, namely dark matter with heavy or ultra-light dark photon mediators and dark matter with an electric dipole moment. In addition, we consider the scenario where only a sub-dominant fraction of dark matter ($f_\chi = 1\%$) interacts with electrons or protons. Our results show complementary of the cosmological probes with the dark matter direct-detection experiments based on electron recoil or nucleon recoil, $N_\text{eff}$ measured from CMB and BBN, stellar cooling constraints, as well as dark matter searches at accelerators or colliders. Besides the CMB+BAO constraint, we also recast the constraints on thermal WDM from Lyman-$\alpha$ forest and MW subhalo data to those of DMb interactions by comparing their transfer functions. We observed different levels of relative sensitivity between the resulting constraints from Lyman-$\alpha$ forest and the MW subhalo data as compared to those from CMB+BAO for the different DMb models we considered, as discussed in Sec.~\ref{sec:res}.

The cosmological probes of dark matter--ordinary matter interactions uniquely capture common signatures of a large set of dark matter models. Sufficiently large interactions affect both the thermal history and the distribution of matter and leave multiple  opportunities for discovery in cosmological observations at different redshifts.
In the future, the Simons Observatory~\cite{SimonsObservatory:2019qwx} and CMB-S4~\cite{Abazajian:2019eic} will bring us more precise measurements of the CMB and better constraints on the dark matter--baryon scattering~\cite{Dvorkin:2020xga}. CMB-HD~\cite{Nguyen:2017zqu,Sehgal:2019ewc,Sehgal:2020yja,Sehgal:2019nmk} will improve on these even further, as well as probe the matter power spectrum to an unprecedented small scale.  In addition, measurements of the 21-cm signal~\cite{Munoz:2019hjh} will probe the matter power spectrum to  small scales.  The measurements of the MW subhalos can be improved through strong gravitational lensing using lens systems from future surveys combined with precise follow-up observations~\cite{Drlica-Wagner:2019xan,Birrer:2020snowmass,Enzi:2020ieg} and astrometric surveys like the {\it Gaia} mission~\cite{refId0, brown2021gaia}. On the theoretical side, future cosmological hydrodynamical simulations will reduce the uncertainties in the predictions of the Lyman-$\alpha$ forest data. Simulations performed specifically with the inclusion of dark matter--baryon interactions will be able to resolve the effects that these interactions have on the halo or galaxy formations, and thus make the constraints coming from MW subhalo data more robust. They may also provide important clues to break the degeneracy between the interacting dark matter effects and the warm dark matter effects.  Finally, future prospects in direct detection, indirect detection, and collider searches, in addition to an increasingly more accurate understanding of the local density and velocity distributions of dark matter, which plays a vital role in direct detection efforts \cite{Savage:2006qr,Freese:2012xd,Necib:2018iwb,OHare:2018trr,Wu:2019nhd,Buckley:2019skk,OHare:2019qxc,Buch:2019aiw,Buch:2020xyt,Radick:2020qip}, could shed light on the mystery of the nature and interactions of dark matter. 

\vskip 8mm

\acknowledgments
We thank Brown University's ``Oscar'' computing resources, made available through the Center for Computation and Visualization (CCV). We also thank Martin Schmaltz and Boston University for their hospitality, and acknowledge the resources offered by the Boston University Shared Computing Cluster (SCC) in the MGHPCC, which were used in the early stages of this work. We also thank Yacine Ali-Ha\"{i}mound, Kimberly Boddy, Andrew Chael, Jens Chluba, Cora Dvorkin, Wolfgang Enzi, Christina Gao, Daniel Gilman, Wayne Hu, Vera Gluscevic, Karime Maamari, Sam McDermott, Riccardo Murgia, Ethan Nadler, David Nguyen, Dimple Sarnaaik, Katelin Schutz, and  Lian-Tao Wang for helpful discussions. Finally, we would like to thank the anonymous Referee, whose comments and suggestions have helped improve this work. MBA is supported by the DOE grant DE-SC-0010010 and NASA grant 80NSSC18K1010. RE acknowledges support from DoE Grant DE-SC0009854, Simons Investigator in Physics Award 623940, and the US-Israel Binational Science Foundation Grant No.~2016153.
DM is supported by the Natural Sciences and Engineering Research Council of Canada and TRIUMF receives federal funding via a contribution agreement with the National Research Council Canada. YZ is supported by the Kavli Institute for Cosmological Physics at the University of Chicago through an endowment from the Kavli Foundation and its founder Fred Kavli.

\appendix

\section{{Constraints Based on Alternative Datasets, Model Assumptions, and Derivation Methods}}
\label{appA}

\subsection{CMB+BAO}
\label{app:cmb}

We mentioned in Secs.~\ref{sec:desc} and \ref{subsubsec:cmb_bao} that dark matter--proton interactions have been previously studied in the literature with Planck 2015 data~\cite{Ade:2015xua} and under slightly different assumptions. For completeness, in this appendix we compare these previous literature with our results for dark matter--proton scattering using Planck 2018 CMB~\cite{Aghanim:2018eyx} + BAO data~\cite{Beutler:2011hx,Ross:2014qpa,Alam:2016hwk}. We present our DMb constraints for two different modalities, based on whether or not we include non-linear effects in the matter power spectrum and thus in the CMB lensing computation. The baseline case, which we used in the main body of our paper, uses {\tt HALOFIT} \cite{Smith:2002dz,Bird:2011rb} to compute this nonlinearities, whereas in the other case we simple ignore any non-linear effects. Despite the fact that {\tt HALOFIT} has been optimized for $\LC$, we have decided to keep it for the main analysis of this work, since \textit{i.}~the transfer function of the $n=0$ DMb model is very similar to that of WDM, and \textit{ii.}~the dark matter sector (for both DMb and WDM) becomes more $\LC$-like at late times,\footnote{For example for the $n=0$ DMb model we found that the transfer function (\Eq{eq:transferDMb}) at $z=6$ (around the time at which the kernel of the lensing power peaks) is virtually the same as that at $z=0$, up to subpercent differences.} when the non-linearities start to become important (as dark matter--proton scatterings become increasingly rare or the dark matter cools down, respectively), which suggests that whatever non-linear effects affect WDM will affect the $n=0$ DMb model as well. Furthermore, we note that the suppression of the WDM matter power spectrum suppression is already typically used in the literature to place bounds on WDM mass, even at small scales where non-linear effects are important; and that these bounds have already been used as benchmarks to place constraints on a plethora of other models (see \Sec{sec:num}).

We find that the impact of the non-linearities is minimal for $n=-2$ and $n=-4$, and our bounds on dark matter--proton interactions are moderately stronger than those in the previous literature simply due to the use of the newest Planck data in combination with BAO measurements, the latter of which breaks degeneracies in the fits of cosmological parameters to CMB data. The exception is $n=0$, for which the inclusion of non-linear effects has a large impact on the CMB lensing computation, resulting in our bounds being stronger than those from previous literature by almost one order of magnitude.

\begin{figure}[t]
  \centering
  \includegraphics[width=0.49\textwidth]{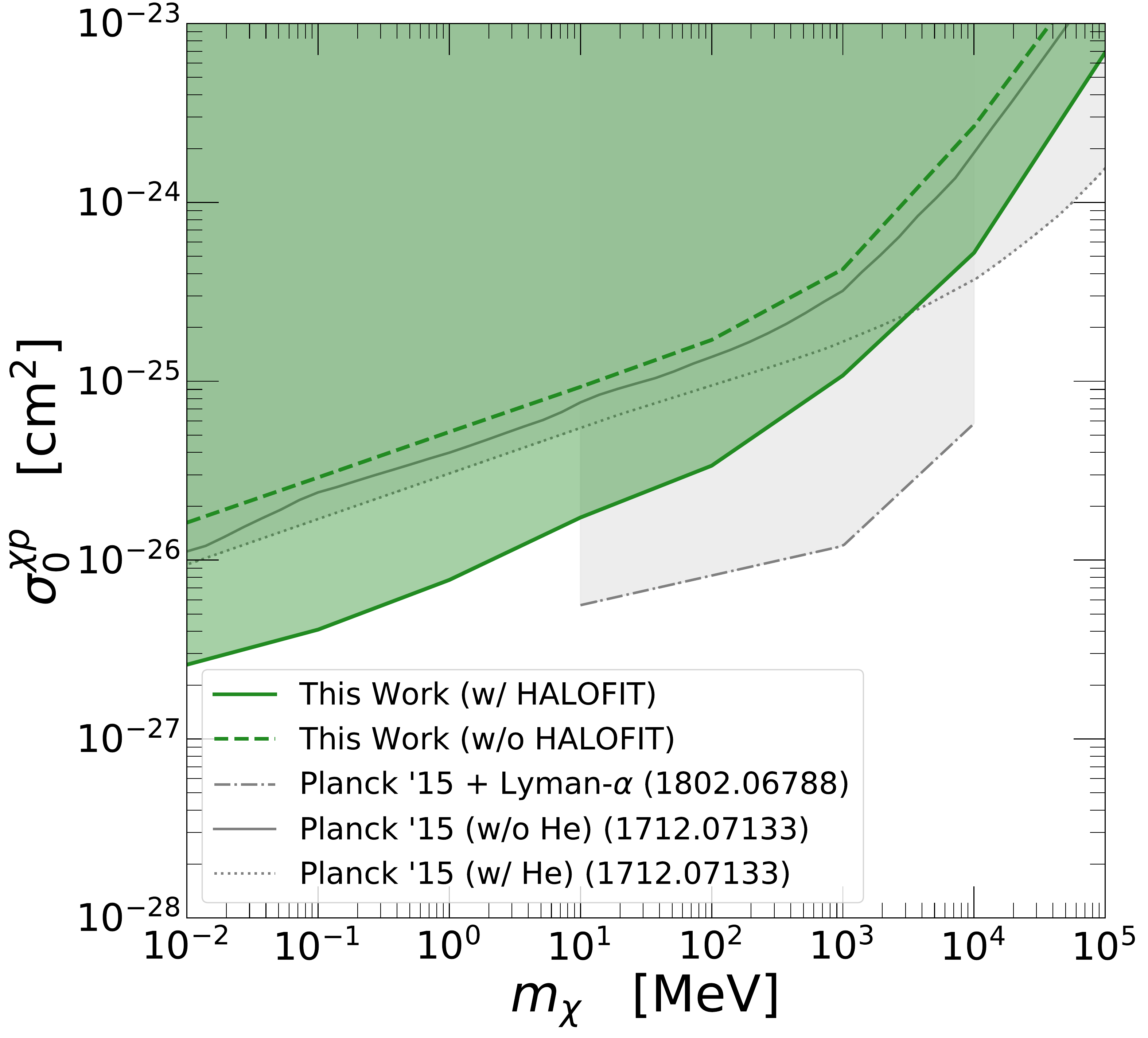}
    \includegraphics[width=0.49\textwidth]{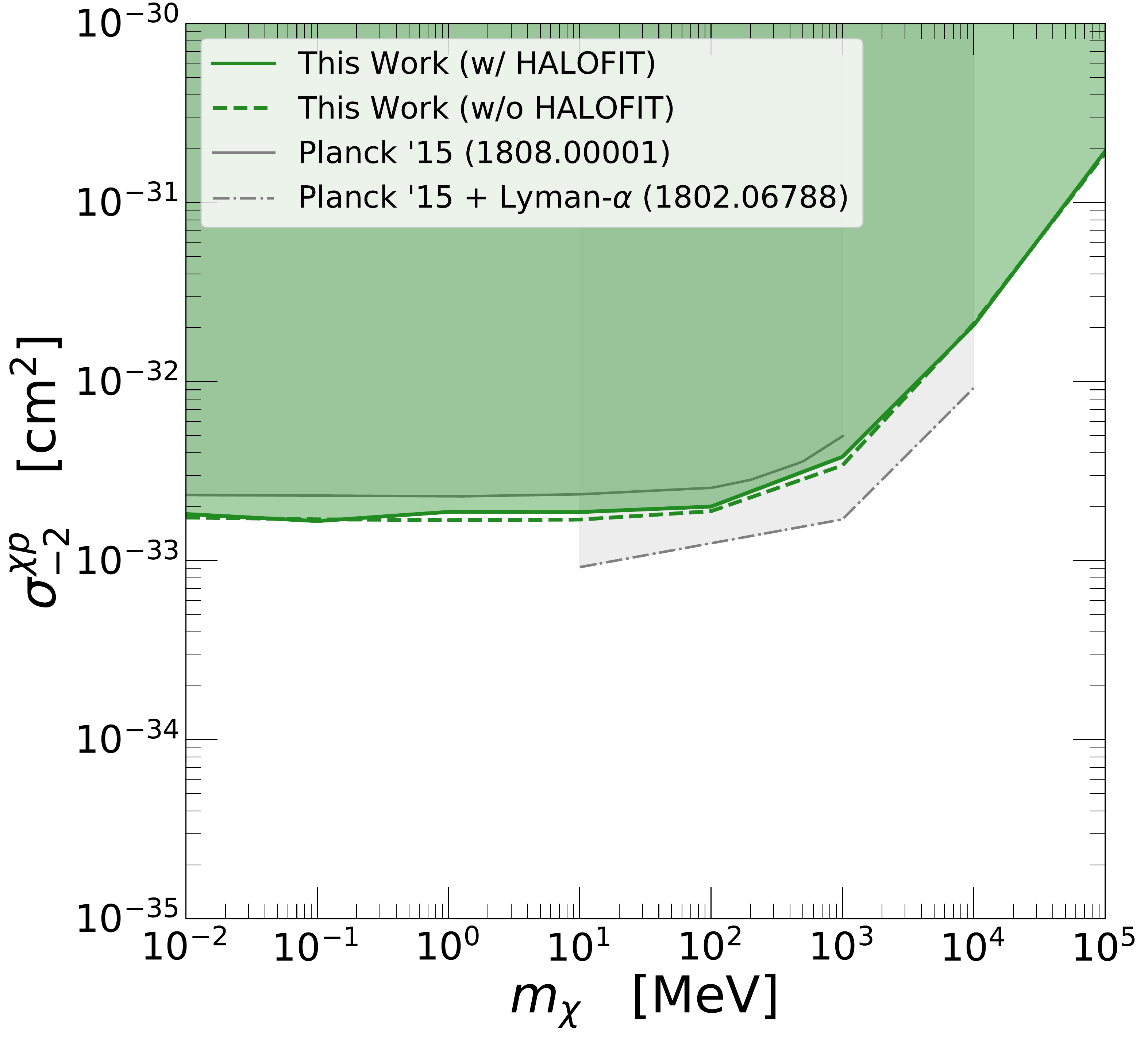} \\
  \includegraphics[width=0.49\textwidth]{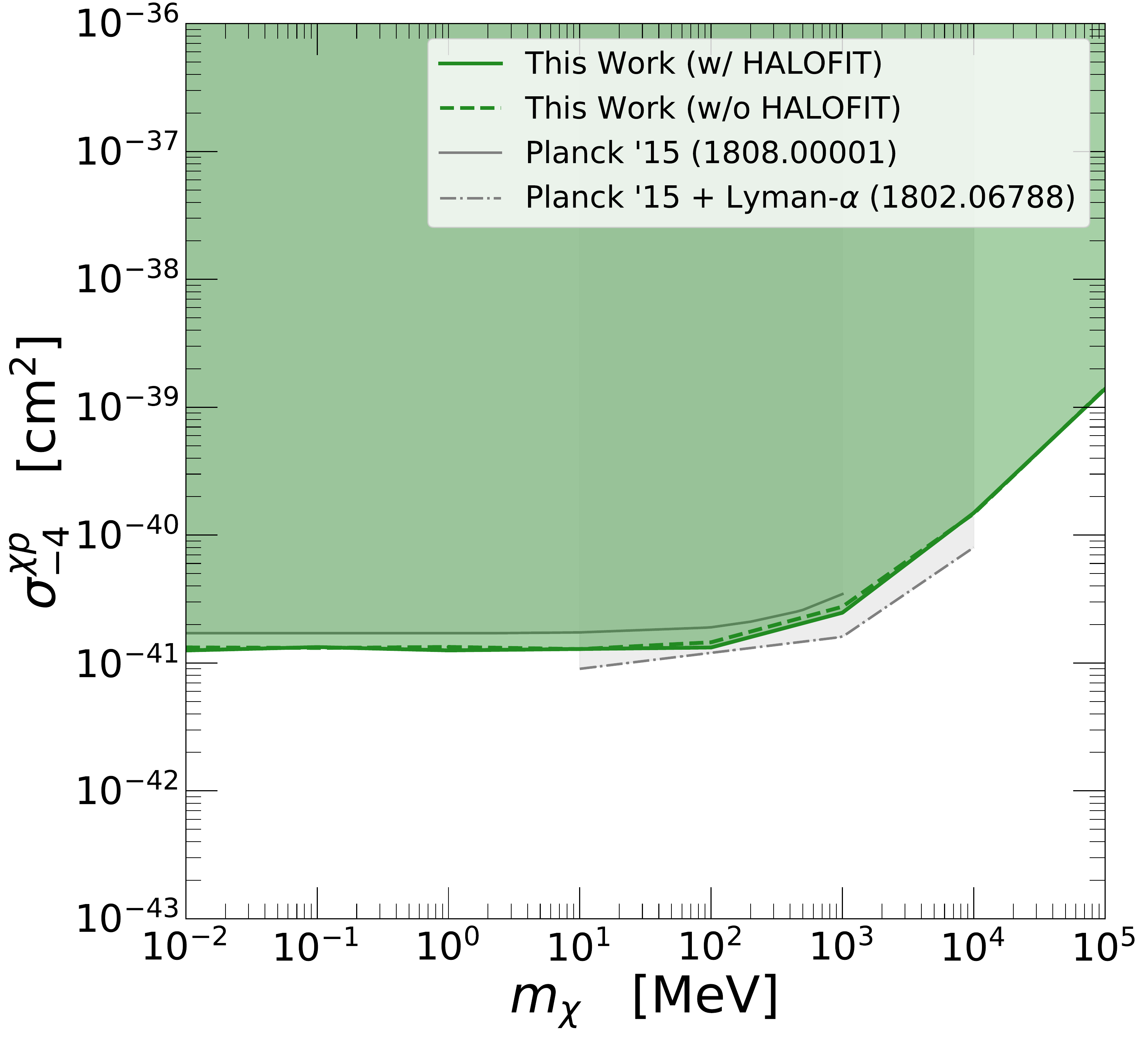}
  \caption{95\% C.L. bounds from CMB (Planck 2018) + BAO ({\bf solid green} lines: with the {\tt HALOFIT} \cite{Smith:2002dz,Bird:2011rb} treatment of nonlinearities; {\bf dashed green} lines: without {\tt HALOFIT}) on the $( \mx, \sn )$ parameter space for dark matter--protons interaction and $n=0$ ({\bf upper}), $n=-2$ ({\bf middle}), and $n=-4$ ({\bf lower}), assuming $f_\chi =100\%$ of dark matter interacts with protons.
  To compare these bounds with those based on Planck 2015 data, we include constraints from~\cite{Gluscevic:2017ywp, Boddy:2018wzy, Xu:2018efh} as {\bf gray} shaded regions (all assume $f_\chi=100\%$). See text for more details.}
      \label{fig:cmb_n024}
\end{figure}

\begin{itemize}
    \item $n=0$: this case was previously studied in~\cite{Gluscevic:2017ywp,Xu:2018efh}.
    Ref.~\cite{Gluscevic:2017ywp} gave constraints for the scenario where dark matter scatters with both protons and helium nuclei. Ref.~\cite{Xu:2018efh} provides separate constraints for (1) dark matter scattering with protons only and (2) dark matter scattering with both protons and helium. Including dark matter--helium interactions is helpful to set stronger bounds on $\sigma^{\chi p}$ for dark matter with mass $\gtrsim \GeV$. However, it also brings complications as we will discuss below. Also note that Ref.~\cite{Xu:2018efh}, as mentioned in \Sec{subsubsec:lya}, included Lyman-$\alpha$ data from the linear matter power spectrum measurements from SDSS-II low resolution, low signal-to-noise quasar spectra~\cite{McDonald:2004eu}. These measurements, however, rely on a $\LC$-based modeling of the matter power spectrum~\cite{McDonald:2004eu,McDonald:2004xn}. This means that the robustness of constraints based on this dataset, for non-$\LC$ models (such as DMb), is not guaranteed.
    
    There are two complications when interpreting constraints from Refs.~\cite{Gluscevic:2017ywp,Xu:2018efh} for the scenario where dark matter--helium scattering is included. For the dark matter--helium scattering process, both references assume dark matter only scatters with the protons (but not with neutrons) in the helium nuclei. Hence the dark matter--helium scattering cross section is related to the dark matter--proton scattering cross section by  $\sigma^{\chi \text{He}} =  4 ({\mu_{\chi \text{He}}}/{\mu_{\chi p}})^2 \sigma^{\chi p}$. The relative strength of dark matter's couplings to the different nucleons is model-dependent; the choice made in~\cite{Gluscevic:2017ywp,Xu:2018efh} leads to \emph{conservative} bounds when scattering on helium is included, compared to the choice of equal couplings to protons and neutrons as is typically done to display results from direct-detection experiments. A second complication to interpreting constraints from dark matter--helium scattering can arise from the breaking down of the scaling relation $\sigma^{\chi \text{He}} =  A^2 ({\mu_{\chi \text{He}}}/{\mu_{\chi p}})^2 \sigma^{\chi N}$ for the parameter region of interest, as mentioned in~\Sec{subsec:pheno}. Ref.~\cite{Digman:2019wdm} shows that $\mathcal{O}(1)$ derivation of the scaling relation happens when $\sigma^{\chi N} \gtrsim 4\times 10^{-28}\,\text{cm}^2$ for $A=4$ (atomic mass number of helium).
    
    The upper-left panel of \Fig{fig:cmb_n024} compares the results from Refs.~\cite{Gluscevic:2017ywp,Xu:2018efh} with our results, for the cases with (solid green lines) and without (dashed green lines) {\tt HALOFIT} \cite{Smith:2002dz,Bird:2011rb} to treat non-linear effects. We can see that non-linearities have an important impact on the strength of the constraints. The reason is due to the Planck lensing likelihood, which depends on the matter power spectrum at small scales, where non-linearities are important. Since for $n=0$ the matter power spectrum is strongly suppressed at small scales (see \Sec{sec:obs}), the inclusion of non-linear effects, and therefore of a stronger lensing, will boost the constraining power of the CMB dataset.
    
    \item $n=-2$: this model was recently considered in Refs.~\cite{Xu:2018efh,Boddy:2018wzy}. The comments from the previous bullet point regarding the use of Lyman-$\alpha$ in Ref.~\cite{Xu:2018efh} also apply to the $n=-2$ case. As mentioned in \Sec{sec:desc}, Ref.~\cite{Munoz:2015bca} and later Ref.~\cite{Boddy:2018wzy} developed an improved treatment of the DMb relative bulk velocity, solving the thermal and perturbation evolution equations in a self-consistent and iterative manner. Since in this accurate treatment the DMb drag tends to decrease the bulk velocity, the constraints are stronger than in the mean-field approximation used in Refs.~\cite{Xu:2018efh,Becker:2020hzj} and used in this work. However the bounds in Ref.~\cite{Boddy:2018wzy} were obtained using Planck 2015 data ~\cite{Ade:2015xua}, while we use the more recent Planck 2018 data in combination with BAO; hence our constraints are stronger overall. The upper-right panel of \Fig{fig:cmb_n024} compares the results from these two references with ours, for the cases with (solid green lines) and without (dashed green lines) {\tt HALOFIT} \cite{Smith:2002dz,Bird:2011rb} to treat non-linear effects. We can see that non-linearities have only a small impact on the constraining power of the CMB dataset. The reason is that for the cross sections being probed the DMb matter power spectrum is not so suppressed as to deviate very strongly from that from $\LC$.
    
    \item $n=-4$: this model was explored in Refs.~\cite{Xu:2018efh,Boddy:2018wzy}. The same comments as in the previous item apply in this case. The lower panel of \Fig{fig:cmb_n024} compares the results from these two references with ours, for the cases with (solid green lines) and without (dashed green lines) {\tt HALOFIT} \cite{Smith:2002dz,Bird:2011rb} to treat non-linear effects.
\end{itemize}

\subsection{Lyman-$\alpha$ forest}
\label{app:lya}

We devoted \Sec{subsubsec:lya} to describing the MIKE/HIRES+XQ-100 Lyman-$\alpha$ dataset~\cite{Viel:2013fqw,Irsic:2017ixq}, as well as the area criterion of \Eq{eq:area_criterion}, first described in Refs.~\cite{Murgia:2017lwo,Murgia:2017cvj,Murgia:2018now}. We took as reference the 95\% C. L. Lyman-$\alpha$ bound on WDM from Ref.~\cite{Irsic:2017ixq}, $\delta A_{\rm ref} = \delta A_{5.3~\keV}^\wdm = 0.31$, and used this criterion to place bounds on the $( \mx, \sn )$ parameter space of the DMb models, for all choices of $B$ and $n$.  The results were shown in Figs.~\ref{fig:n0_bounds}, \ref{fig:n-2_bounds}, and \ref{fig:n-4_bounds} of \Sec{sec:res}. 

\begin{figure}[t]
  \centering
  \includegraphics[width=0.45\textwidth]{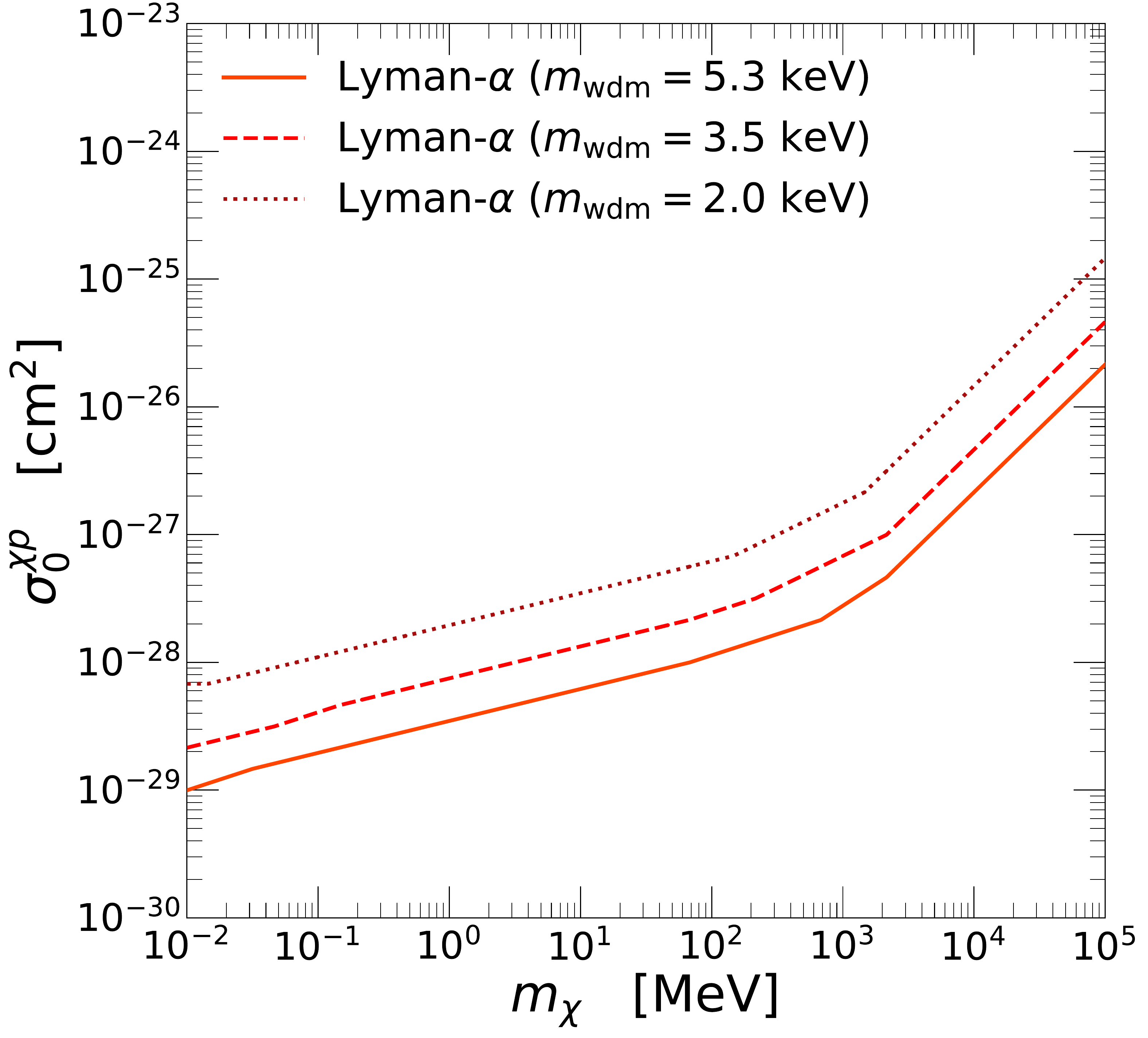}
  \includegraphics[width=0.45\textwidth]{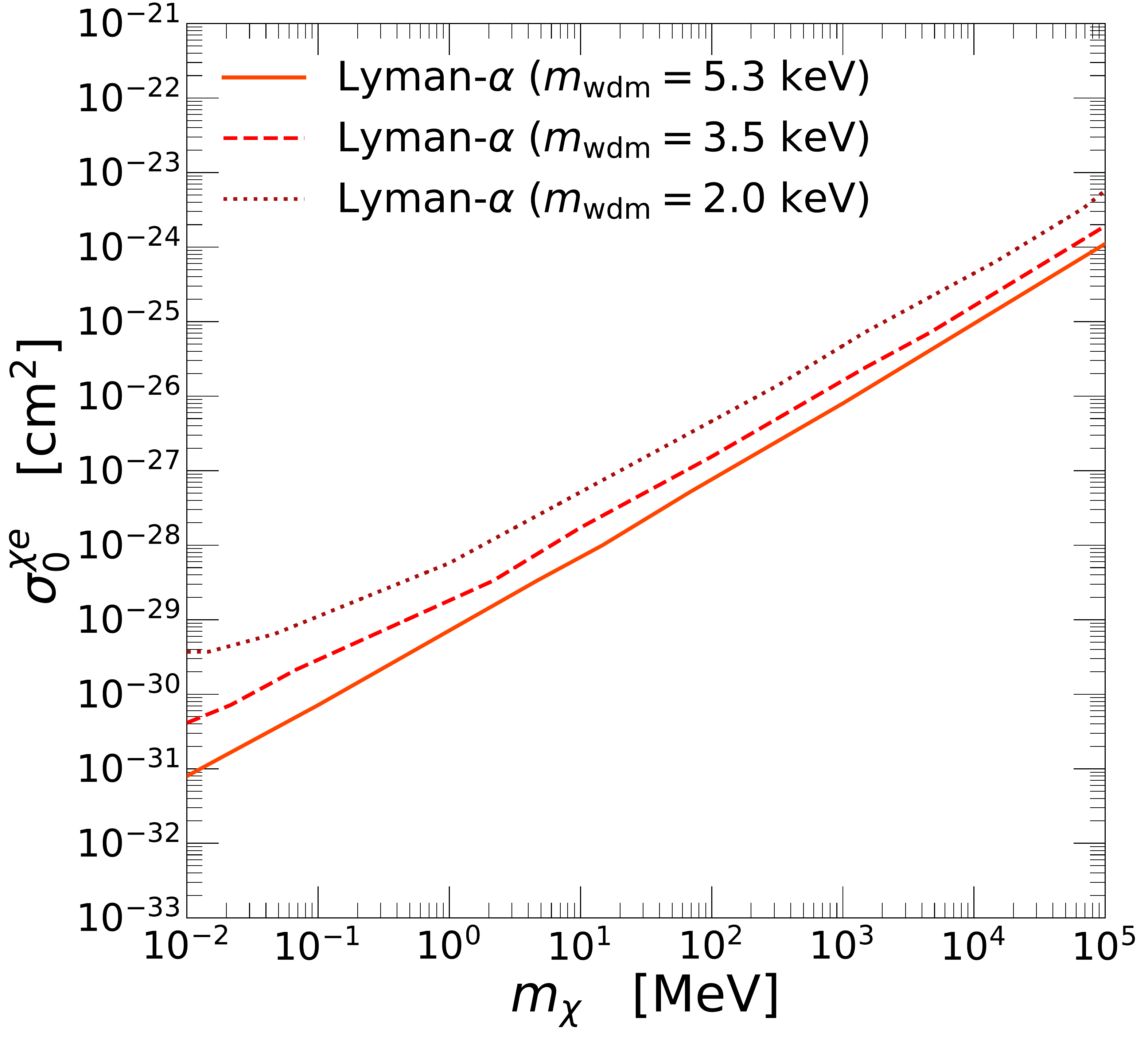}
  \includegraphics[width=0.45\textwidth]{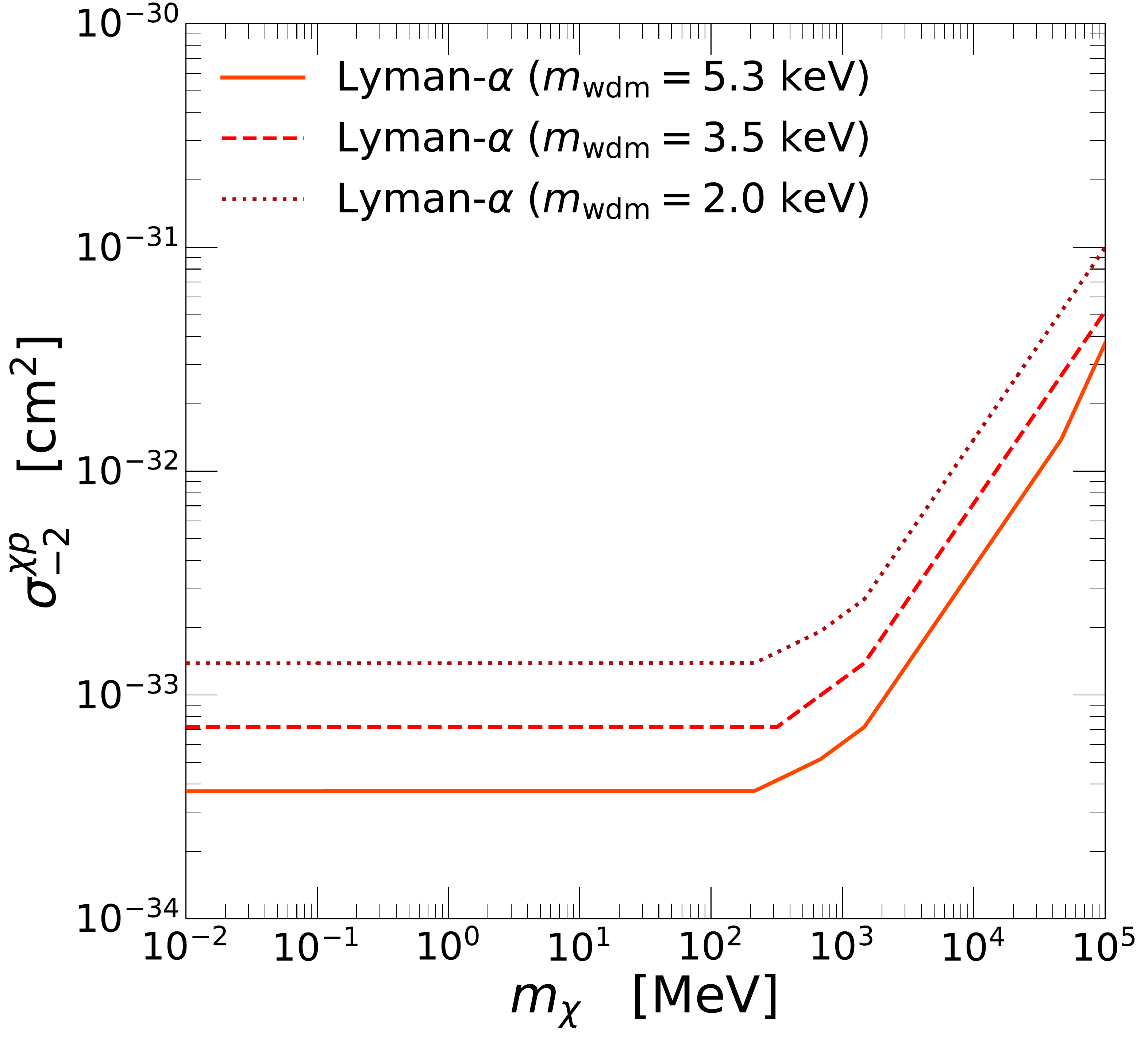}
  \includegraphics[width=0.45\textwidth]{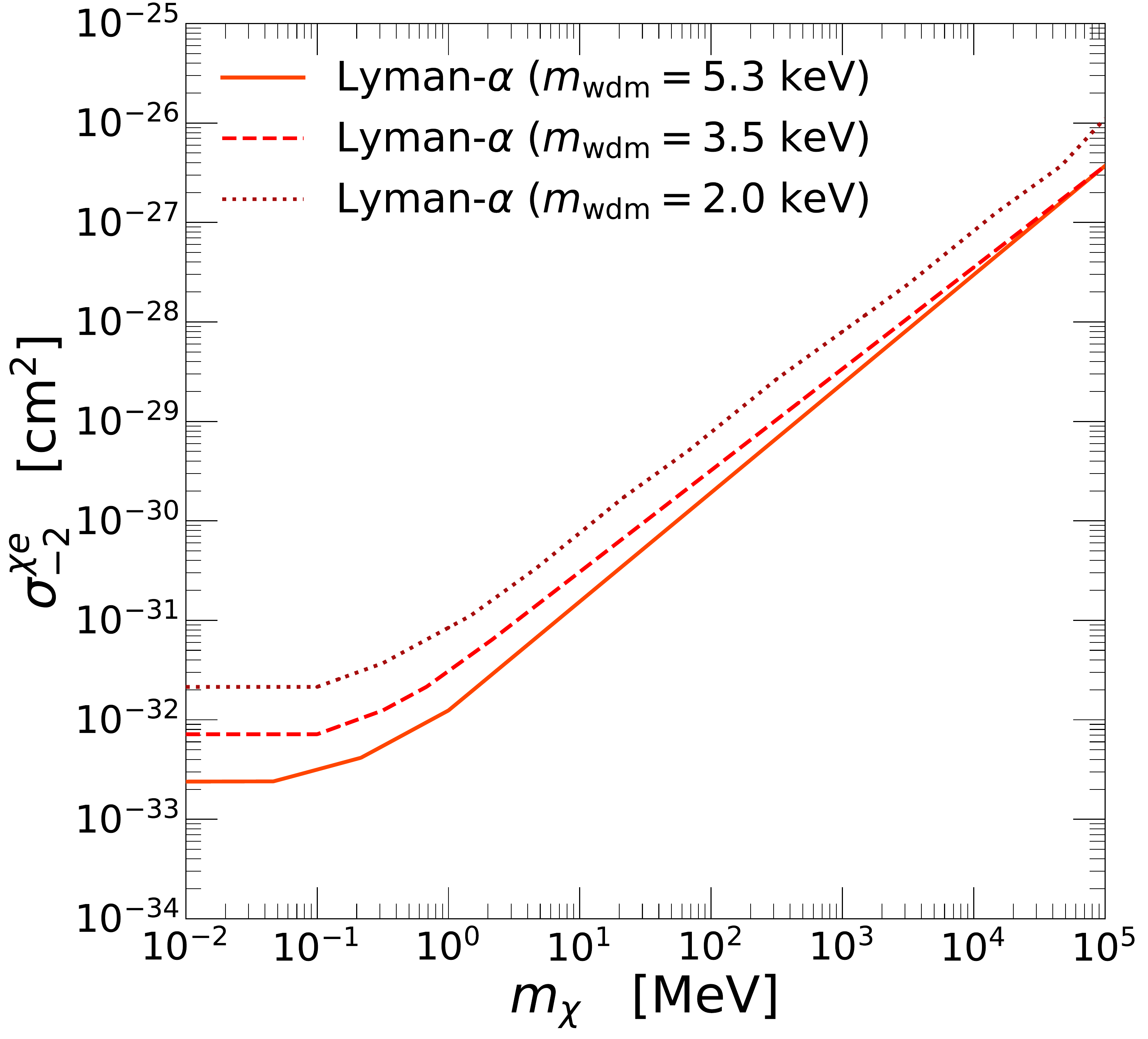}
  \includegraphics[width=0.45\textwidth]{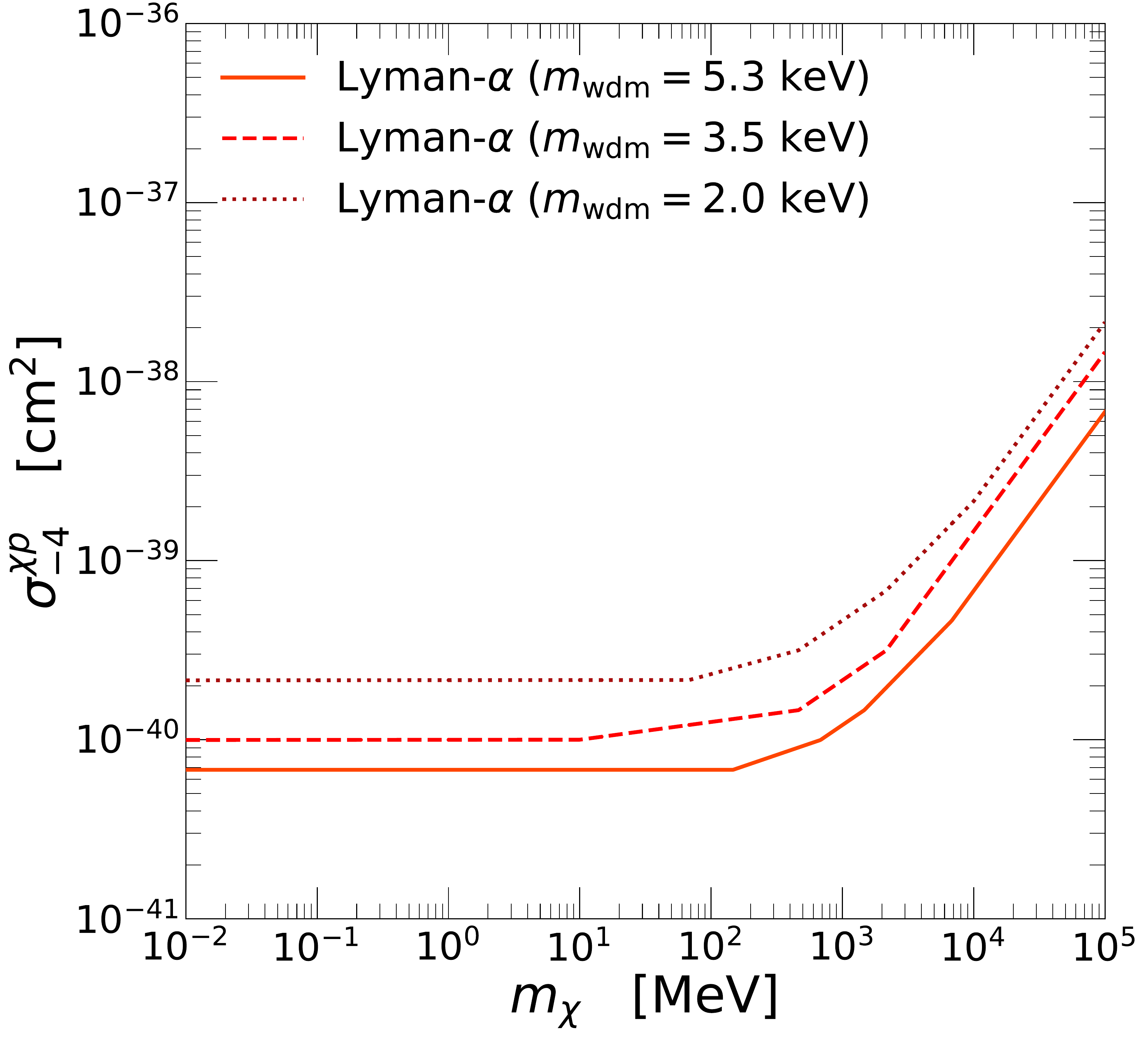}
  \includegraphics[width=0.45\textwidth]{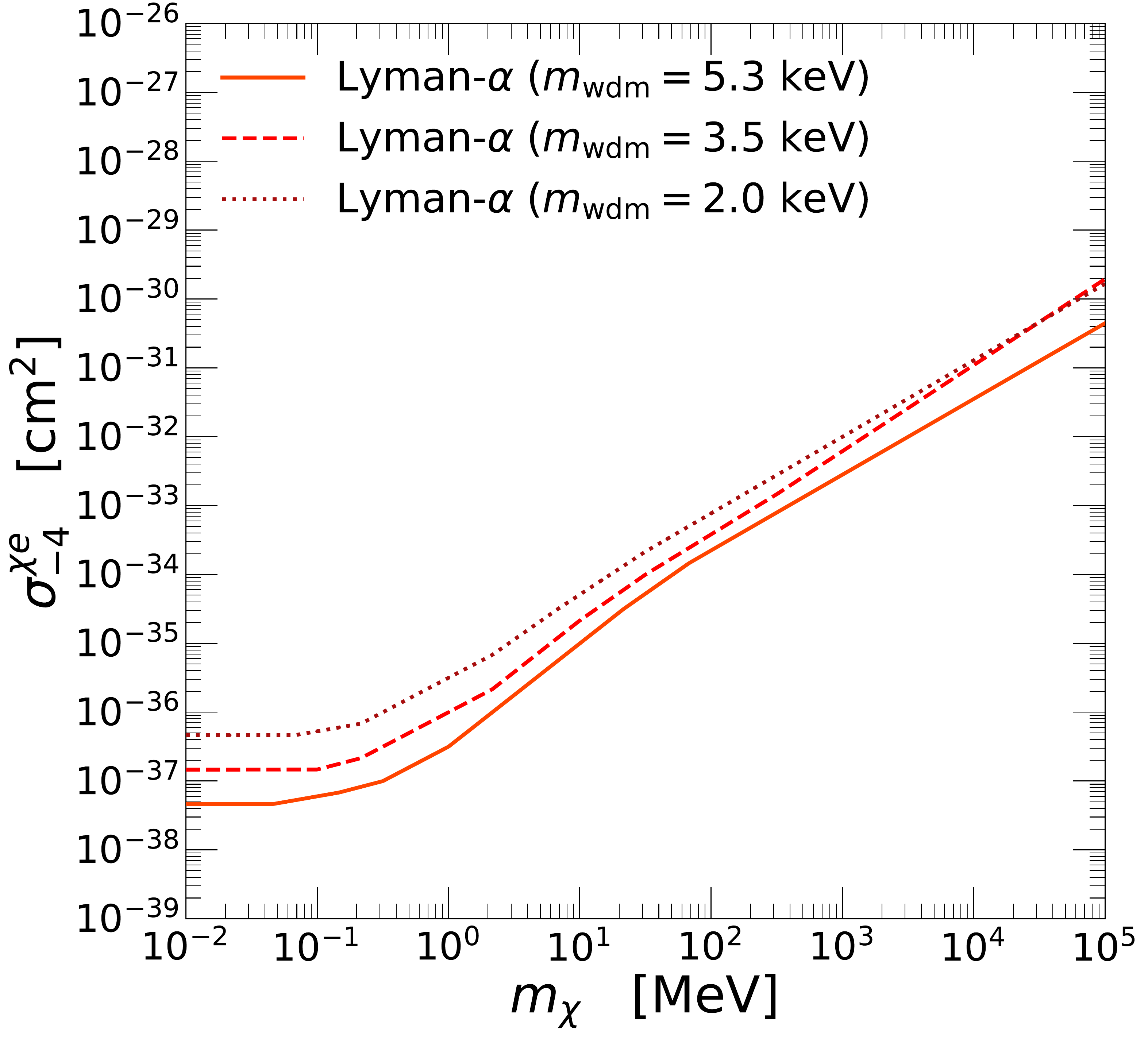}
  \caption{95\% C.L. bounds from Lyman-$\alpha$ on the DMb $( \mx, \sn )$ parameter space for $n=0$ (\textbf{top row}), $n=-2$ (\textbf{middle row}), and $n=-4$ (\textbf{bottom row}); and for dark matter--proton (\textbf{left column}) or dark matter--electron (\textbf{right column}) scattering, with both the tight ($\delta A_{5.3~\keV}^\wdm = 0.31$, {\bf solid} line) and conservative ($\delta A_{3.5~\keV}^\wdm = 0.46$, {\bf dashed} line) reference values for the area criterion. We have also included, for illustrative purposes, bounds from the most conservative WDM masses considered in this paper ($m_\wdm = 2.02~\keV$ \cite{Newton:2020cog}, yielding $\delta A_{2.02~\keV}^\wdm = 0.67$, {\bf dotted} line) which were not derived from Lyman-$\alpha$ data but Milky Way luminous satellites (see \Tab{tab:mw-limit}).}
  \label{fig:lya}
\end{figure}

It is worth noting that the $m_\wdm > 5.3~\keV$ 95\% C.L. bound in Ref.~\cite{Irsic:2017ixq} was obtained by using the standard smooth power-law temperature evolution of the intergalactic medium, similar to what was employed in Ref.~\cite{Palanque-Delabrouille:2019iyz}. Assuming instead a non-smooth evolution with sudden changes of up to $\Delta T = 50,000~\mathrm{K}$ between contiguous redshift bins (10 bins for the combined MIKE/HIRES+XQ-100 analysis), Ref.~\cite{Irsic:2017ixq} found a looser, conservative constraint of $m_\wdm > 3.5~\keV$ on the WDM mass. For this WDM bound the reference area suppression is $\delta A_{\rm ref} = \delta A_{3.5~\keV}^\wdm = 0.46$. Shown in \Fig{fig:lya} are the constraints on the DMb models assuming this reference value for the area criterion instead, compared with the bounds shown in the main text of our work.\footnote{For visual clarity, we only show the boundaries of the exclusion regions in \Fig{fig:lya} without shading parameter space above (also for~\Fig{fig:mws}).} Overall the resulting constraints differ by a factor of a few.

\subsection{Milky Way subhalos}
\label{app:mwsg}
As discussed in Sec.~\ref{subsubsec:joint}, various studies draw constraints on the transfer function of WDM by analyzing the abundance of MW subhalos. Those bounds are then translated into constraints on the DMb interaction, assuming there are no additional late-time baryonic effects other than enhanced subhalo disruptions, by matching the transfer function of the DMb model, $T_\text{DMb}$, to that of WDM, $T_\text{WDM}$, using two matching criteria, half-mode and fixed $k$ (see \Sec{subsubsec:joint}). In the main text, we adopted the critical WDM mass of $m_\text{WDM} > 6.5\,\keV$~\cite{Nadler:2020prv} (95\% C.L.) to set bounds on DMb. Here we explore the resulting constraints from other critical WDM mass values ($m_\text{WDM} > 2.02\,\keV$~\cite{Newton:2020cog}, $m_\text{WDM} > 6.7\,\keV$~\cite{Enzi:2020ieg} and $m_\text{WDM} > 9.7\,\keV$~\cite{Nadler:2020prv}) that have appeared in the literature and summarized in~\Tab{tab:mw-limit}. In particular, it is interesting to see the change in DMb bound based on the WDM constraint from~\cite{Newton:2020cog} ($m_\text{WDM} > 2.02\,\keV$), which used a different galaxy formation modeling from that of~\cite{Nadler:2020prv,Nadler:2021dft} in the derivation of WDM constraints.

Note that for the fixed $k$ matching criteria, we again choose a comoving wavenumber for matching that has a transfer function value $T_\text{WDM, crit} =T_\text{WDM} (k_\text{match}) \approx 2\%$. To be more concrete, we choose $k_\text{match} = 35\,h/\Mpc$ for the $m_\text{WDM} = 2.02\,\keV$ case, $k_\text{match} = 135\,h/\Mpc$ for the $m_\text{WDM} = 6.7\,\keV$ case, and $k_\text{match} = 200\,h/\Mpc$ for the $m_\text{WDM} = 9.7\,\keV$ case. {Since both the DMb and WDM transfer functions are near-vanishing at large $k$, the DMb constraints resulting from this scheme will not change significantly if we adopt a different comoving wavenumber for matching, as long as the fixed $k$ requirement $T_\text{WDM, crit} =T_\text{WDM} (k_\text{match}) \lesssim 2\%$ is satisfied.}

\begin{figure}[!h]
  \centering
  \includegraphics[width=0.45\textwidth]{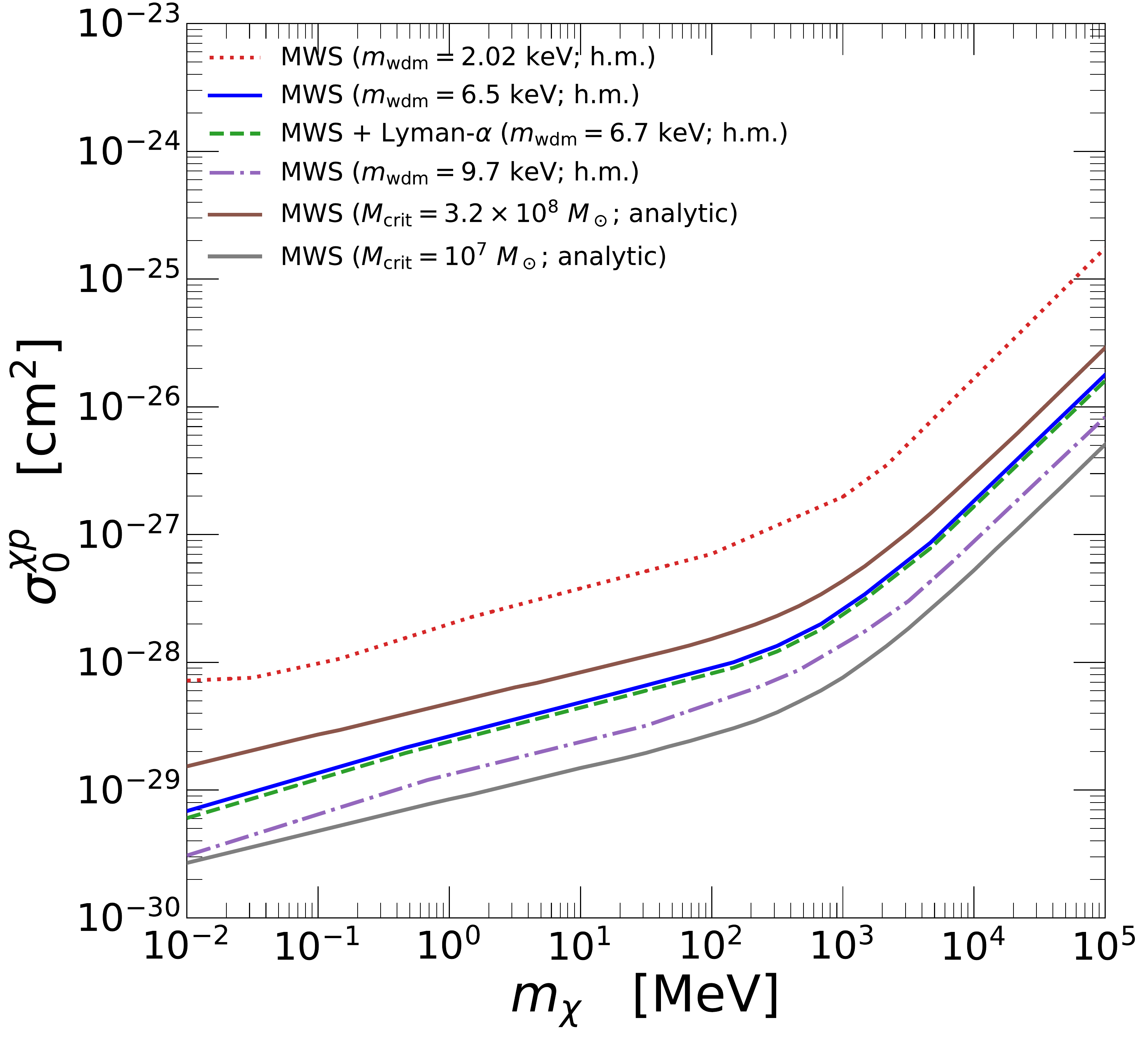}
  \includegraphics[width=0.45\textwidth]{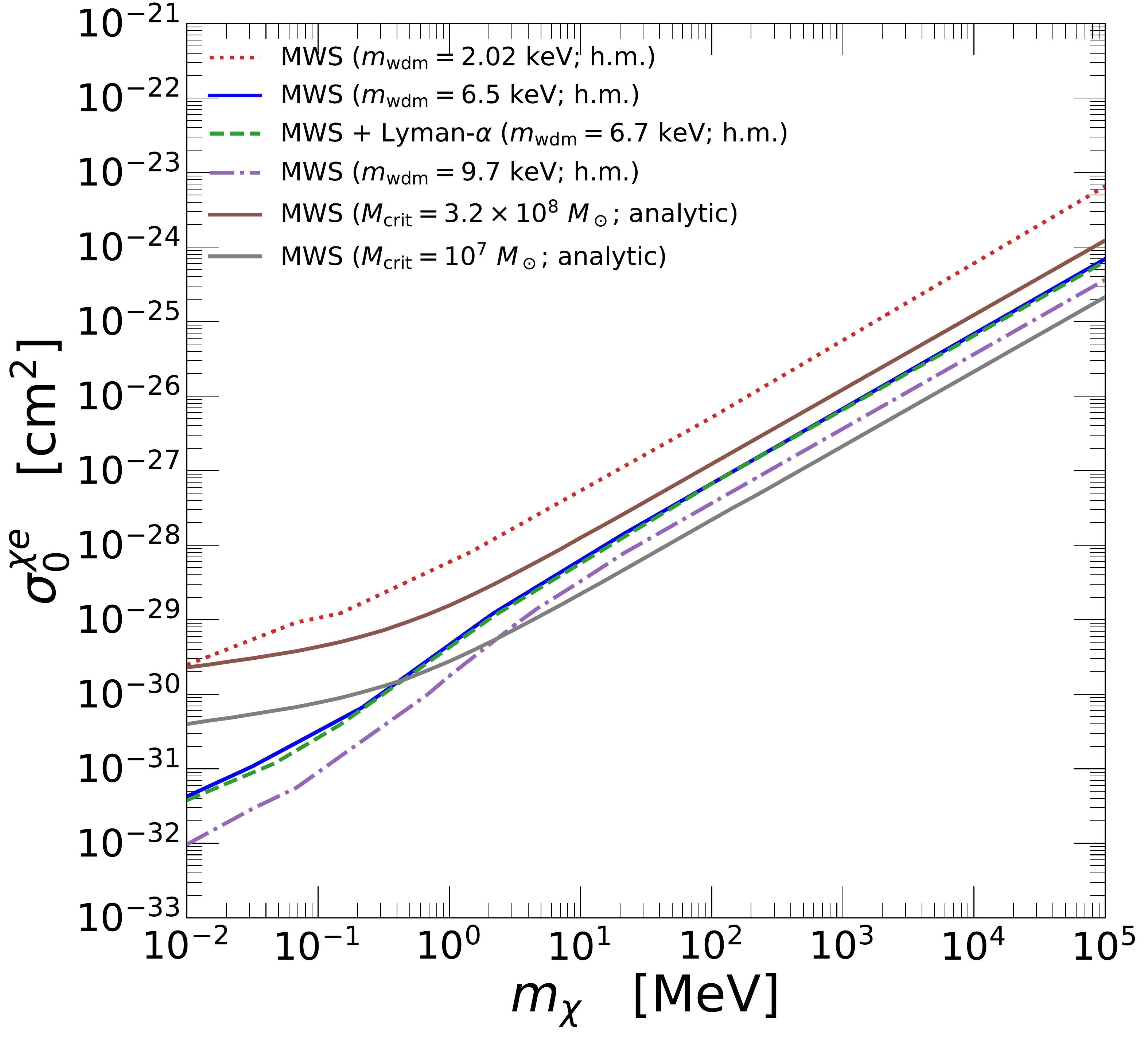}
  \includegraphics[width=0.45\textwidth]{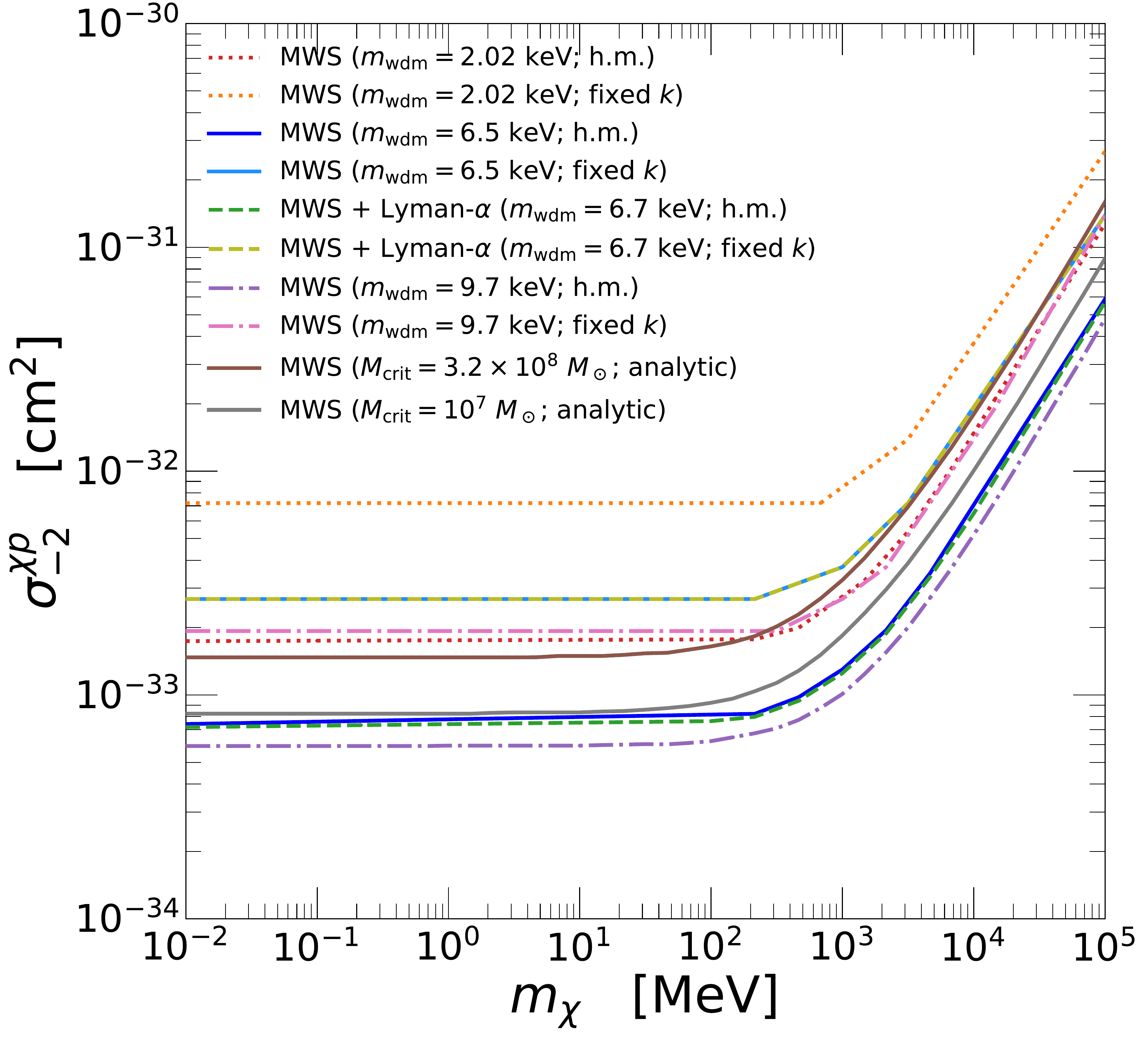}
  \includegraphics[width=0.45\textwidth]{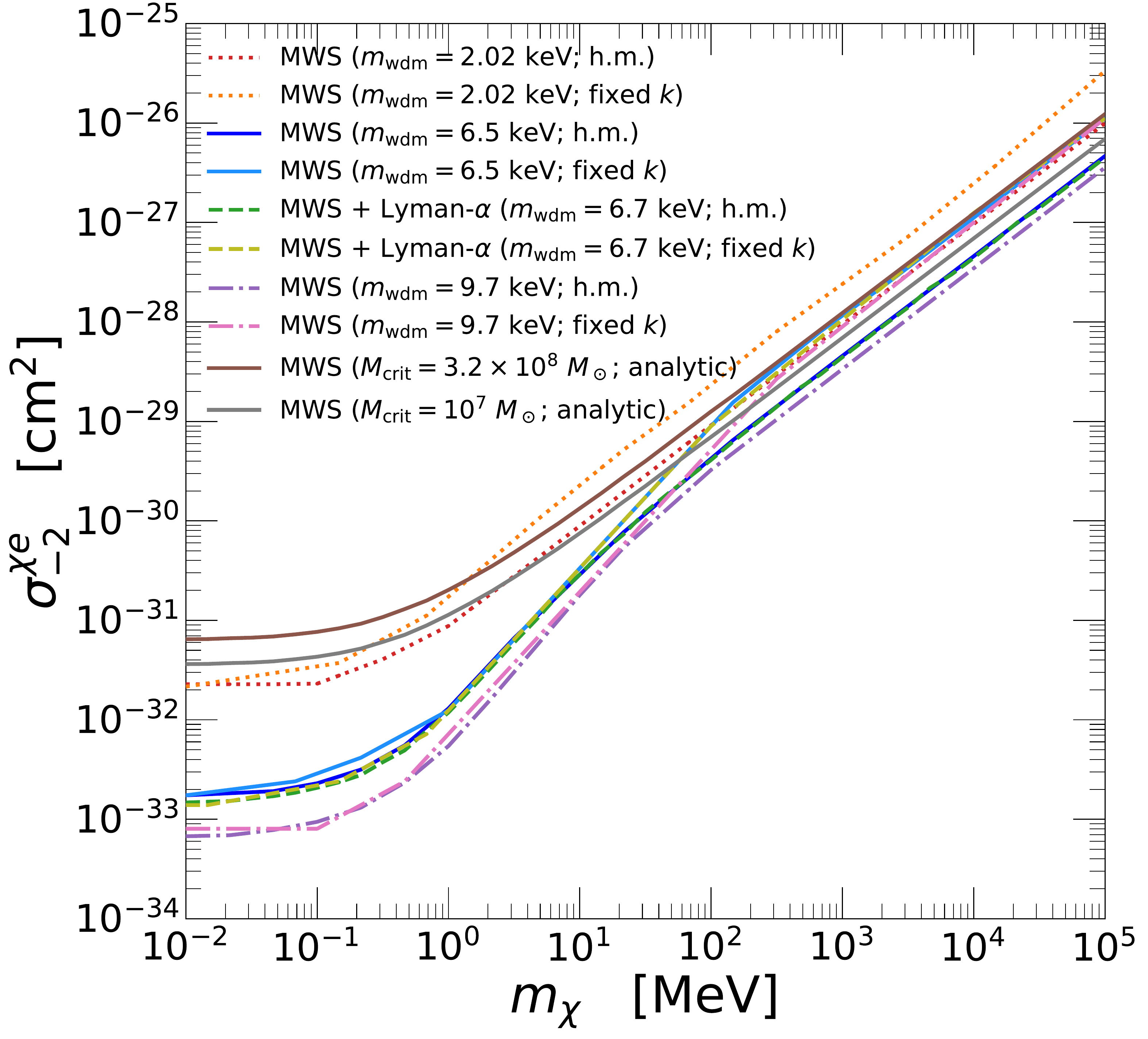}
  \includegraphics[width=0.45\textwidth]{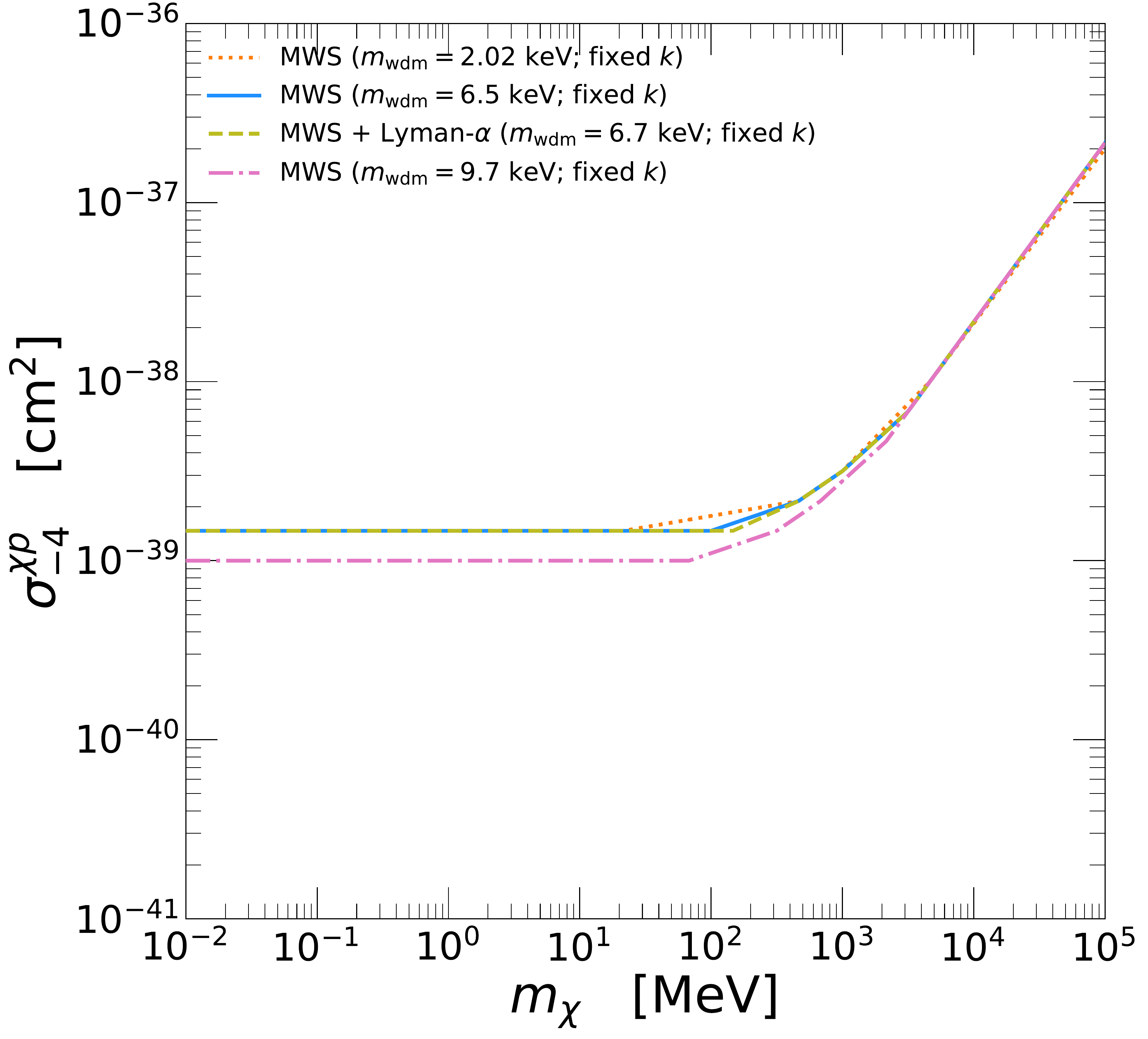}
  \includegraphics[width=0.45\textwidth]{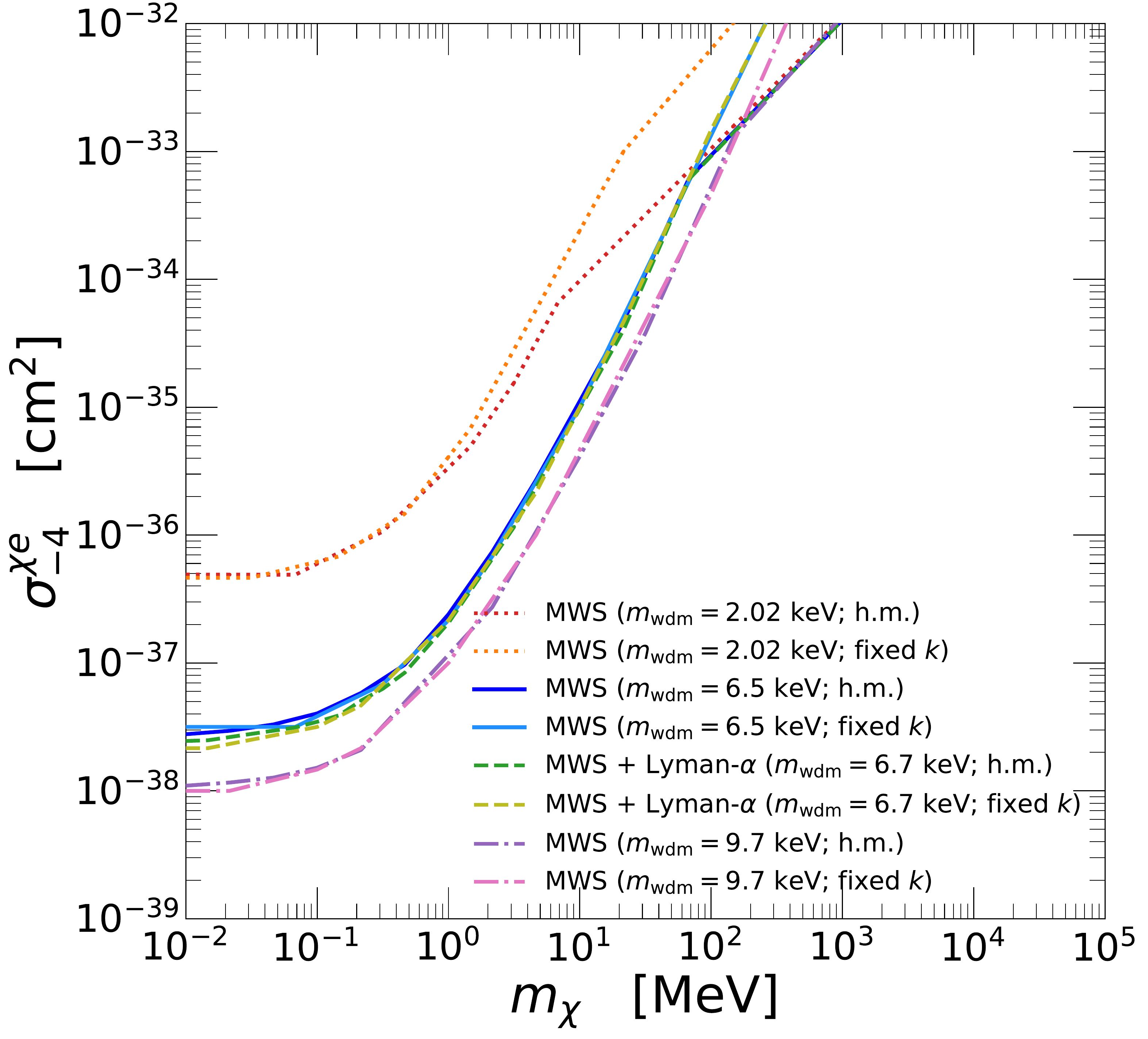}
  \caption{95\% C.L. bounds from MW subhalos on the $( \mx, \sn )$ parameter space for $n=0$ (\textbf{top row}), $n=-2$ (\textbf{middle row}), $n=-4$ (\textbf{bottom row}), and dark matter--proton (\textbf{left column}) or dark matter--electron (\textbf{right column}) scattering. Note that the WDM bound from~\cite{Enzi:2020ieg} includes data from Lyman-$\alpha$ forest besides the MW subhalos. For $n=0$, since the DMb and WDM transfer functions are very similar, the constraints from the two numerical matching criteria are almost identical, and we only show the bound from the half-mode matching criterion. For $n=0$ and $n=-2$, we also include the constraints from the analytic method described in this appendix. Finally, since for $n=-4$ dark matter--proton interactions the DMb transfer function is very different from that of the WDM, we only show the constraint from the fixed $k$ matching criterion.}
  \label{fig:mws}
\end{figure}

The resulting bounds on $n=0$, $n=-2$, and $n=-4$ type DMb interactions are shown in \Fig{fig:mws}, where we also show the constraints from $m_\text{WDM} > 6.5\,\keV$ that we used in main text for comparison. For $n=0$, the transfer functions of DMb are very similar to that of WDM, and the constraints from the two matching criteria are almost identical. Therefore we only show the constraints from the half-mode matching scheme in~\Fig{fig:mws} for $n=0$. The similarity in the transfer functions also occurs for $n=-2$ dark matter--electron interactions with dark matter mass smaller than 10 MeV, and for $n=-4$ dark matter--electron interactions with dark matter mass smaller than 100 MeV. However, the shape of the transfer functions for $n=-2,-4$ dark matter--electron interactions with heavier dark matter masses, and those for $n=-2,-4$ dark matter--proton interactions, are different from that of the corresponding WDM transfer functions. As a consequence, the resulting constraints from our fixed $k$ matching criterion become much weaker than those from our half-mode matching criterion. In the case of $n=-4$ dark matter--proton interaction, the transfer function of DMb becomes almost flat  for $10^{-1} h/\text{Mpc} \leq k \leq 10^2 h/\text{Mpc}$. Consequently, half-mode matching to the WDM transfer function no longer guarantees that the DMb is ruled out at small scales ($k> k_\text{half-mode}$). Therefore we only use the fixed $k$ matching criterion for dark matter--proton scattering with $n=-4$, as shown in \Fig{fig:mws}.

For completeness, we also include a simple analytic method to derive upper limits based on the smallest halo mass, described in Ref.~\cite{Nadler:2019zrb} and summarized here. For $n=0$ and $n=-2$ the ratio of the DMb momentum-transfer rate $\Rx$ (defined in \Eq{eq:Rx}) to the conformal Hubble expansion rate $\mH$ decreases with time. We can then define a critical redshift $z_{\rm kin}$ at which both are equal, which corresponds to the kinetic decoupling of the dark matter fluid from the baryons:
\beq
    \mH \big\vert_{z_{\rm kin}} = \Rx \big\vert_{z_{\rm kin}} \quad \text{(kinetic decoupling)} \ .
\eeq
To obtain $z_{\rm kin}$ we first need to solve for the thermal decoupling redshift $z_{\rm th}$ at which the heat-transfer rate $\Rxp$ (defined in \Eq{eq:Rxp}):
\beq
    \mH \big\vert_{z_{\rm th}} = \Rxp \big\vert_{z_{\rm th}} \quad \text{(thermal decoupling)} \ ,
\eeq
and then substitute it in a simplified analytic formula for the evolution of the dark matter temperature $T_\chi$, which accounts for the tightly-coupled and adiabatic regimes:
\beqa
    T_\chi & = & T_0 (1+z) \quad \text{for } z > z_{\rm th} \nonumber\\
    & = & T_0 (1+z_{\rm th}) \bl( \frac{1+z}{1+z_{\rm th}} \br)^2  \quad \text{for } z \leq z_{\rm th} \ ,
\eeqa
with $T_0$ being the CMB temperature today.

Having thus found $z_{\rm kin}$, we compute the critical comoving wavenumber $k_{\rm crit}$ that enters the horizon just as the dark matter fluid is decoupling from the baryons, its associated wavelength $\lambda_{\rm crit}$, and the mass $M_{\rm crit}$ of the halos that correspond to the $k_{\rm crit}$ perturbation modes:
\beqa
    k_{\rm crit} & \equiv & 2 \mH \big\vert_{z_{\rm kin}} \ , \\
    \lambda_{\rm crit} & \equiv &  2\pi/k_{\rm crit} \ , \\
    M_{\rm crit} & \equiv & \frac{4 \pi}{3} \rho_\mat \bl( \frac{\lambda_{\rm crit}}{2} \br)^3 \ .
\eeqa
$M_{\rm crit}$ can then be taken as a proxy for the mass of the smallest halos that can be formed in the context of a particular choice of parameters for the DMb model. Our analytic bounds, shown in \Fig{fig:mws} for $n=0$ and $n=-2$, are then obtained by contrasting $M_{\rm crit}$ with the mass $M_{\rm min}$ of the smallest halos observed. We use two different values from the literature: $M_{\rm min} = 3.2 \times 10^8 ~ M_\odot$~\cite{Maamari:2020aqz} and $M_{\rm min} = M_{\rm h.m.} = 10^7~ M_\odot$~\cite{Nadler:2021dft}. This latter value corresponds to the mass of halos associated to the half-mode of the WDM transfer function with $m_{\rm wdm} = 9.7~\keV$, which we include for comparison with our other matching schemes.

\section{The momentum-transfer cross section for dark matter--ordinary matter interaction with ultra-light dark photon mediation.}
\label{appB}

In Sec.~\ref{subsubsec:darkphoton}, we introduce two prescriptions to regulate the  momentum-transfer cross sections for dark matter--ordinary matter interaction with ultra-light dark photon mediation: \textit{i.} the Debye-length screening and \textit{ii.} the Debye-mass screening. Here we give detailed derivation for the momentum-transfer cross section for each prescription. 

\subsubsection*{\textit{i.} Debye-length screening} 
The prescription regulates the Coulomb-like interaction, \Eq{eq:massless}, at the phase-space integration level by setting  the maximal impact parameter to be the Debye screening length of the charged plasma~\cite{McDermott:2010pa},
\beq
\lambda_\text{D} = \sqrt{\frac{ T_e}{e^2 n_e}} \ ,
\eeq
where $T_e$ and $n_e$ are the temperature and number density of the electrons, respectively. In our case, the Standard Model baryonic fluid contains both electrons and protons, which have unit charge, the Debye screen length should be generalized to 
\beq
\lambda_\text{D} = \sqrt{ \frac{1}{e^2 (n_e/T_e +n_p/T_p)}} = \sqrt{ \frac{T_b}{2 e^2  x_e n_b}} \ ,
\eeq 
where $n_p$, $T_p$, $n_b$, $T_b$, $x_e = n_e/n_b$ are the proton number density, proton temperature, baryon number density, baryon temperature, and ionization fraction, respectively.  We take $n_e = n_p = x_e n_b$ and $T_e = T_p = T_b$ in the second equality. For Coulomb-like interactions, the maximal impact parameter can be translated into a minimal scattering angle, 
\beq\label{eq:theta_min}
\theta_\text{min} = 2 \arctan \frac{\epsilon e g_\chi }{4\pi \mu_{\chi b} v_\text{rel}^2 \lambda_\text{D}} \approx \frac{\epsilon  e g_\chi}{6 \pi T_b \lambda_\text{D}} \ ,
\eeq
where we assume $\theta_{\min} \ll 1$ and apply the equipartition relation $\mu_{\chi B} v_\text{rel}^2/2 \approx 3 T_b/2$ in the second step. Note that this cutoff contains a coupling dependence on $\epsilon \alpha$. Given the regularization, we can write the momentum-transfer cross section in terms of the scattering angle cutoff from \Eq{eq:theta_min},\footnote{The formula can be applied to millicharge dark matter by setting $\alpha_\chi=\alpha$. It is similar to those used in Refs.~\cite{Kovetz:2018zan, Xu:2018efh} except for a factor of 2 difference in the denominator of the logarithmic factor, taking into account photon scattering off both electrons and protons.} 
\beq
\begin{aligned}
\sigma_{\rm T}^{\chi B} &= \frac{2\pi \epsilon^2 \alpha \alpha_\chi}{\mu_{\chi B}^2 v_\text{rel}^4 }\ln \left(\csc^2 \frac{\theta_\text{min}}{2}\right) \approx \frac{2\pi \epsilon^2 \alpha \alpha_\chi}{\mu_{\chi B}^2 v_\text{rel}^4} \ln \left(\frac{9 T_b^3}{8\pi \epsilon^2 \alpha^2 \alpha_\chi x_e n_b}\right).
\end{aligned},
\eeq
\ie, \Eq{eq:ultralight2}.

\subsubsection*{\textit{ii.} Debye mass screening} 
The prescription regulates the Coulomb-like interaction at the level of the matrix element. The spin-averaged amplitude squared for the dark matter--baryon interaction, including the in-medium effects , becomes~\cite{Chang:2018rso}
\beq
\overline{|\mathcal M|^2} = \frac{\epsilon^2 e^2 g_\chi^2}{4} \left|\frac{q^2}{q^2 -m_{A'}^2+i m_{A'} \Gamma_{A'}+\Pi_D} \sum_{\rm spin} J_{B,\mu} D^{\mu\nu} (q) J_{\chi,\nu} \right|^2 \ ,
\label{eq:screen}
\eeq
with the Standard Model baryonic current $J_{B,\mu} = \bar{B} \gamma_\mu B$, the dark current $J_\chi =  \bar \chi \gamma_\mu \chi$, and where $q=(\omega_q, \vec q)$ is the 4-vector of the momentum transfer. In addition, $m_A'$, $\Gamma_{A'}$, and $\Pi_D$ are the mass, decay width, and self-energy of the dark photon, respectively, which we assume are all negligibly small, and $D^{\mu\nu}$ is the effective in-medium propagator. We take the effective propagator from~\cite{Braaten:1993jw}, which is given under the Coulomb gauge $\nabla \cdot \vec A = 0$.  The non-zero components of $D^{\mu \nu}$ are
\beq
D^{00}(\omega, \vec k) =\frac{1}{|\vec k|^2 - \Pi_L(\omega, |\vec k|)}, \quad D^{ij}(\omega, \vec k) =\frac{1}{\omega^2-|\vec{k}|^2-\Pi_T(\omega, |\vec k|)}\left(\delta^{ij}- \frac{k^i  k^j}{|\vec k|^2}\right) \ ,
\eeq
where $i,j=1,2,3$ and $\Pi_{L,T}$ are the transverse and longitudinal projections of the photon polarization tensor $\Pi^{\mu\nu}(\omega, \vec k)$, due to the Compton scattering of in-medium photons with charged particles~\cite{Braaten:1993jw}. They are given by
\begin{align}
\Pi_T (\omega, |\vec k|) ={}& \frac{4\alpha}{\pi} \int_0^\infty \dd |\vec p| \frac{|\vec p|^2}{E} \left[\frac{\omega^2}{|\vec k|^2}-\frac{\omega(\omega^2 -|\vec k|^2)}{2v |\vec k|^3}\ln\left(\frac{\omega +v |\vec k|}{\omega-v|\vec k|}\right)\right] f_E \ ,\label{eq:trans}\\
\Pi_L (\omega, |\vec k|) ={}&\frac{4\alpha}{\pi} \int_0^\infty \dd |\vec p| \frac{|\vec p|^2}{E} \left[-1 -\frac{\omega^2-|\vec k|^2}{\omega^2-v^2|\vec k|^2}+ \frac{\omega}{v |\vec k|}\ln \left(\frac{\omega + v |\vec k|}{\omega-v|\vec k|}\right)\right] f_E \ , \label{eq:long}
\end{align}
where $p=(E, \vec p)$ is the 4-momentum of the charged particle with mass $m$, and $v=|\vec p|/E$ is the magnitude of its 3-velocity, $f_E =\frac{1}{e^{(E-\mu)/T}+1}+\frac{1}{e^{(E+\mu)/ T}+1}$ is the sum of the distributions of charged particles and anti-particles, where $\mu$ is the chemical potential. The baryonic fluid we consider is in the classical limit, \ie, $ T \ll m$ and $m-\mu \gg  T$, under which the anti-particle contribution to $f_E$ can be ignored and the charged particle distribution is approximated by $f_E \approx e^{\frac{\mu-E}{ T}}\approx e^{\frac{\mu-m}{ T}} e^{-\frac{|\vec p|^2}{2  m T}}$. The number density of the charged particles in this limit is therefore given by
\beq
n_c \approx \frac{1}{\pi^2} \int_0^\infty \dd |\vec p| |\vec p|^2 f_E = e^{\frac{\mu-m}{ T}}\frac{(m  T)^{3/2}}{2^{1/2} \pi^{3/2}} \ .
\eeq

For the Coulomb-like interaction of the dark matter  and the classical baryonic fluid, we take the static limit for the four-momentum transfer~\cite{Raffelt:1996wa, Chu:2019rok}, $\omega_q=0$. In the static limit, the transverse polarization of the in-medium photon (\Eq{eq:trans}) vanishes,
\beq
\Pi_T(0, |\vec q|) = 0 \ ,
\label{eq:static1}
\eeq
while its longitudinal polarization (\Eq{eq:long}) approaches the Debye mass squared,
\beq
\begin{aligned}
 \Pi_L (0, |\vec q|) &= -\frac{4\alpha}{\pi} \int_0^\infty \dd |\vec p| |\vec p| ( v + v^{-1}) f_E 
 \\
 &\approx -2 \alpha e^{\frac{\mu-m}{ T}} \sqrt{\frac{2m^3  T}{\pi}} = -\frac{e^2 n_c}{ T}\equiv -m^2_\text{D} \ .
\label{eq:static2}
\end{aligned}
\eeq

From \Eqs{eq:static1}{eq:static2}, only the longitudinal polarization is thermally screened, while there is no screening for the transverse mode. The above derivation for the polarization modes focus on a single charged species. For the baryonic fluid we consider, both electrons and protons can Compton scatter with the in-medium photons and their number densities are equal. Hence $\Pi^{\mu\nu} = \Pi^{\mu\nu}_e + \Pi^{\mu\nu}_p$ and \Eq{eq:static2} should be generalized as
\beq
\Pi_L (0, |\vec q|) = \Pi_{L,e} (0, |\vec q|) + \Pi_{L,p} (0, |\vec q|) \approx - \frac{e^2 (n_e + n_p)}{ T} = - \frac{2 e^2 x_e n_b}{T_b} \equiv - m^2_\text{D} \ ,
\label{eq:static3}
\eeq
where in the last step we have generalized the definition of the Debye mass to include both electron and proton contributions, and taken $T = T_e = T_p = T_b$. Numerically, the in-medium photon mass is about $m_\text{D}\approx 1.3\,\mu\text{eV} \left(\frac{1+z}{1100}\right)$ prior to recombination. Substituting \Eqs{eq:static1}{eq:static3} in \Eq{eq:screen} yields 
\begin{align}
\label{eq:regu}
&\overline{|\mathcal M|^2} \approx{} \frac{\epsilon^2 e^2 g_\chi^2}{4} \left|\sum_{\rm spin} \left[ \frac{J_{B, 0} J_{\chi,0}}{|\vec q|^2 + m_\text{D}^2} + \frac{J_{B,i} J_{\chi, j}}{-|\vec q|^2}\left(\delta^{ij} - \hat q^i \hat q^j \right)  \right]\right|^2 \\
=&\frac{\epsilon^2 e^2 g_\chi^2}{4} \left|\sum_{\rm spin} \left[ \frac{J_{B, 0} J_{\chi,0}}{2p_\text{CM}^2(1-\cos \theta_\text{CM}) + m_\text{D}^2} + \frac{J_{B,i} J_{\chi, j}}{- 2 p^2_\text{CM}(1-\cos \theta_\text{CM})}\left(\delta^{ij} - \hat q^i \hat q^j \right)  \right]\right|^2 \ , \nonumber
\end{align}
where $\hat q^i \equiv q^i/|\vec q|$. The Debye mass regulates the forward scattering singularity in the denominator of the term $|\sum_\text{spin} J_{B,0} J_{\chi,0}|^2$ of the matrix element, as well as that of the cross terms, $\text{Re}[\sum_\text{spin}J_{B,0} J_{\chi,0} J^*_{B,i} J^*_{\chi,j}]$. However, the forward scattering singularity in the term $|\sum_\text{spin} J_{B,i} J_{\chi,j}|^2$ remains unchanged. This divergence might be cured if we go beyond the static limit $\omega_q=0$; we leave a detailed investigation for the future. Nevertheless, this uncertainty in the regulation may turn to be unimportant. The numerator of the term  $J_{B,0} J_{\chi,0}$ is of order $m_\chi m_B$, while that of $J_{B,i} J_{\chi,j}$ is of order $p^2_\text{CM}$. Given $p_\text{CM}^2 \ll m_\chi m_B$, the term $|\sum_\text{spin} J_{B,0} J_{\chi,0}|^2$  dominates over the cross term. Furthermore, since $p_\text{CM}^4 \ll m_\chi^2 m_B^2$, we expect it to dominate also over the $|\sum_\text{spin} J_{B,i} J_{\chi,j}|^2$ term, despite the latter terms possibly containing a larger logarithmic factor after regulation.\footnote{To be more concrete, if the transverse polarization is regulated by the Debye screening length, it receives a logarithmic factor  like that in~\Eq{eq:ultralight}, which is $\mathcal{O}(10)$ for the parameter space we consider.} 

To summarize, in the non-relativistic limit, the averaged amplitude squared can be approximated by the term $|J_{B,0} J_{\chi,0}|^2$,
\beq
\overline{|\mathcal M|^2} \approx \frac{16 \epsilon^2 e^2 g_\chi^2  m_\chi^2 m_B^2}{\left[2p_\text{CM}^2(1-\cos \theta_\text{CM}) + m_\text{D}^2\right]^2},
\eeq
\ie\!, \Eq{eq:ultralight2me}, and the corresponding momentum-transfer cross section is given by~\Eq{eq:ultralight2}.

\bibliography{ref}

\providecommand{\href}[2]{#2}\begingroup\raggedright\begin{thebibliography}{100}

\bibitem{McDermott:2010pa}
S.~D. McDermott, H.-B. Yu and K.~M. Zurek, \emph{{Turning off the Lights: How
  Dark is Dark Matter?}},
  \href{https://doi.org/10.1103/PhysRevD.83.063509}{\emph{Phys. Rev. D}
  {\bfseries 83} (2011) 063509},
  [\href{https://arxiv.org/abs/1011.2907}{{\ttfamily 1011.2907}}].

\bibitem{Dvorkin:2013cea}
C.~Dvorkin, K.~Blum and M.~Kamionkowski, \emph{{Constraining Dark Matter-Baryon
  Scattering with Linear Cosmology}},
  \href{https://doi.org/10.1103/PhysRevD.89.023519}{\emph{Phys. Rev. D}
  {\bfseries 89} (2014) 023519},
  [\href{https://arxiv.org/abs/1311.2937}{{\ttfamily 1311.2937}}].

\bibitem{Munoz:2015bca}
J.~B. Mu\~noz, E.~D. Kovetz and Y.~Ali-Ha\"\i{}moud, \emph{{Heating of Baryons
  due to Scattering with Dark Matter During the Dark Ages}},
  \href{https://doi.org/10.1103/PhysRevD.92.083528}{\emph{Phys. Rev. D}
  {\bfseries 92} (2015) 083528},
  [\href{https://arxiv.org/abs/1509.00029}{{\ttfamily 1509.00029}}].

\bibitem{Gluscevic:2017ywp}
V.~Gluscevic and K.~K. Boddy, \emph{{Constraints on Scattering of
  keV\textendash{}TeV Dark Matter with Protons in the Early Universe}},
  \href{https://doi.org/10.1103/PhysRevLett.121.081301}{\emph{Phys. Rev. Lett.}
  {\bfseries 121} (2018) 081301},
  [\href{https://arxiv.org/abs/1712.07133}{{\ttfamily 1712.07133}}].

\bibitem{Xu:2018efh}
W.~L. Xu, C.~Dvorkin and A.~Chael, \emph{{Probing sub-GeV Dark Matter-Baryon
  Scattering with Cosmological Observables}},
  \href{https://doi.org/10.1103/PhysRevD.97.103530}{\emph{Phys. Rev. D}
  {\bfseries 97} (2018) 103530},
  [\href{https://arxiv.org/abs/1802.06788}{{\ttfamily 1802.06788}}].

\bibitem{Slatyer:2018aqg}
T.~R. Slatyer and C.-L. Wu, \emph{{Early-Universe constraints on dark
  matter-baryon scattering and their implications for a global 21 cm signal}},
  \href{https://doi.org/10.1103/PhysRevD.98.023013}{\emph{Phys. Rev. D}
  {\bfseries 98} (2018) 023013},
  [\href{https://arxiv.org/abs/1803.09734}{{\ttfamily 1803.09734}}].

\bibitem{Boddy:2018wzy}
K.~K. Boddy, V.~Gluscevic, V.~Poulin, E.~D. Kovetz, M.~Kamionkowski and
  R.~Barkana, \emph{{Critical assessment of CMB limits on dark matter-baryon
  scattering: New treatment of the relative bulk velocity}},
  \href{https://doi.org/10.1103/PhysRevD.98.123506}{\emph{Phys. Rev. D}
  {\bfseries 98} (2018) 123506},
  [\href{https://arxiv.org/abs/1808.00001}{{\ttfamily 1808.00001}}].

\bibitem{Nadler:2019zrb}
E.~O. Nadler, V.~Gluscevic, K.~K. Boddy and R.~H. Wechsler, \emph{{Constraints
  on Dark Matter Microphysics from the Milky Way Satellite Population}},
  \href{https://doi.org/10.3847/2041-8213/ab1eb2}{\emph{Astrophys. J. Lett.}
  {\bfseries 878} (2019) 32},
  [\href{https://arxiv.org/abs/1904.10000}{{\ttfamily 1904.10000}}].

\bibitem{Maamari:2020aqz}
K.~Maamari, V.~Gluscevic, K.~K. Boddy, E.~O. Nadler and R.~H. Wechsler,
  \emph{{Bounds on velocity-dependent dark matter-proton scattering from Milky
  Way satellite abundance}},
  \href{https://doi.org/10.3847/2041-8213/abd807}{\emph{Astrophys. J. Lett.}
  {\bfseries 907} (2021) L46},
  [\href{https://arxiv.org/abs/2010.02936}{{\ttfamily 2010.02936}}].

\bibitem{Becker:2020hzj}
N.~Becker, D.~C. Hooper, F.~Kahlhoefer, J.~Lesgourgues and N.~Sch\"oneberg,
  \emph{{Cosmological constraints on multi-interacting dark matter}},
  \href{https://arxiv.org/abs/2010.04074}{{\ttfamily 2010.04074}}.

\bibitem{Wadekar:2019xnf}
D.~Wadekar and G.~R. Farrar, \emph{{First astrophysical constraints on dark
  matter interactions with ordinary matter at low relative velocity}},
  \href{https://arxiv.org/abs/1903.12190}{{\ttfamily 1903.12190}}.

\bibitem{Ali-Haimoud:2021lka}
Y.~Ali-Ha\"\i{}moud, \emph{{Testing dark matter interactions with CMB spectral
  distortions}}, \href{https://doi.org/10.1103/PhysRevD.103.043541}{\emph{Phys.
  Rev. D} {\bfseries 103} (2021) 043541},
  [\href{https://arxiv.org/abs/2101.04070}{{\ttfamily 2101.04070}}].

\bibitem{Nguyen:2021cnb}
D.~Nguyen, D.~Sarnaaik, K.~K. Boddy, E.~O. Nadler and V.~Gluscevic,
  \emph{{Observational constraints on dark matter scattering with electrons}},
  \href{https://arxiv.org/abs/2107.12380}{{\ttfamily 2107.12380}}.

\bibitem{dePutter:2018xte}
R.~de~Putter, O.~Dor\'e, J.~Gleyzes, D.~Green and J.~Meyers, \emph{{Dark Matter
  Interactions, Helium, and the Cosmic Microwave Background}},
  \href{https://doi.org/10.1103/PhysRevLett.122.041301}{\emph{Phys. Rev. Lett.}
  {\bfseries 122} (2019) 041301},
  [\href{https://arxiv.org/abs/1805.11616}{{\ttfamily 1805.11616}}].

\bibitem{Chen:2002yh}
X.-l. Chen, S.~Hannestad and R.~J. Scherrer, \emph{{Cosmic microwave background
  and large scale structure limits on the interaction between dark matter and
  baryons}}, \href{https://doi.org/10.1103/PhysRevD.65.123515}{\emph{Phys. Rev.
  D} {\bfseries 65} (2002) 123515},
  [\href{https://arxiv.org/abs/astro-ph/0202496}{{\ttfamily
  astro-ph/0202496}}].

\bibitem{Sigurdson:2004zp}
K.~Sigurdson, M.~Doran, A.~Kurylov, R.~R. Caldwell and M.~Kamionkowski,
  \emph{{Dark-matter electric and magnetic dipole moments}},
  \href{https://doi.org/10.1103/PhysRevD.70.083501}{\emph{Phys. Rev. D}
  {\bfseries 70} (2004) 083501},
  [\href{https://arxiv.org/abs/astro-ph/0406355}{{\ttfamily
  astro-ph/0406355}}].

\bibitem{Melchiorri:2007sq}
A.~Melchiorri, A.~Polosa and A.~Strumia, \emph{{New bounds on millicharged
  particles from cosmology}},
  \href{https://doi.org/10.1016/j.physletb.2007.05.042}{\emph{Phys. Lett. B}
  {\bfseries 650} (2007) 416--420},
  [\href{https://arxiv.org/abs/hep-ph/0703144}{{\ttfamily hep-ph/0703144}}].

\bibitem{Ma:1995ey}
C.-P. Ma and E.~Bertschinger, \emph{{Cosmological perturbation theory in the
  synchronous and conformal Newtonian gauges}},
  \href{https://doi.org/10.1086/176550}{\emph{Astrophys. J.} {\bfseries 455}
  (1995) 7--25}, [\href{https://arxiv.org/abs/astro-ph/9506072}{{\ttfamily
  astro-ph/9506072}}].

\bibitem{Shen:2021frv}
X.~Shen, P.~F. Hopkins, L.~Necib, F.~Jiang, M.~Boylan-Kolchin and A.~Wetzel,
  \emph{{Dissipative Dark Matter on FIRE: I. Structural and kinematic
  properties of dwarf galaxies}},
  \href{https://arxiv.org/abs/2102.09580}{{\ttfamily 2102.09580}}.

\bibitem{Ali-Haimoud:2018dvo}
Y.~Ali-Ha\"\i{}moud, \emph{{Boltzmann-Fokker-Planck formalism for
  dark-matter--baryon scattering}},
  \href{https://doi.org/10.1103/PhysRevD.99.023523}{\emph{Phys. Rev. D}
  {\bfseries 99} (2019) 023523},
  [\href{https://arxiv.org/abs/1811.09903}{{\ttfamily 1811.09903}}].

\bibitem{Dvorkin:2019zdi}
C.~Dvorkin, T.~Lin and K.~Schutz, \emph{{Making dark matter out of light:
  freeze-in from plasma effects}},
  \href{https://doi.org/10.1103/PhysRevD.99.115009}{\emph{Phys. Rev. D}
  {\bfseries 99} (2019) 115009},
  [\href{https://arxiv.org/abs/1902.08623}{{\ttfamily 1902.08623}}].

\bibitem{Dvorkin:2020xga}
C.~Dvorkin, T.~Lin and K.~Schutz, \emph{{The cosmology of sub-MeV dark matter
  freeze-in}},  \href{https://arxiv.org/abs/2011.08186}{{\ttfamily
  2011.08186}}.

\bibitem{Tulin:2017ara}
S.~Tulin and H.-B. Yu, \emph{{Dark Matter Self-interactions and Small Scale
  Structure}}, \href{https://doi.org/10.1016/j.physrep.2017.11.004}{\emph{Phys.
  Rept.} {\bfseries 730} (2018) 1--57},
  [\href{https://arxiv.org/abs/1705.02358}{{\ttfamily 1705.02358}}].

\bibitem{Aghanim:2018eyx}
{\scshape Planck} collaboration, N.~Aghanim et~al., \emph{{Planck 2018 results.
  VI. Cosmological parameters}},
  \href{https://doi.org/10.1051/0004-6361/201833910}{\emph{Astron. Astrophys.}
  {\bfseries 641} (2020) A6},
  [\href{https://arxiv.org/abs/1807.06209}{{\ttfamily 1807.06209}}].

\bibitem{Cyr-Racine:2015ihg}
F.-Y. Cyr-Racine, K.~Sigurdson, J.~Zavala, T.~Bringmann, M.~Vogelsberger and
  C.~Pfrommer, \emph{{ETHOS\textemdash{}an effective theory of structure
  formation: From dark particle physics to the matter distribution of the
  Universe}}, \href{https://doi.org/10.1103/PhysRevD.93.123527}{\emph{Phys.
  Rev. D} {\bfseries 93} (2016) 123527},
  [\href{https://arxiv.org/abs/1512.05344}{{\ttfamily 1512.05344}}].

\bibitem{1953ApJ...117..134L}
D.~N. {Limber}, \emph{{The Analysis of Counts of the Extragalactic Nebulae in
  Terms of a Fluctuating Density Field.}},
  \href{https://doi.org/10.1086/145672}{\emph{Astrophys. J.} {\bfseries 117}
  (Jan., 1953) 134}.

\bibitem{Limber:1954zz}
D.~N. Limber, \emph{{The Analysis of Counts of the Extragalactic Nebulae in
  Terms of a Fluctuating Density Field. II}},
  \href{https://doi.org/10.1086/145870}{\emph{Astrophys. J.} {\bfseries 119}
  (1954) 655}.

\bibitem{Pan:2014xua}
Z.~Pan, L.~Knox and M.~White, \emph{{Dependence of the Cosmic Microwave
  Background Lensing Power Spectrum on the Matter Density}},
  \href{https://doi.org/10.1093/mnras/stu1971}{\emph{Mon. Not. Roy. Astron.
  Soc.} {\bfseries 445} (2014) 2941--2945},
  [\href{https://arxiv.org/abs/1406.5459}{{\ttfamily 1406.5459}}].

\bibitem{Buen-Abad:2017gxg}
M.~A. Buen-Abad, M.~Schmaltz, J.~Lesgourgues and T.~Brinckmann,
  \emph{{Interacting Dark Sector and Precision Cosmology}},
  \href{https://doi.org/10.1088/1475-7516/2018/01/008}{\emph{JCAP} {\bfseries
  01} (2018) 008}, [\href{https://arxiv.org/abs/1708.09406}{{\ttfamily
  1708.09406}}].

\bibitem{Zaldarriaga:1996xe}
M.~Zaldarriaga and U.~Seljak, \emph{{An all sky analysis of polarization in the
  microwave background}},
  \href{https://doi.org/10.1103/PhysRevD.55.1830}{\emph{Phys. Rev. D}
  {\bfseries 55} (1997) 1830--1840},
  [\href{https://arxiv.org/abs/astro-ph/9609170}{{\ttfamily
  astro-ph/9609170}}].

\bibitem{Lesgourgues:2013qba}
J.~Lesgourgues, \emph{{Cosmological Perturbations}},  in \emph{{Theoretical
  Advanced Study Institute in Elementary Particle Physics}: {Searching for New
  Physics at Small and Large Scales}}, 2, 2013,
  \href{https://arxiv.org/abs/1302.4640}{{\ttfamily 1302.4640}},
  \href{https://doi.org/10.1142/9789814525220_0002}{DOI}.

\bibitem{Blas:2011rf}
D.~Blas, J.~Lesgourgues and T.~Tram, \emph{{The Cosmic Linear Anisotropy
  Solving System (CLASS) II: Approximation schemes}},
  \href{https://doi.org/10.1088/1475-7516/2011/07/034}{\emph{JCAP} {\bfseries
  07} (2011) 034}, [\href{https://arxiv.org/abs/1104.2933}{{\ttfamily
  1104.2933}}].

\bibitem{Seager:1999bc}
S.~Seager, D.~D. Sasselov and D.~Scott, \emph{{A new calculation of the
  recombination epoch}}, \href{https://doi.org/10.1086/312250}{\emph{Astrophys.
  J. Lett.} {\bfseries 523} (1999) L1--L5},
  [\href{https://arxiv.org/abs/astro-ph/9909275}{{\ttfamily
  astro-ph/9909275}}].

\bibitem{Beutler:2011hx}
F.~Beutler, C.~Blake, M.~Colless, D.~Jones, L.~Staveley-Smith, L.~Campbell
  et~al., \emph{{The 6dF Galaxy Survey: Baryon Acoustic Oscillations and the
  Local Hubble Constant}},
  \href{https://doi.org/10.1111/j.1365-2966.2011.19250.x}{\emph{Mon. Not. Roy.
  Astron. Soc.} {\bfseries 416} (2011) 3017--3032},
  [\href{https://arxiv.org/abs/1106.3366}{{\ttfamily 1106.3366}}].

\bibitem{Ross:2014qpa}
A.~J. Ross, L.~Samushia, C.~Howlett, W.~J. Percival, A.~Burden and M.~Manera,
  \emph{{The clustering of the SDSS DR7 main Galaxy sample \textendash I. A
  4~per cent distance measure at $z~=~0.15$}},
  \href{https://doi.org/10.1093/mnras/stv154}{\emph{Mon. Not. Roy. Astron.
  Soc.} {\bfseries 449} (2015) 835--847},
  [\href{https://arxiv.org/abs/1409.3242}{{\ttfamily 1409.3242}}].

\bibitem{Alam:2016hwk}
{\scshape BOSS} collaboration, S.~Alam et~al., \emph{{The clustering of
  galaxies in the completed SDSS-III Baryon Oscillation Spectroscopic Survey:
  cosmological analysis of the DR12 galaxy sample}},
  \href{https://doi.org/10.1093/mnras/stx721}{\emph{Mon. Not. Roy. Astron.
  Soc.} {\bfseries 470} (2017) 2617--2652},
  [\href{https://arxiv.org/abs/1607.03155}{{\ttfamily 1607.03155}}].

\bibitem{Brinckmann:2018cvx}
T.~Brinckmann and J.~Lesgourgues, \emph{{MontePython 3: boosted MCMC sampler
  and other features}},  \href{https://arxiv.org/abs/1804.07261}{{\ttfamily
  1804.07261}}.

\bibitem{Audren:2012wb}
B.~Audren, J.~Lesgourgues, K.~Benabed and S.~Prunet, \emph{{Conservative
  Constraints on Early Cosmology: an illustration of the Monte Python
  cosmological parameter inference code}},
  \href{https://doi.org/10.1088/1475-7516/2013/02/001}{\emph{JCAP} {\bfseries
  1302} (2013) 001}, [\href{https://arxiv.org/abs/1210.7183}{{\ttfamily
  1210.7183}}].

\bibitem{10.2307/2246093}
A.~Gelman and D.~B. Rubin, \emph{Inference from iterative simulation using
  multiple sequences}, {\emph{Statistical Science} {\bfseries 7} (1992)
  457--472}.

\bibitem{Ade:2015xua}
{\scshape Planck} collaboration, P.~A.~R. Ade et~al., \emph{{Planck 2015
  results. XIII. Cosmological parameters}},
  \href{https://doi.org/10.1051/0004-6361/201525830}{\emph{Astron. Astrophys.}
  {\bfseries 594} (2016) A13},
  [\href{https://arxiv.org/abs/1502.01589}{{\ttfamily 1502.01589}}].

\bibitem{McQuinn:2015icp}
M.~McQuinn, \emph{{The Evolution of the Intergalactic Medium}},
  \href{https://doi.org/10.1146/annurev-astro-082214-122355}{\emph{Ann. Rev.
  Astron. Astrophys.} {\bfseries 54} (2016) 313--362},
  [\href{https://arxiv.org/abs/1512.00086}{{\ttfamily 1512.00086}}].

\bibitem{Viel:2013fqw}
M.~Viel, G.~D. Becker, J.~S. Bolton and M.~G. Haehnelt, \emph{{Warm dark matter
  as a solution to the small scale crisis: New constraints from high redshift
  Lyman-\ensuremath{\alpha} forest data}},
  \href{https://doi.org/10.1103/PhysRevD.88.043502}{\emph{Phys. Rev. D}
  {\bfseries 88} (2013) 043502},
  [\href{https://arxiv.org/abs/1306.2314}{{\ttfamily 1306.2314}}].

\bibitem{Irsic:2017ixq}
V.~Ir\v{s}i\v{c} et~al., \emph{{New Constraints on the free-streaming of warm
  dark matter from intermediate and small scale Lyman-$\alpha$ forest data}},
  \href{https://doi.org/10.1103/PhysRevD.96.023522}{\emph{Phys. Rev. D}
  {\bfseries 96} (2017) 023522},
  [\href{https://arxiv.org/abs/1702.01764}{{\ttfamily 1702.01764}}].

\bibitem{Murgia:2017lwo}
R.~Murgia, A.~Merle, M.~Viel, M.~Totzauer and A.~Schneider, \emph{{''Non-cold''
  dark matter at small scales: a general approach}},
  \href{https://doi.org/10.1088/1475-7516/2017/11/046}{\emph{JCAP} {\bfseries
  11} (2017) 046}, [\href{https://arxiv.org/abs/1704.07838}{{\ttfamily
  1704.07838}}].

\bibitem{Murgia:2017cvj}
R.~Murgia, \emph{{A general approach for testing non-cold dark matter at small
  cosmological scales}},
  \href{https://doi.org/10.1088/1742-6596/956/1/012005}{\emph{J. Phys. Conf.
  Ser.} {\bfseries 956} (2018) 012005},
  [\href{https://arxiv.org/abs/1712.04810}{{\ttfamily 1712.04810}}].

\bibitem{Murgia:2018now}
R.~Murgia, V.~Ir\v{s}i\v{c} and M.~Viel, \emph{{Novel constraints on noncold,
  nonthermal dark matter from Lyman- \ensuremath{\alpha} forest data}},
  \href{https://doi.org/10.1103/PhysRevD.98.083540}{\emph{Phys. Rev. D}
  {\bfseries 98} (2018) 083540},
  [\href{https://arxiv.org/abs/1806.08371}{{\ttfamily 1806.08371}}].

\bibitem{Viel:2005qj}
M.~Viel, J.~Lesgourgues, M.~G. Haehnelt, S.~Matarrese and A.~Riotto,
  \emph{{Constraining warm dark matter candidates including sterile neutrinos
  and light gravitinos with WMAP and the Lyman-alpha forest}},
  \href{https://doi.org/10.1103/PhysRevD.71.063534}{\emph{Phys. Rev. D}
  {\bfseries 71} (2005) 063534},
  [\href{https://arxiv.org/abs/astro-ph/0501562}{{\ttfamily
  astro-ph/0501562}}].

\bibitem{DEramo:2020gpr}
F.~D'Eramo and A.~Lenoci, \emph{{Lower Mass Bounds on FIMPs}},
  \href{https://arxiv.org/abs/2012.01446}{{\ttfamily 2012.01446}}.

\bibitem{Egana-Ugrinovic:2021gnu}
D.~Egana-Ugrinovic, R.~Essig, D.~Gift and M.~LoVerde, \emph{{The Cosmological
  Evolution of Self-interacting Dark Matter}},
  \href{https://doi.org/10.1088/1475-7516/2021/05/013}{\emph{JCAP} {\bfseries
  05} (2021) 013}, [\href{https://arxiv.org/abs/2102.06215}{{\ttfamily
  2102.06215}}].

\bibitem{Palanque-Delabrouille:2019iyz}
N.~Palanque-Delabrouille, C.~Y\`eche, N.~Sch\"oneberg, J.~Lesgourgues,
  M.~Walther, S.~Chabanier et~al., \emph{{Hints, neutrino bounds and WDM
  constraints from SDSS DR14 Lyman-$\alpha$ and Planck full-survey data}},
  \href{https://doi.org/10.1088/1475-7516/2020/04/038}{\emph{JCAP} {\bfseries
  04} (2020) 038}, [\href{https://arxiv.org/abs/1911.09073}{{\ttfamily
  1911.09073}}].

\bibitem{Chabanier:2018rga}
S.~Chabanier et~al., \emph{{The one-dimensional power spectrum from the SDSS
  DR14 Ly$\alpha$ forests}},
  \href{https://doi.org/10.1088/1475-7516/2019/07/017}{\emph{JCAP} {\bfseries
  07} (2019) 017}, [\href{https://arxiv.org/abs/1812.03554}{{\ttfamily
  1812.03554}}].

\bibitem{Dawson2012}
K.~S. Dawson, D.~J. Schlegel, C.~P. Ahn, S.~F. Anderson, E.~Aubourg, S.~Bailey
  et~al., \emph{The baryon oscillation spectroscopic survey of sdss-iii},
  \href{https://doi.org/10.1088/0004-6256/145/1/10}{\emph{The Astronomical
  Journal} {\bfseries 145} (Dec, 2012) 10}.

\bibitem{Dawson:2015wdb}
K.~S. Dawson et~al., \emph{{The SDSS-IV extended Baryon Oscillation
  Spectroscopic Survey: Overview and Early Data}},
  \href{https://doi.org/10.3847/0004-6256/151/2/44}{\emph{Astron. J.}
  {\bfseries 151} (2016) 44},
  [\href{https://arxiv.org/abs/1508.04473}{{\ttfamily 1508.04473}}].

\bibitem{Archidiacono:2019wdp}
M.~Archidiacono, D.~C. Hooper, R.~Murgia, S.~Bohr, J.~Lesgourgues and M.~Viel,
  \emph{{Constraining Dark Matter-Dark Radiation interactions with CMB, BAO,
  and Lyman-$\alpha$}},
  \href{https://doi.org/10.1088/1475-7516/2019/10/055}{\emph{JCAP} {\bfseries
  10} (2019) 055}, [\href{https://arxiv.org/abs/1907.01496}{{\ttfamily
  1907.01496}}].

\bibitem{McDonald:2004eu}
{\scshape SDSS} collaboration, P.~McDonald et~al., \emph{{The Lyman-alpha
  forest power spectrum from the Sloan Digital Sky Survey}},
  \href{https://doi.org/10.1086/444361}{\emph{Astrophys. J. Suppl.} {\bfseries
  163} (2006) 80--109},
  [\href{https://arxiv.org/abs/astro-ph/0405013}{{\ttfamily
  astro-ph/0405013}}].

\bibitem{McDonald:2004xn}
{\scshape SDSS} collaboration, P.~McDonald et~al., \emph{{The Linear theory
  power spectrum from the Lyman-alpha forest in the Sloan Digital Sky Survey}},
  \href{https://doi.org/10.1086/497563}{\emph{Astrophys. J.} {\bfseries 635}
  (2005) 761--783}, [\href{https://arxiv.org/abs/astro-ph/0407377}{{\ttfamily
  astro-ph/0407377}}].

\bibitem{Hui:2016ltb}
L.~Hui, J.~P. Ostriker, S.~Tremaine and E.~Witten, \emph{{Ultralight scalars as
  cosmological dark matter}},
  \href{https://doi.org/10.1103/PhysRevD.95.043541}{\emph{Phys. Rev. D}
  {\bfseries 95} (2017) 043541},
  [\href{https://arxiv.org/abs/1610.08297}{{\ttfamily 1610.08297}}].

\bibitem{Munoz:2017qpy}
J.~B. Mu\~noz and A.~Loeb, \emph{{Constraints on Dark Matter-Baryon Scattering
  from the Temperature Evolution of the Intergalactic Medium}},
  \href{https://doi.org/10.1088/1475-7516/2017/11/043}{\emph{JCAP} {\bfseries
  11} (2017) 043}, [\href{https://arxiv.org/abs/1708.08923}{{\ttfamily
  1708.08923}}].

\bibitem{Bode:2000gq}
P.~Bode, J.~P. Ostriker and N.~Turok, \emph{{Halo formation in warm dark matter
  models}}, \href{https://doi.org/10.1086/321541}{\emph{Astrophys. J.}
  {\bfseries 556} (2001) 93--107},
  [\href{https://arxiv.org/abs/astro-ph/0010389}{{\ttfamily
  astro-ph/0010389}}].

\bibitem{Schneider:2011yu}
A.~Schneider, R.~E. Smith, A.~V. Maccio and B.~Moore, \emph{{Nonlinear
  Evolution of Cosmological Structures in Warm Dark Matter Models}},
  \href{https://doi.org/10.1111/j.1365-2966.2012.21252.x}{\emph{Mon. Not. Roy.
  Astron. Soc.} {\bfseries 424} (2012) 684},
  [\href{https://arxiv.org/abs/1112.0330}{{\ttfamily 1112.0330}}].

\bibitem{Nadler:2020prv}
{\scshape DES} collaboration, E.~O. Nadler et~al., \emph{{Milky Way Satellite
  Census. III. Constraints on Dark Matter Properties from Observations of Milky
  Way Satellite Galaxies}},
  \href{https://doi.org/10.1103/PhysRevLett.126.091101}{\emph{Phys. Rev. Lett.}
  {\bfseries 126} (2021) 091101},
  [\href{https://arxiv.org/abs/2008.00022}{{\ttfamily 2008.00022}}].

\bibitem{Escudero:2018thh}
M.~Escudero, L.~Lopez-Honorez, O.~Mena, S.~Palomares-Ruiz and
  P.~Villanueva-Domingo, \emph{{A fresh look into the interacting dark matter
  scenario}}, \href{https://doi.org/10.1088/1475-7516/2018/06/007}{\emph{JCAP}
  {\bfseries 06} (2018) 007},
  [\href{https://arxiv.org/abs/1803.08427}{{\ttfamily 1803.08427}}].

\bibitem{Newton:2020cog}
O.~Newton, M.~Leo, M.~Cautun, A.~Jenkins, C.~S. Frenk, M.~R. Lovell et~al.,
  \emph{{Constraints on the properties of warm dark matter using the satellite
  galaxies of the Milky Way}},
  \href{https://arxiv.org/abs/2011.08865}{{\ttfamily 2011.08865}}.

\bibitem{Enzi:2020ieg}
W.~Enzi et~al., \emph{{Joint constraints on thermal relic dark matter from a
  selection of astrophysical probes}},
  \href{https://arxiv.org/abs/2010.13802}{{\ttfamily 2010.13802}}.

\bibitem{Wechsler:2018pic}
R.~H. Wechsler and J.~L. Tinker, \emph{{The Connection between Galaxies and
  their Dark Matter Halos}},
  \href{https://doi.org/10.1146/annurev-astro-081817-051756}{\emph{Ann. Rev.
  Astron. Astrophys.} {\bfseries 56} (2018) 435--487},
  [\href{https://arxiv.org/abs/1804.03097}{{\ttfamily 1804.03097}}].

\bibitem{Banik:2018pjp}
N.~Banik, G.~Bertone, J.~Bovy and N.~Bozorgnia, \emph{{Probing the nature of
  dark matter particles with stellar streams}},
  \href{https://doi.org/10.1088/1475-7516/2018/07/061}{\emph{JCAP} {\bfseries
  07} (2018) 061}, [\href{https://arxiv.org/abs/1804.04384}{{\ttfamily
  1804.04384}}].

\bibitem{Bonaca:2018fek}
A.~Bonaca, D.~W. Hogg, A.~M. Price-Whelan and C.~Conroy, \emph{{The Spur and
  the Gap in GD-1: Dynamical evidence for a dark substructure in the Milky Way
  halo}},  \href{https://arxiv.org/abs/1811.03631}{{\ttfamily 1811.03631}}.

\bibitem{Banik:2019smi}
N.~Banik, J.~Bovy, G.~Bertone, D.~Erkal and T.~J.~L. de~Boer, \emph{{Novel
  constraints on the particle nature of dark matter from stellar streams}},
  \href{https://arxiv.org/abs/1911.02663}{{\ttfamily 1911.02663}}.

\bibitem{Dalal:2020mjw}
N.~Dalal, J.~Bovy, L.~Hui and X.~Li, \emph{{Don't cross the streams: caustics
  from Fuzzy Dark Matter}},
  \href{https://doi.org/10.1088/1475-7516/2021/03/076}{\emph{JCAP} {\bfseries
  03} (2021) 076}, [\href{https://arxiv.org/abs/2011.13141}{{\ttfamily
  2011.13141}}].

\bibitem{Birrer:2017rpp}
S.~Birrer, A.~Amara and A.~Refregier, \emph{{Lensing substructure
  quantification in RXJ1131-1231: A 2 keV lower bound on dark matter thermal
  relic mass}},
  \href{https://doi.org/10.1088/1475-7516/2017/05/037}{\emph{JCAP} {\bfseries
  05} (2017) 037}, [\href{https://arxiv.org/abs/1702.00009}{{\ttfamily
  1702.00009}}].

\bibitem{Vegetti:2018dly}
S.~Vegetti, G.~Despali, M.~R. Lovell and W.~Enzi, \emph{{Constraining sterile
  neutrino cosmologies with strong gravitational lensing observations at
  redshift z \ensuremath{\sim} 0.2}},
  \href{https://doi.org/10.1093/mnras/sty2393}{\emph{Mon. Not. Roy. Astron.
  Soc.} {\bfseries 481} (2018) 3661--3669},
  [\href{https://arxiv.org/abs/1801.01505}{{\ttfamily 1801.01505}}].

\bibitem{Ritondale:2018cvp}
E.~Ritondale, S.~Vegetti, G.~Despali, M.~W. Auger, L.~V.~E. Koopmans and J.~P.
  McKean, \emph{{Low-mass halo perturbations in strong gravitational lenses at
  redshift z \ensuremath{\sim} 0.5 are consistent with CDM}},
  \href{https://doi.org/10.1093/mnras/stz464}{\emph{Mon. Not. Roy. Astron.
  Soc.} {\bfseries 485} (2019) 2179--2193},
  [\href{https://arxiv.org/abs/1811.03627}{{\ttfamily 1811.03627}}].

\bibitem{gilman2019probing}
D.~Gilman, S.~Birrer, T.~Treu, A.~Nierenberg and A.~Benson, \emph{Probing dark
  matter structure down to 107 solar masses: flux ratio statistics in
  gravitational lenses with line-of-sight haloes}, {\emph{Monthly Notices of
  the Royal Astronomical Society} {\bfseries 487} (2019) 5721--5738}.

\bibitem{Hsueh:2019ynk}
J.-W. Hsueh, W.~Enzi, S.~Vegetti, M.~Auger, C.~D. Fassnacht, G.~Despali et~al.,
  \emph{{SHARP \textendash{} VII. New constraints on the dark matter
  free-streaming properties and substructure abundance from gravitationally
  lensed quasars}}, \href{https://doi.org/10.1093/mnras/stz3177}{\emph{Mon.
  Not. Roy. Astron. Soc.} {\bfseries 492} (2020) 3047--3059},
  [\href{https://arxiv.org/abs/1905.04182}{{\ttfamily 1905.04182}}].

\bibitem{gilman2020warm}
D.~Gilman, S.~Birrer, A.~Nierenberg, T.~Treu, X.~Du and A.~Benson, \emph{Warm
  dark matter chills out: constraints on the halo mass function and the
  free-streaming length of dark matter with eight quadruple-image strong
  gravitational lenses}, {\emph{Monthly Notices of the Royal Astronomical
  Society} {\bfseries 491} (2020) 6077--6101}.

\bibitem{Nadler:2021dft}
E.~O. Nadler, S.~Birrer, D.~Gilman, R.~H. Wechsler, X.~Du, A.~Benson et~al.,
  \emph{{Dark Matter Constraints from a Unified Analysis of Strong
  Gravitational Lenses and Milky Way Satellite Galaxies}},
  \href{https://arxiv.org/abs/2101.07810}{{\ttfamily 2101.07810}}.

\bibitem{Gilman:2021sdr}
D.~Gilman, J.~Bovy, T.~Treu, A.~Nierenberg, S.~Birrer, A.~Benson et~al.,
  \emph{{Strong lensing signatures of self-interacting dark matter in low-mass
  halos}},  \href{https://arxiv.org/abs/2105.05259}{{\ttfamily 2105.05259}}.

\bibitem{hinshaw2013nine}
G.~Hinshaw, D.~Larson, E.~Komatsu, D.~N. Spergel, C.~Bennett, J.~Dunkley
  et~al., \emph{Nine-year wilkinson microwave anisotropy probe (wmap)
  observations: cosmological parameter results}, {\emph{The Astrophysical
  Journal Supplement Series} {\bfseries 208} (2013) 19}.

\bibitem{Bringmann:2006mu}
T.~Bringmann and S.~Hofmann, \emph{{Thermal decoupling of WIMPs from first
  principles}},
  \href{https://doi.org/10.1088/1475-7516/2007/04/016}{\emph{JCAP} {\bfseries
  04} (2007) 016}, [\href{https://arxiv.org/abs/hep-ph/0612238}{{\ttfamily
  hep-ph/0612238}}].

\bibitem{kasahara2009neutralino}
J.~Kasahara, \emph{Neutralino dark matter: the mass of the smallest halo and
  the golden region}.
\newblock 2009.

\bibitem{Binder:2016pnr}
T.~Binder, L.~Covi, A.~Kamada, H.~Murayama, T.~Takahashi and N.~Yoshida,
  \emph{{Matter Power Spectrum in Hidden Neutrino Interacting Dark Matter
  Models: A Closer Look at the Collision Term}},
  \href{https://doi.org/10.1088/1475-7516/2016/11/043}{\emph{JCAP} {\bfseries
  11} (2016) 043}, [\href{https://arxiv.org/abs/1602.07624}{{\ttfamily
  1602.07624}}].

\bibitem{Gondolo:2012vh}
P.~Gondolo, J.~Hisano and K.~Kadota, \emph{{The Effect of quark interactions on
  dark matter kinetic decoupling and the mass of the smallest dark halos}},
  \href{https://doi.org/10.1103/PhysRevD.86.083523}{\emph{Phys. Rev. D}
  {\bfseries 86} (2012) 083523},
  [\href{https://arxiv.org/abs/1205.1914}{{\ttfamily 1205.1914}}].

\bibitem{Bertoni:2014mva}
B.~Bertoni, S.~Ipek, D.~McKeen and A.~E. Nelson, \emph{{Constraints and
  consequences of reducing small scale structure via large dark matter-neutrino
  interactions}}, \href{https://doi.org/10.1007/JHEP04(2015)170}{\emph{JHEP}
  {\bfseries 04} (2015) 170},
  [\href{https://arxiv.org/abs/1412.3113}{{\ttfamily 1412.3113}}].

\bibitem{Boehm:2014vja}
C.~Boehm, J.~A. Schewtschenko, R.~J. Wilkinson, C.~M. Baugh and S.~Pascoli,
  \emph{{Using the Milky Way satellites to study interactions between cold dark
  matter and radiation}},
  \href{https://doi.org/10.1093/mnrasl/slu115}{\emph{Mon. Not. Roy. Astron.
  Soc.} {\bfseries 445} (2014) L31--L35},
  [\href{https://arxiv.org/abs/1404.7012}{{\ttfamily 1404.7012}}].

\bibitem{Gondolo:2016mrz}
P.~Gondolo and K.~Kadota, \emph{{Late Kinetic Decoupling of Light Magnetic
  Dipole Dark Matter}},
  \href{https://doi.org/10.1088/1475-7516/2016/06/012}{\emph{JCAP} {\bfseries
  06} (2016) 012}, [\href{https://arxiv.org/abs/1603.05783}{{\ttfamily
  1603.05783}}].

\bibitem{Chu:2018qrm}
X.~Chu, J.~Pradler and L.~Semmelrock, \emph{{Light dark states with
  electromagnetic form factors}},
  \href{https://doi.org/10.1103/PhysRevD.99.015040}{\emph{Phys. Rev. D}
  {\bfseries 99} (2019) 015040},
  [\href{https://arxiv.org/abs/1811.04095}{{\ttfamily 1811.04095}}].

\bibitem{Angloher:2002in}
G.~Angloher et~al., \emph{{Limits on WIMP dark matter using sapphire cryogenic
  detectors}},
  \href{https://doi.org/10.1016/S0927-6505(02)00111-1}{\emph{Astropart. Phys.}
  {\bfseries 18} (2002) 43--55}.

\bibitem{Akerib:2003px}
{\scshape CDMS} collaboration, D.~S. Akerib et~al., \emph{{New results from the
  cryogenic dark matter search experiment}},
  \href{https://doi.org/10.1103/PhysRevD.68.082002}{\emph{Phys. Rev. D}
  {\bfseries 68} (2003) 082002},
  [\href{https://arxiv.org/abs/hep-ex/0306001}{{\ttfamily hep-ex/0306001}}].

\bibitem{Abdelhameed:2019hmk}
{\scshape CRESST} collaboration, A.~H. Abdelhameed et~al., \emph{{First results
  from the CRESST-III low-mass dark matter program}},
  \href{https://doi.org/10.1103/PhysRevD.100.102002}{\emph{Phys. Rev. D}
  {\bfseries 100} (2019) 102002},
  [\href{https://arxiv.org/abs/1904.00498}{{\ttfamily 1904.00498}}].

\bibitem{Abdelhameed:2019mac}
{\scshape CRESST} collaboration, A.~H. Abdelhameed et~al., \emph{{Description
  of CRESST-III Data}},  \href{https://arxiv.org/abs/1905.07335}{{\ttfamily
  1905.07335}}.

\bibitem{Angloher:2017sxg}
{\scshape CRESST} collaboration, G.~Angloher et~al., \emph{{Results on
  MeV-scale dark matter from a gram-scale cryogenic calorimeter operated above
  ground}}, \href{https://doi.org/10.1140/epjc/s10052-017-5223-9}{\emph{Eur.
  Phys. J. C} {\bfseries 77} (2017) 637},
  [\href{https://arxiv.org/abs/1707.06749}{{\ttfamily 1707.06749}}].

\bibitem{Aprile:2017iyp}
{\scshape XENON} collaboration, E.~Aprile et~al., \emph{{First Dark Matter
  Search Results from the XENON1T Experiment}},
  \href{https://doi.org/10.1103/PhysRevLett.119.181301}{\emph{Phys. Rev. Lett.}
  {\bfseries 119} (2017) 181301},
  [\href{https://arxiv.org/abs/1705.06655}{{\ttfamily 1705.06655}}].

\bibitem{Armengaud:2019kfj}
{\scshape EDELWEISS} collaboration, E.~Armengaud et~al., \emph{{Searching for
  low-mass dark matter particles with a massive Ge bolometer operated
  above-ground}}, \href{https://doi.org/10.1103/PhysRevD.99.082003}{\emph{Phys.
  Rev. D} {\bfseries 99} (2019) 082003},
  [\href{https://arxiv.org/abs/1901.03588}{{\ttfamily 1901.03588}}].

\bibitem{Essig:2012yx}
R.~Essig, A.~Manalaysay, J.~Mardon, P.~Sorensen and T.~Volansky, \emph{{First
  Direct Detection Limits on sub-GeV Dark Matter from XENON10}},
  \href{https://doi.org/10.1103/PhysRevLett.109.021301}{\emph{Phys. Rev. Lett.}
  {\bfseries 109} (2012) 021301},
  [\href{https://arxiv.org/abs/1206.2644}{{\ttfamily 1206.2644}}].

\bibitem{Essig:2017kqs}
R.~Essig, T.~Volansky and T.-T. Yu, \emph{{New Constraints and Prospects for
  sub-GeV Dark Matter Scattering off Electrons in Xenon}},
  \href{https://doi.org/10.1103/PhysRevD.96.043017}{\emph{Phys. Rev. D}
  {\bfseries 96} (2017) 043017},
  [\href{https://arxiv.org/abs/1703.00910}{{\ttfamily 1703.00910}}].

\bibitem{Angle:2011th}
{\scshape XENON10} collaboration, J.~Angle et~al., \emph{{A search for light
  dark matter in XENON10 data}},
  \href{https://doi.org/10.1103/PhysRevLett.107.051301}{\emph{Phys. Rev. Lett.}
  {\bfseries 107} (2011) 051301},
  [\href{https://arxiv.org/abs/1104.3088}{{\ttfamily 1104.3088}}].

\bibitem{Aprile:2016wwo}
{\scshape XENON} collaboration, E.~Aprile et~al., \emph{{Low-mass dark matter
  search using ionization signals in XENON100}},
  \href{https://doi.org/10.1103/PhysRevD.94.092001}{\emph{Phys. Rev. D}
  {\bfseries 94} (2016) 092001},
  [\href{https://arxiv.org/abs/1605.06262}{{\ttfamily 1605.06262}}].

\bibitem{Agnes:2018oej}
{\scshape DarkSide} collaboration, P.~Agnes et~al., \emph{{Constraints on
  Sub-GeV Dark-Matter\textendash{}Electron Scattering from the DarkSide-50
  Experiment}},
  \href{https://doi.org/10.1103/PhysRevLett.121.111303}{\emph{Phys. Rev. Lett.}
  {\bfseries 121} (2018) 111303},
  [\href{https://arxiv.org/abs/1802.06998}{{\ttfamily 1802.06998}}].

\bibitem{Agnese:2018col}
{\scshape SuperCDMS} collaboration, R.~Agnese et~al., \emph{{First Dark Matter
  Constraints from a SuperCDMS Single-Charge Sensitive Detector}},
  \href{https://doi.org/10.1103/PhysRevLett.121.051301}{\emph{Phys. Rev. Lett.}
  {\bfseries 121} (2018) 051301},
  [\href{https://arxiv.org/abs/1804.10697}{{\ttfamily 1804.10697}}].

\bibitem{Crisler:2018gci}
{\scshape SENSEI} collaboration, M.~Crisler, R.~Essig, J.~Estrada,
  G.~Fernandez, J.~Tiffenberg, M.~Sofo~haro et~al., \emph{{SENSEI: First
  Direct-Detection Constraints on sub-GeV Dark Matter from a Surface Run}},
  \href{https://doi.org/10.1103/PhysRevLett.121.061803}{\emph{Phys. Rev. Lett.}
  {\bfseries 121} (2018) 061803},
  [\href{https://arxiv.org/abs/1804.00088}{{\ttfamily 1804.00088}}].

\bibitem{Abramoff:2019dfb}
{\scshape SENSEI} collaboration, O.~Abramoff et~al., \emph{{SENSEI:
  Direct-Detection Constraints on Sub-GeV Dark Matter from a Shallow
  Underground Run Using a Prototype Skipper-CCD}},
  \href{https://doi.org/10.1103/PhysRevLett.122.161801}{\emph{Phys. Rev. Lett.}
  {\bfseries 122} (2019) 161801},
  [\href{https://arxiv.org/abs/1901.10478}{{\ttfamily 1901.10478}}].

\bibitem{Barak:2020fql}
{\scshape SENSEI} collaboration, L.~Barak et~al., \emph{{SENSEI:
  Direct-Detection Results on sub-GeV Dark Matter from a New Skipper-CCD}},
  \href{https://doi.org/10.1103/PhysRevLett.125.171802}{\emph{Phys. Rev. Lett.}
  {\bfseries 125} (2020) 171802},
  [\href{https://arxiv.org/abs/2004.11378}{{\ttfamily 2004.11378}}].

\bibitem{Arina:2020mxo}
C.~Arina, A.~Cheek, K.~Mimasu and L.~Pagani, \emph{{Light and Darkness:
  consistently coupling dark matter to photons via effective operators}},
  \href{https://doi.org/10.1140/epjc/s10052-021-09010-1}{\emph{Eur. Phys. J. C}
  {\bfseries 81} (2021) 223},
  [\href{https://arxiv.org/abs/2005.12789}{{\ttfamily 2005.12789}}].

\bibitem{McCammon:2002gb}
D.~McCammon et~al., \emph{{A High spectral resolution observation of the soft
  x-ray diffuse background with thermal detectors}},
  \href{https://doi.org/10.1086/341727}{\emph{Astrophys. J.} {\bfseries 576}
  (2002) 188--203}, [\href{https://arxiv.org/abs/astro-ph/0205012}{{\ttfamily
  astro-ph/0205012}}].

\bibitem{Erickcek:2007jv}
A.~L. Erickcek, P.~J. Steinhardt, D.~McCammon and P.~C. McGuire,
  \emph{{Constraints on the Interactions between Dark Matter and Baryons from
  the X-ray Quantum Calorimetry Experiment}},
  \href{https://doi.org/10.1103/PhysRevD.76.042007}{\emph{Phys. Rev. D}
  {\bfseries 76} (2007) 042007},
  [\href{https://arxiv.org/abs/0704.0794}{{\ttfamily 0704.0794}}].

\bibitem{Mahdawi:2018euy}
M.~S. Mahdawi and G.~R. Farrar, \emph{{Constraints on Dark Matter with a
  moderately large and velocity-dependent DM-nucleon cross-section}},
  \href{https://doi.org/10.1088/1475-7516/2018/10/007}{\emph{JCAP} {\bfseries
  10} (2018) 007}, [\href{https://arxiv.org/abs/1804.03073}{{\ttfamily
  1804.03073}}].

\bibitem{Yin:2018yjn}
W.~Yin, \emph{{Highly-boosted dark matter and cutoff for cosmic-ray neutrinos
  through neutrino portal}},
  \href{https://doi.org/10.1051/epjconf/201920804003}{\emph{EPJ Web Conf.}
  {\bfseries 208} (2019) 04003},
  [\href{https://arxiv.org/abs/1809.08610}{{\ttfamily 1809.08610}}].

\bibitem{Bringmann:2018cvk}
T.~Bringmann and M.~Pospelov, \emph{{Novel direct detection constraints on
  light dark matter}},
  \href{https://doi.org/10.1103/PhysRevLett.122.171801}{\emph{Phys. Rev. Lett.}
  {\bfseries 122} (2019) 171801},
  [\href{https://arxiv.org/abs/1810.10543}{{\ttfamily 1810.10543}}].

\bibitem{Ema:2018bih}
Y.~Ema, F.~Sala and R.~Sato, \emph{{Light Dark Matter at Neutrino
  Experiments}},
  \href{https://doi.org/10.1103/PhysRevLett.122.181802}{\emph{Phys. Rev. Lett.}
  {\bfseries 122} (2019) 181802},
  [\href{https://arxiv.org/abs/1811.00520}{{\ttfamily 1811.00520}}].

\bibitem{Cappiello:2019qsw}
C.~Cappiello and J.~F. Beacom, \emph{{Strong New Limits on Light Dark Matter
  from Neutrino Experiments}},
  \href{https://doi.org/10.1103/PhysRevD.100.103011}{\emph{Phys. Rev. D}
  {\bfseries 100} (2019) 103011},
  [\href{https://arxiv.org/abs/1906.11283}{{\ttfamily 1906.11283}}].

\bibitem{Ema:2020ulo}
Y.~Ema, F.~Sala and R.~Sato, \emph{{Neutrino experiments probe hadrophilic
  light dark matter}},
  \href{https://doi.org/10.21468/SciPostPhys.10.3.072}{\emph{SciPost Phys.}
  {\bfseries 10} (2021) 072},
  [\href{https://arxiv.org/abs/2011.01939}{{\ttfamily 2011.01939}}].

\bibitem{An:2017ojc}
H.~An, M.~Pospelov, J.~Pradler and A.~Ritz, \emph{{Directly Detecting MeV-scale
  Dark Matter via Solar Reflection}},
  \href{https://doi.org/10.1103/PhysRevLett.120.141801}{\emph{Phys. Rev. Lett.}
  {\bfseries 120} (2018) 141801},
  [\href{https://arxiv.org/abs/1708.03642}{{\ttfamily 1708.03642}}].

\bibitem{Emken:2021lgc}
T.~Emken, \emph{{Solar reflection of light dark matter with heavy mediators}},
  \href{https://arxiv.org/abs/2102.12483}{{\ttfamily 2102.12483}}.

\bibitem{Fixsen:1996nj}
D.~J. Fixsen, E.~S. Cheng, J.~M. Gales, J.~C. Mather, R.~A. Shafer and E.~L.
  Wright, \emph{{The Cosmic Microwave Background spectrum from the full COBE
  FIRAS data set}}, \href{https://doi.org/10.1086/178173}{\emph{Astrophys. J.}
  {\bfseries 473} (1996) 576},
  [\href{https://arxiv.org/abs/astro-ph/9605054}{{\ttfamily
  astro-ph/9605054}}].

\bibitem{Ali-Haimoud:2015pwa}
Y.~Ali-Ha\"\i{}moud, J.~Chluba and M.~Kamionkowski, \emph{{Constraints on Dark
  Matter Interactions with Standard Model Particles from Cosmic Microwave
  Background Spectral Distortions}},
  \href{https://doi.org/10.1103/PhysRevLett.115.071304}{\emph{Phys. Rev. Lett.}
  {\bfseries 115} (2015) 071304},
  [\href{https://arxiv.org/abs/1506.04745}{{\ttfamily 1506.04745}}].

\bibitem{Digman:2019wdm}
M.~C. Digman, C.~V. Cappiello, J.~F. Beacom, C.~M. Hirata and A.~H.~G. Peter,
  \emph{{Not as big as a barn: Upper bounds on dark matter-nucleus cross
  sections}}, \href{https://doi.org/10.1103/PhysRevD.100.063013}{\emph{Phys.
  Rev. D} {\bfseries 100} (2019) 063013},
  [\href{https://arxiv.org/abs/1907.10618}{{\ttfamily 1907.10618}}].

\bibitem{Cappiello:2020lbk}
C.~V. Cappiello, J.~I. Collar and J.~F. Beacom, \emph{{New experimental
  constraints in a new landscape for composite dark matter}},
  \href{https://doi.org/10.1103/PhysRevD.103.023019}{\emph{Phys. Rev. D}
  {\bfseries 103} (2021) 023019},
  [\href{https://arxiv.org/abs/2008.10646}{{\ttfamily 2008.10646}}].

\bibitem{Rich:1987st}
J.~Rich, R.~Rocchia and M.~Spiro, \emph{{A Search for Strongly Interacting Dark
  Matter}}, \href{https://doi.org/10.1016/0370-2693(87)90788-X}{\emph{Phys.
  Lett. B} {\bfseries 194} (1987) 173}.

\bibitem{Emken:2019tni}
T.~Emken, R.~Essig, C.~Kouvaris and M.~Sholapurkar, \emph{{Direct Detection of
  Strongly Interacting Sub-GeV Dark Matter via Electron Recoils}},
  \href{https://doi.org/10.1088/1475-7516/2019/09/070}{\emph{JCAP} {\bfseries
  09} (2019) 070}, [\href{https://arxiv.org/abs/1905.06348}{{\ttfamily
  1905.06348}}].

\bibitem{Sabti:2019mhn}
N.~Sabti, J.~Alvey, M.~Escudero, M.~Fairbairn and D.~Blas, \emph{{Refined
  Bounds on MeV-scale Thermal Dark Sectors from BBN and the CMB}},
  \href{https://doi.org/10.1088/1475-7516/2020/01/004}{\emph{JCAP} {\bfseries
  01} (2020) 004}, [\href{https://arxiv.org/abs/1910.01649}{{\ttfamily
  1910.01649}}].

\bibitem{Chu:2019rok}
X.~Chu, J.-L. Kuo, J.~Pradler and L.~Semmelrock, \emph{{Stellar probes of dark
  sector-photon interactions}},
  \href{https://doi.org/10.1103/PhysRevD.100.083002}{\emph{Phys. Rev. D}
  {\bfseries 100} (2019) 083002},
  [\href{https://arxiv.org/abs/1908.00553}{{\ttfamily 1908.00553}}].

\bibitem{Creque-Sarbinowski:2019mcm}
C.~Creque-Sarbinowski, L.~Ji, E.~D. Kovetz and M.~Kamionkowski, \emph{{Direct
  millicharged dark matter cannot explain the EDGES signal}},
  \href{https://doi.org/10.1103/PhysRevD.100.023528}{\emph{Phys. Rev. D}
  {\bfseries 100} (2019) 023528},
  [\href{https://arxiv.org/abs/1903.09154}{{\ttfamily 1903.09154}}].

\bibitem{Munoz:2018pzp}
J.~B. Mu\~noz and A.~Loeb, \emph{{A small amount of mini-charged dark matter
  could cool the baryons in the early Universe}},
  \href{https://doi.org/10.1038/s41586-018-0151-x}{\emph{Nature} {\bfseries
  557} (2018) 684}, [\href{https://arxiv.org/abs/1802.10094}{{\ttfamily
  1802.10094}}].

\bibitem{Vogel:2013raa}
H.~Vogel and J.~Redondo, \emph{{Dark Radiation constraints on minicharged
  particles in models with a hidden photon}},
  \href{https://doi.org/10.1088/1475-7516/2014/02/029}{\emph{JCAP} {\bfseries
  02} (2014) 029}, [\href{https://arxiv.org/abs/1311.2600}{{\ttfamily
  1311.2600}}].

\bibitem{Chang:2018rso}
J.~H. Chang, R.~Essig and S.~D. McDermott, \emph{{Supernova 1987A Constraints
  on Sub-GeV Dark Sectors, Millicharged Particles, the QCD Axion, and an
  Axion-like Particle}},
  \href{https://doi.org/10.1007/JHEP09(2018)051}{\emph{JHEP} {\bfseries 09}
  (2018) 051}, [\href{https://arxiv.org/abs/1803.00993}{{\ttfamily
  1803.00993}}].

\bibitem{Fortin:2011hv}
J.-F. Fortin and T.~M.~P. Tait, \emph{{Collider Constraints on
  Dipole-Interacting Dark Matter}},
  \href{https://doi.org/10.1103/PhysRevD.85.063506}{\emph{Phys. Rev. D}
  {\bfseries 85} (2012) 063506},
  [\href{https://arxiv.org/abs/1103.3289}{{\ttfamily 1103.3289}}].

\bibitem{Prinz:1998ua}
A.~A. Prinz et~al., \emph{{Search for millicharged particles at SLAC}},
  \href{https://doi.org/10.1103/PhysRevLett.81.1175}{\emph{Phys. Rev. Lett.}
  {\bfseries 81} (1998) 1175--1178},
  [\href{https://arxiv.org/abs/hep-ex/9804008}{{\ttfamily hep-ex/9804008}}].

\bibitem{Acciarri:2019jly}
{\scshape ArgoNeuT} collaboration, R.~Acciarri et~al., \emph{{Improved Limits
  on Millicharged Particles Using the ArgoNeuT Experiment at Fermilab}},
  \href{https://doi.org/10.1103/PhysRevLett.124.131801}{\emph{Phys. Rev. Lett.}
  {\bfseries 124} (2020) 131801},
  [\href{https://arxiv.org/abs/1911.07996}{{\ttfamily 1911.07996}}].

\bibitem{Liang:2019zkb}
J.~Liang, Z.~Liu, Y.~Ma and Y.~Zhang, \emph{{Millicharged particles at electron
  colliders}}, \href{https://doi.org/10.1103/PhysRevD.102.015002}{\emph{Phys.
  Rev. D} {\bfseries 102} (2020) 015002},
  [\href{https://arxiv.org/abs/1909.06847}{{\ttfamily 1909.06847}}].

\bibitem{Plestid:2020kdm}
R.~Plestid, V.~Takhistov, Y.-D. Tsai, T.~Bringmann, A.~Kusenko and M.~Pospelov,
  \emph{{New Constraints on Millicharged Particles from Cosmic-ray
  Production}}, \href{https://doi.org/10.1103/PhysRevD.102.115032}{\emph{Phys.
  Rev. D} {\bfseries 102} (2020) 115032},
  [\href{https://arxiv.org/abs/2002.11732}{{\ttfamily 2002.11732}}].

\bibitem{Magill:2018tbb}
G.~Magill, R.~Plestid, M.~Pospelov and Y.-D. Tsai, \emph{{Millicharged
  particles in neutrino experiments}},
  \href{https://doi.org/10.1103/PhysRevLett.122.071801}{\emph{Phys. Rev. Lett.}
  {\bfseries 122} (2019) 071801},
  [\href{https://arxiv.org/abs/1806.03310}{{\ttfamily 1806.03310}}].

\bibitem{Davidson:1991si}
S.~Davidson, B.~Campbell and D.~C. Bailey, \emph{{Limits on particles of small
  electric charge}},
  \href{https://doi.org/10.1103/PhysRevD.43.2314}{\emph{Phys. Rev. D}
  {\bfseries 43} (1991) 2314--2321}.

\bibitem{Battaglieri:2017aum}
M.~Battaglieri et~al., \emph{{US Cosmic Visions: New Ideas in Dark Matter 2017:
  Community Report}},  in \emph{{U.S. Cosmic Visions: New Ideas in Dark
  Matter}}, 7, 2017, \href{https://arxiv.org/abs/1707.04591}{{\ttfamily
  1707.04591}}.

\bibitem{Hambye:2019dwd}
T.~Hambye, M.~H.~G. Tytgat, J.~Vandecasteele and L.~Vanderheyden, \emph{{Dark
  matter from dark photons: a taxonomy of dark matter production}},
  \href{https://doi.org/10.1103/PhysRevD.100.095018}{\emph{Phys. Rev. D}
  {\bfseries 100} (2019) 095018},
  [\href{https://arxiv.org/abs/1908.09864}{{\ttfamily 1908.09864}}].

\bibitem{Boehm:2013jpa}
C.~Boehm, M.~J. Dolan and C.~McCabe, \emph{{A Lower Bound on the Mass of Cold
  Thermal Dark Matter from Planck}},
  \href{https://doi.org/10.1088/1475-7516/2013/08/041}{\emph{JCAP} {\bfseries
  08} (2013) 041}, [\href{https://arxiv.org/abs/1303.6270}{{\ttfamily
  1303.6270}}].

\bibitem{Nollett:2014lwa}
K.~M. Nollett and G.~Steigman, \emph{{BBN And The CMB Constrain Neutrino
  Coupled Light WIMPs}},
  \href{https://doi.org/10.1103/PhysRevD.91.083505}{\emph{Phys. Rev. D}
  {\bfseries 91} (2015) 083505},
  [\href{https://arxiv.org/abs/1411.6005}{{\ttfamily 1411.6005}}].

\bibitem{Chang:2019xva}
J.~H. Chang, R.~Essig and A.~Reinert, \emph{{Light(ly)-coupled Dark Matter in
  the keV Range: Freeze-In and Constraints}},
  \href{https://doi.org/10.1007/JHEP03(2021)141}{\emph{JHEP} {\bfseries 03}
  (2021) 141}, [\href{https://arxiv.org/abs/1911.03389}{{\ttfamily
  1911.03389}}].

\bibitem{Lambiase:2021xcj}
G.~Lambiase, S.~Mohanty, A.~Nautiyal and S.~Rao, \emph{{Constraints on
  electromagnetic form factors of sub-GeV dark matter from the Cosmic Microwave
  Background anisotropy}},  \href{https://arxiv.org/abs/2102.04840}{{\ttfamily
  2102.04840}}.

\bibitem{Daci:2015hca}
N.~Daci, I.~De~Bruyn, S.~Lowette, M.~H.~G. Tytgat and B.~Zaldivar,
  \emph{{Simplified SIMPs and the LHC}},
  \href{https://doi.org/10.1007/JHEP11(2015)108}{\emph{JHEP} {\bfseries 11}
  (2015) 108}, [\href{https://arxiv.org/abs/1503.05505}{{\ttfamily
  1503.05505}}].

\bibitem{Hambye:2018dpi}
T.~Hambye, M.~H.~G. Tytgat, J.~Vandecasteele and L.~Vanderheyden, \emph{{Dark
  matter direct detection is testing freeze-in}},
  \href{https://doi.org/10.1103/PhysRevD.98.075017}{\emph{Phys. Rev. D}
  {\bfseries 98} (2018) 075017},
  [\href{https://arxiv.org/abs/1807.05022}{{\ttfamily 1807.05022}}].

\bibitem{HERA:2021noe}
{\scshape HERA} collaboration, Z.~Abdurashidova et~al., \emph{{HERA Phase I
  Limits on the Cosmic 21 cm Signal: Constraints on Astrophysics and Cosmology
  during the Epoch of Reionization}},
  \href{https://doi.org/10.3847/1538-4357/ac2ffc}{\emph{Astrophys. J.}
  {\bfseries 924} (2022) 51},
  [\href{https://arxiv.org/abs/2108.07282}{{\ttfamily 2108.07282}}].

\bibitem{Essig:2011nj}
R.~Essig, J.~Mardon and T.~Volansky, \emph{{Direct Detection of Sub-GeV Dark
  Matter}}, \href{https://doi.org/10.1103/PhysRevD.85.076007}{\emph{Phys. Rev.
  D} {\bfseries 85} (2012) 076007},
  [\href{https://arxiv.org/abs/1108.5383}{{\ttfamily 1108.5383}}].

\bibitem{Chu:2011be}
X.~Chu, T.~Hambye and M.~H.~G. Tytgat, \emph{{The Four Basic Ways of Creating
  Dark Matter Through a Portal}},
  \href{https://doi.org/10.1088/1475-7516/2012/05/034}{\emph{JCAP} {\bfseries
  05} (2012) 034}, [\href{https://arxiv.org/abs/1112.0493}{{\ttfamily
  1112.0493}}].

\bibitem{SimonsObservatory:2019qwx}
{\scshape Simons Observatory} collaboration, M.~H. Abitbol et~al., \emph{{The
  Simons Observatory: Astro2020 Decadal Project Whitepaper}}, {\emph{Bull. Am.
  Astron. Soc.} {\bfseries 51} (2019) 147},
  [\href{https://arxiv.org/abs/1907.08284}{{\ttfamily 1907.08284}}].

\bibitem{Abazajian:2019eic}
K.~Abazajian et~al., \emph{{CMB-S4 Science Case, Reference Design, and Project
  Plan}},  \href{https://arxiv.org/abs/1907.04473}{{\ttfamily 1907.04473}}.

\bibitem{Nguyen:2017zqu}
H.~N. Nguyen, N.~Sehgal and M.~Madhavacheril, \emph{{Measuring the Small-Scale
  Matter Power Spectrum with High-Resolution CMB Lensing}},
  \href{https://doi.org/10.1103/PhysRevD.99.023502}{\emph{Phys. Rev. D}
  {\bfseries 99} (2019) 023502},
  [\href{https://arxiv.org/abs/1710.03747}{{\ttfamily 1710.03747}}].

\bibitem{Sehgal:2019ewc}
N.~Sehgal et~al., \emph{{CMB-HD: An Ultra-Deep, High-Resolution Millimeter-Wave
  Survey Over Half the Sky}},
  \href{https://arxiv.org/abs/1906.10134}{{\ttfamily 1906.10134}}.

\bibitem{Sehgal:2020yja}
N.~Sehgal et~al., \emph{{CMB-HD: Astro2020 RFI Response}},
  \href{https://arxiv.org/abs/2002.12714}{{\ttfamily 2002.12714}}.

\bibitem{Sehgal:2019nmk}
N.~Sehgal et~al., \emph{{Science from an Ultra-Deep, High-Resolution
  Millimeter-Wave Survey}},  \href{https://arxiv.org/abs/1903.03263}{{\ttfamily
  1903.03263}}.

\bibitem{Munoz:2019hjh}
J.~B. Mu\~noz, C.~Dvorkin and F.-Y. Cyr-Racine, \emph{{Probing the Small-Scale
  Matter Power Spectrum with Large-Scale 21-cm Data}},
  \href{https://doi.org/10.1103/PhysRevD.101.063526}{\emph{Phys. Rev. D}
  {\bfseries 101} (2020) 063526},
  [\href{https://arxiv.org/abs/1911.11144}{{\ttfamily 1911.11144}}].

\bibitem{Drlica-Wagner:2019xan}
{\scshape LSST Dark Matter Group} collaboration, A.~Drlica-Wagner et~al.,
  \emph{{Probing the Fundamental Nature of Dark Matter with the Large Synoptic
  Survey Telescope}},  \href{https://arxiv.org/abs/1902.01055}{{\ttfamily
  1902.01055}}.

\bibitem{Birrer:2020snowmass}
``Strong lensing probes of dark matter.''
  \url{https://www.snowmass21.org/docs/files/summaries/CF/SNOWMASS21-CF3_CF7-TF8_TF9-CompF2_CompF0_Simon_Birrer-037.pdf}.

\bibitem{refId0}
{Gaia Collaboration}, {Helmi, A.}, {van Leeuwen, F.}, {McMillan, P. J.},
  {Massari, D.}, {Antoja, T.} et~al., \emph{Gaia data release 2 - kinematics of
  globular clusters and dwarf galaxies around the milky way},
  \href{https://doi.org/10.1051/0004-6361/201832698}{\emph{A\&A} {\bfseries
  616} (2018) A12}.

\bibitem{brown2021gaia}
A.~G. Brown, A.~Vallenari, T.~Prusti, J.~De~Bruijne, C.~Babusiaux, M.~Biermann
  et~al., \emph{Gaia early data release 3-summary of the contents and survey
  properties}, {\emph{Astronomy \& Astrophysics} {\bfseries 649} (2021) A1}.

\bibitem{Savage:2006qr}
C.~Savage, K.~Freese and P.~Gondolo, \emph{{Annual Modulation of Dark Matter in
  the Presence of Streams}},
  \href{https://doi.org/10.1103/PhysRevD.74.043531}{\emph{Phys. Rev. D}
  {\bfseries 74} (2006) 043531},
  [\href{https://arxiv.org/abs/astro-ph/0607121}{{\ttfamily
  astro-ph/0607121}}].

\bibitem{Freese:2012xd}
K.~Freese, M.~Lisanti and C.~Savage, \emph{{Colloquium: Annual modulation of
  dark matter}}, \href{https://doi.org/10.1103/RevModPhys.85.1561}{\emph{Rev.
  Mod. Phys.} {\bfseries 85} (2013) 1561--1581},
  [\href{https://arxiv.org/abs/1209.3339}{{\ttfamily 1209.3339}}].

\bibitem{Necib:2018iwb}
L.~Necib, M.~Lisanti and V.~Belokurov, \emph{{Inferred Evidence For Dark Matter
  Kinematic Substructure with SDSS-Gaia}},
  \href{https://arxiv.org/abs/1807.02519}{{\ttfamily 1807.02519}}.

\bibitem{OHare:2018trr}
C.~A.~J. O'Hare, C.~McCabe, N.~W. Evans, G.~Myeong and V.~Belokurov,
  \emph{{Dark matter hurricane: Measuring the S1 stream with dark matter
  detectors}}, \href{https://doi.org/10.1103/PhysRevD.98.103006}{\emph{Phys.
  Rev. D} {\bfseries 98} (2018) 103006},
  [\href{https://arxiv.org/abs/1807.09004}{{\ttfamily 1807.09004}}].

\bibitem{Wu:2019nhd}
Y.~Wu, K.~Freese, C.~Kelso, P.~Stengel and M.~Valluri, \emph{{Uncertainties in
  Direct Dark Matter Detection in Light of Gaia's Escape Velocity
  Measurements}},
  \href{https://doi.org/10.1088/1475-7516/2019/10/034}{\emph{JCAP} {\bfseries
  10} (2019) 034}, [\href{https://arxiv.org/abs/1904.04781}{{\ttfamily
  1904.04781}}].

\bibitem{Buckley:2019skk}
M.~R. Buckley, G.~Mohlabeng and C.~W. Murphy, \emph{{Direct Detection Anomalies
  in light of $Gaia$ Data}},
  \href{https://doi.org/10.1103/PhysRevD.100.055039}{\emph{Phys. Rev. D}
  {\bfseries 100} (2019) 055039},
  [\href{https://arxiv.org/abs/1905.05189}{{\ttfamily 1905.05189}}].

\bibitem{OHare:2019qxc}
C.~A.~J. O'Hare, N.~W. Evans, C.~McCabe, G.~Myeong and V.~Belokurov,
  \emph{{Velocity substructure from Gaia and direct searches for dark matter}},
  \href{https://doi.org/10.1103/PhysRevD.101.023006}{\emph{Phys. Rev. D}
  {\bfseries 101} (2020) 023006},
  [\href{https://arxiv.org/abs/1909.04684}{{\ttfamily 1909.04684}}].

\bibitem{Buch:2019aiw}
J.~Buch, J.~Fan and J.~S.~C. Leung, \emph{{Implications of the Gaia Sausage for
  Dark Matter Nuclear Interactions}},
  \href{https://doi.org/10.1103/PhysRevD.101.063026}{\emph{Phys. Rev. D}
  {\bfseries 101} (2020) 063026},
  [\href{https://arxiv.org/abs/1910.06356}{{\ttfamily 1910.06356}}].

\bibitem{Buch:2020xyt}
J.~Buch, M.~A. Buen-Abad, J.~Fan and J.~S.~C. Leung, \emph{{Dark Matter
  Substructure under the Electron Scattering Lamppost}},
  \href{https://doi.org/10.1103/PhysRevD.102.083010}{\emph{Phys. Rev. D}
  {\bfseries 102} (2020) 083010},
  [\href{https://arxiv.org/abs/2007.13750}{{\ttfamily 2007.13750}}].

\bibitem{Radick:2020qip}
A.~Radick, A.-M. Taki and T.-T. Yu, \emph{{Dependence of Dark Matter - Electron
  Scattering on the Galactic Dark Matter Velocity Distribution}},
  \href{https://doi.org/10.1088/1475-7516/2021/02/004}{\emph{JCAP} {\bfseries
  02} (2021) 004}, [\href{https://arxiv.org/abs/2011.02493}{{\ttfamily
  2011.02493}}].

\bibitem{Smith:2002dz}
{\scshape VIRGO Consortium} collaboration, R.~E. Smith, J.~A. Peacock,
  A.~Jenkins, S.~D.~M. White, C.~S. Frenk, F.~R. Pearce et~al., \emph{{Stable
  clustering, the halo model and nonlinear cosmological power spectra}},
  \href{https://doi.org/10.1046/j.1365-8711.2003.06503.x}{\emph{Mon. Not. Roy.
  Astron. Soc.} {\bfseries 341} (2003) 1311},
  [\href{https://arxiv.org/abs/astro-ph/0207664}{{\ttfamily
  astro-ph/0207664}}].

\bibitem{Bird:2011rb}
S.~Bird, M.~Viel and M.~G. Haehnelt, \emph{{Massive Neutrinos and the
  Non-linear Matter Power Spectrum}},
  \href{https://doi.org/10.1111/j.1365-2966.2011.20222.x}{\emph{Mon. Not. Roy.
  Astron. Soc.} {\bfseries 420} (2012) 2551--2561},
  [\href{https://arxiv.org/abs/1109.4416}{{\ttfamily 1109.4416}}].

\bibitem{Kovetz:2018zan}
E.~D. Kovetz, V.~Poulin, V.~Gluscevic, K.~K. Boddy, R.~Barkana and
  M.~Kamionkowski, \emph{{Tighter limits on dark matter explanations of the
  anomalous EDGES 21 cm signal}},
  \href{https://doi.org/10.1103/PhysRevD.98.103529}{\emph{Phys. Rev. D}
  {\bfseries 98} (2018) 103529},
  [\href{https://arxiv.org/abs/1807.11482}{{\ttfamily 1807.11482}}].

\bibitem{Braaten:1993jw}
E.~Braaten and D.~Segel, \emph{{Neutrino energy loss from the plasma process at
  all temperatures and densities}},
  \href{https://doi.org/10.1103/PhysRevD.48.1478}{\emph{Phys. Rev. D}
  {\bfseries 48} (1993) 1478--1491},
  [\href{https://arxiv.org/abs/hep-ph/9302213}{{\ttfamily hep-ph/9302213}}].

\bibitem{Raffelt:1996wa}
G.~G. Raffelt, \emph{{Stars as laboratories for fundamental physics}: {The
  astrophysics of neutrinos, axions, and other weakly interacting particles}}.
\newblock 5, 1996.

\end{thebibliography}\endgroup
\bibliographystyle{JHEP}

\end{document}